\documentclass[aps,prd,amsmath,twocolumn,groupedaddress,showpacs,showkeys,
preprintnumbers]{revtex4}
 
 \def\square{\hbox{\vrule\vbox{\hrule\phantom{o}\hrule}\vrule}}
 \renewcommand{\theequation}{\arabic{section}.\arabic{equation}}

 \usepackage[dvips]{graphicx}
 \usepackage{dcolumn}
 \usepackage{longtable}
\begin{document}

 \title{More examples of structure formation in the Lema\^{\i}tre-Tolman
model}

 \author{Andrzej Krasi\'{n}ski}
 \thanks{This research was supported by the Polish Research Committee
grant no 2 P03B 12 724}
 \affiliation{N. Copernicus Astronomical Center,
 Polish Academy of Sciences, \\
 Bartycka 18, 00 716 Warszawa, Poland}
 \email{akr@camk.edu.pl}

 \author{Charles Hellaby}
 \thanks{and by a grant from the South African National Research
Foundation}
 \affiliation{Department of Mathematics and Applied Mathematics, \\
 University of Cape Town, Rondebosch 7701, South Africa}
 \email{cwh@maths.uct.ac.za}

 \date {5/3/2003}

 \begin{abstract}
In continuing our earlier research, we find the formulae needed to determine 
the arbitrary functions in the Lema\^{\i}tre-Tolman model when the evolution 
proceeds from a given initial velocity distribution to a final state that is 
determined either by a density distribution or by a velocity distribution. In 
each case the initial and final distributions uniquely determine the L-T model 
that evolves between them, and the sign of the energy-function is determined by 
a simple inequality. We also show how the final density profile can be more 
accurately fitted to observational data than was done in our previous paper. We 
work out new numerical examples of the evolution: the creation of a galaxy 
cluster out of different velocity distributions, reflecting the current data on 
temperature anisotropies of CMB, the creation of the same out of different 
density distributions, and the creation of a void. The void in its present 
state is surrounded by a nonsingular wall of high density.

 \begin{center}

 {Short title:~~
 More examples of structure formation in LT
 }
 \\[4mm]

 {\it Phys. Rev. D, Submitted 5/3/2003, Accepted datedate} \\[1mm]

 \end{center}

 \end{abstract}



 \preprint{gr-qc/0303016}

 \maketitle

 \section{Scope}

 \setcounter{equation}{0}

In a previous paper \cite{KrHe2001}, which we shall call Paper I, we showed 
that one can uniquely define the Lema\^{\i}tre-Tolman cosmological model 
\cite{Lema1933, Tolm1934} by specifying an initial density profile (i.e. the 
mass-density as a function of the radial coordinate at an initial instant $t = 
t_1$) and a final density profile. The formulae defining the L-T functions 
$E(M)$ and $t_B(M)$ (where $M$ is the active gravitational mass, used here as a 
radial coordinate) are implicit but unique, and can be solved for $E$ and $t_B$ 
numerically. (For definitions of $E$ and $t_B$ see eqs. (\ref{LTds2}) and (\ref 
{Rtsq}).) We also worked out a numerical example in which a galaxy-cluster-like 
final profile was created out of an initial profile whose density amplitude and 
linear size were small.

   In the present paper, we develop that study for new elements: we show that
instead of a density distribution, one can specify a velocity distribution
(strictly speaking, this is $R_{,t}/M^{1/3}$ --- a measure of the velocity) at
either the initial instant or the final instant or both. We prove a theorem,
analogous to the one proven in paper I: given the initial and the final
profile, the L-T model that evolves between them is uniquely determined. We also
show how to adapt the initial and final density profiles to the astrophysical
data more precisely than it was done in paper I. We provide numerical examples
of L-T evolution between an initial profile (of density or velocity) consistent
with the implications of the CMB measurements, and a final profile that
corresponds either to a galaxy cluster or to a void.

   The paper is arranged as follows. We recall the basic properties of the L-T
model in sec. 2. In sec. 3 we describe how the final profile can be adapted to
observational data more exactly than in Paper I. In sec. 4 we find the implicit
formulae to define the L-T functions $E(M)$ and $t_B(M)$ when the initial state
is specified by a velocity distribution and the final state by a density
distribution. In sec. 5 we do the same for both states being specified by
velocity distributions. In sec. 6 we deduce the amplitude of the velocity
perturbation at $t_1$ allowed by observations of the CMB radiation. In sec. 7 we
specify our choices of density and velocity profiles for numerical
investigations. Secs. 8 and 9 contain the presentation of numerical results in
several figures. The results are summarized and interpreted in sec. 10, and
brief conclusions follow in sec. 11.

   The approach presented in this and in paper I is more suited to astrophysical
practice than the traditional approach, in which we first try to deduce the
initial state of the Universe from various kinds of data, and then proceed to
calculate the evolution of the cosmological model, in order to compare its final
state with the observations of the current state of the actual Universe. Our
approach allows one to make simultaneous use of the data on the initial and on
the final state of the Universe --- the real astronomical data is indeed such a
mixture.

   Naturally the L-T model can describe the actual astrophysical process of
structure formation only approximately. The obvious limitation is spherical
symmetry, in consequence of which we cannot take into account the rotation of
the objects formed. Thus, no matter how well we reproduce the profile of a
galaxy or cluster, the L-T model will continue to evolve, and the profile will
look quite different after $10^6$ years or less, whereas real galaxies and
clusters are fairly stable over $10^9$ years. However, we hope that the method
presented here will be a starting point for generalisations that will be done
once (and if) more general exact cosmological models are discovered.

 \section{Basic properties of the Lema\^{\i}tre-Tolman model.}

 \setcounter{equation}{0}

   The Lema\^{\i}tre-Tolman (L-T) model \cite{Lema1933, Tolm1934} is a spherically
symmetric nonstatic solution of the Einstein equations with a dust source. Its
metric is:
 \begin{equation} \label{LTds2}
 \text{d}s^2 = \text{d}t^2 - \frac {{R,_r}^2}{1 + 2E(r)}\text{d}r^2
 - R^2(t,r)(\text{d}\vartheta^2 + \sin^2\vartheta\text{d}\varphi^2),
 \end{equation}
 where $E(r)$ is an arbitrary function (arising as an integration
constant from the Einstein equations), $R,_r = \partial R/ \partial r$, and $R$
obeys
 \begin{equation}   \label{Rtsq}
{R_{,t}}^2 = 2E + 2M/R + \frac 13 \Lambda R^2,
 \end{equation}
 where $\Lambda$ is the cosmological constant. Eq.~(\ref{Rtsq}) is a
first integral of the Einstein equations, and $M(r)$ is another arbitrary
function that arises as an integration constant. The mass-density is:
 \begin{equation}   \label{rhoLT}
\kappa \rho = \frac {2M,_r}{R^2R,_r}, \qquad \text{where}\ \kappa = \frac {8\pi
G} {c^4}.
 \end{equation}
 See \cite{Kras1997} for an extensive list of properties and other work on this
model. In the following, we will assume $\Lambda = 0$. Then eq.~(\ref{Rtsq}) 
can be solved explicitly. The solutions are the following.

   Elliptic case%
 \footnote{
 More correctly, the elliptic, parabolic and hyperbolic conditions are
$E/M^{2/3} < 0$, $= 0$, and $> 0$ respectively, as both $E$ \& $M$ go to
zero at an origin.
 }%
 , $E < 0$:
 \begin{subequations}
   \label{EllEv}
 \begin{equation}
 R(t,r) = - \frac{M}{2E}(1 - \cos\eta),
 \end{equation}
 \begin{equation}
 \eta - \sin\eta = \frac{(-2E)^{3/2}}{M} (t - t_B(r)).
 \end{equation}
 \end{subequations}
 where $\eta$ is a parameter;

   Parabolic case, $E = 0$:
 \begin{equation}   \label{ParEv}
R(t,r) = \left[ \frac{9}{2} M (t - t_B(r))^2\right]^{1/3},
 \end{equation}

   Hyperbolic case, $E > 0$:
 \begin{subequations}
   \label{HypEv}
 \begin{equation}
 R(t,r) = \frac{M}{2E}(\cosh\eta - 1),
 \end{equation}
 \begin{equation}
 \sinh\eta - \eta = \frac{(2E)^{(3/2)}}{M} (t - t_B(r)),
 \end{equation}
 \end{subequations}
 where $t_B(r)$ is one more arbitrary function (the bang time). Note
that all the formulae given so far are covariant under arbitrary coordinate
transformations $r = g(r')$, and so $r$ can be chosen at will. This means one
of the three functions $E(r)$, $M(r)$ and $t_B(r)$ can be fixed at our
convenience by the appropriate choice of $g$.

   The Friedmann models are contained in the Lema\^{\i}tre-Tolman class as the
limit:
 \begin{equation}
t_B = \text{ const}, \qquad |E|^{3/2}/M = \text{ const},
 \end{equation}
 and one of the standard radial coordinates for the Friedmann model
results if, in addition, the coordinates in (\ref{EllEv}) -- (\ref{HypEv})
are chosen so that:
 \begin{equation}
M = M_0 r^3,
 \end{equation}
 where $M_0$ is an arbitrary constant.  This implies $E / r^2 =
\text{ const} := - k/2,\ k$ being the
 Robertson-Walker curvature index.

   It will be convenient in most of what follows to use $M(r)$ as the radial
coordinate (i.e. $r' = M(r)$) because, in the structure formation context,
one does not expect any ``necks" or ``bellies" where $M,_r = 0$, and so
$M(r)$ should be a strictly increasing function in the whole region under
consideration. Then:
 \begin{equation}   \label{rhoRM}
\kappa \rho = 2/(R^2R,_M) \equiv 6/(R^3),_M,
 \end{equation}
 from which we find
 \begin{equation}   \label{RcubedInt}
   R^3(M) - R_0^3 = \frac{3}{4 \pi} \int_{M_0}^{M} \frac{du}{\rho(u)},
 \end{equation}
 where $R_0$ and $M_0$ will commonly be zero.

   In the present paper we will apply the L-T model to problems of a similar kind
to that considered in Paper I: Connecting an initial state of the Universe
(defined by {\it either} a mass-density distribution {\it or} a velocity
distribution) to a final state (also defined by one of these distributions) by
an L-T evolution, and in particular to the formation of galaxy-cluster-like and
void-like objects out of initial perturbations of density or velocity that are
small in amplitude and in some cases small in mass compared to the final object.

 \section{Modelling the final density profile.}

 \setcounter{equation}{0}

We will now incorporate in our models the observational data on mass 
distribution in galaxy clusters in a more detailed way than in Paper I. 

The quantities of interest in the profile are the following: the maximal 
density (with the shapes we assume below and later, this will be at the center 
of the object), the radius of the object (assumed spherically symmetric), the 
mass of the object, the average density of the cosmic background, and the 
compensation radius (defined below).

   We define the following parameters.

 \begin{itemize}

 \item $M_m$ --- the mass of a galaxy cluster, out to some radius; to be
taken from astronomical tables.

 \item $R_m$ --- the radius within which the mass $M_m$ is contained; also to be
taken from astronomical tables. In fact there may be two or more $M_m$--$R_m$
pairs available for some clusters (see section
\ref{AbellClusterProfileChoices}).

\item $\rho(M_m) = \rho_b d$ --- this is the geometrical definition of the mass 
$M_m$ ($M = M_m$ at that value of $R = R_m$ at which $\rho$ equals a certain 
specified multiple of the background density).
 \footnote{One of the definitions of $R_m$ used in astronomy is: The size 
$R_m$ of a galaxy cluster is that radius, at which the {\it average} 
density within $R_m$ equals a specified multiple of the background density 
$\rho_b$ (e.g. 200 \cite{NaFW1996}). With this definition, the procedure 
of determining $\mu_2$ is different from the one presented in eqs. 
(\ref{finRdef}) -- (\ref{ProfFitB2Eq}).}
 We do not assume any value of $d$ at this point yet, except that $1 \leq 
d < \rho_{\text{max}}/\rho_b$. 

 \item The compensation radius $R = R_c$, at which the total mass within $R =
R_c$ is the same as the background mass would be if no inhomogeneity were
created. This is needed to let us know roughly where our inhomogeneity is
matched to the Friedmann background. See subsection \ref{comprad}.

 \end{itemize}

 \subsection{Compensation Radius \& Mass}\label{comprad}

   The matching of contained mass at $R_c$ is not sufficient for a
 Swiss-cheese type matching, as we have not required the time since the
bang to match up at this location, and therefore is technically the wrong
definition%
 \footnote{
 A proper matching of first and second fundamental forms would have
matched masses, energies and ages, and be a comoving surface $M =
M_{matched}$.  If our `compensation' procedure were executed at each time,
the resulting surface would not be properly matched, nor would it be
comoving.
 }%
 .  However, it has the advantage that it can be calculated knowing only
the density profile at a given time, and under the circumstances used here
is a fairly good estimate. All our models are then calculated out to
$M_c$, i.e $0 \leq M \leq M_c$.

   We can put upper and lower limits on the compensation radius and mass.  For a
condensation of measured mass $M_m$, the $M_c$ value obviously cannot be less
than this.  In fact the region around the visible condensation, though of low
density is of large volume, and will add noticeably to the mass to be
`compensated for'.  Therefore,
 \begin{equation}
   M_{c, \text{min}} > M_m~,~~~~~~~~
   R_{c, \text{min}} >
   \left( \frac{3 M_m}{4 \pi \rho_{b, \text{today}}} \right)^{1/3},
 \end{equation}
 where eq. (\ref{rho_b(t)}) gives $\rho_{b, \text{today}} = 4.075 \times
10^{-30}$~g/cc. in the chosen parabolic background, or more conveniently
 \begin{equation}
   R_{c, \text{min}} > 1.58 \times 10^{-4} (M_m/M_\odot)^{1/3}~\text{in Mpc}.
 \end{equation}

   On the other hand, the observed average separation of condensations puts
an upper limit on $R_c$.  Since the contents of the universe are a mixed bag of
galaxy clusters, rich clusters, superclusters, field galaxies, voids, walls,
etc, the average separation of rich clusters say is not meaningful for this
purpose. So we instead argue that there are around $3 \times 10^6$ large
galaxies \& $3 \times 10^7$ dwarf galaxies within $10^9$ light years.  At
say $10^{11}~M_\odot$ and $10^{10}~M_\odot$, respectively, this gives a
mean density of around $4.97 \times 10^9~M_\odot/$Mpc$^3 = 3.36 \times
10^{-28}$~kg/m$^3 = 0.0826 \rho_{b,\text{today}}$. Therefore an object of
mass $M_m$ in $M_\odot$ should on average occupy a volume of radius about
 \begin{equation}
   R_{c, \text{max}} = 3.64 \times 10^{-4} (M_m/M_\odot)^{1/3}~\text{in Mpc}.
 \end{equation}

   For a galaxy of $10^{11}~M_\odot$, these two limits are
$0.735 < R_c < 1.69$~Mpc, whereas for an Abell Cluster of $10^{15}~M_\odot$,
these two limits are $15.8 < R_c < 36.4$~Mpc.

For a void, the interior density will not be zero, but is not well known, and 
the radius is of the order of 60~Mpc.  Only by including some of the galaxies 
in the surrounding walls can one bring the average density up to the background 
value.  So $R_c > 60$~Mpc, which at background density gives $M_c > 5.5 \times 
10^{16}~M_\odot$.

 \subsection{An example of fitting a profile.}

   Since little is known about mass-distribution within galaxy clusters, we cannot
attempt to model any actual cluster to any significant accuracy. However, we
wish to show that such modelling is in principle possible. Therefore, for the
beginning, we will use the profile from Paper I and will show below that its
free parameters can be made equal to some observed/observable quantities.

   Let $\rho_b$ be the background density.  We choose the profile at $t_2$
to be
 \begin{equation}\label{finrhodef}
   \rho(M) = B_2\rho_b \text{e}^{-(M/\mu_2)^2},
 \end{equation}
 where $B_2$ and $\mu_2$ are free parameters to be adapted to the
observational constraints.

   Given two pairs of $(M_m, R_m)$ data, $(M_a, R_a)$ \& $(M_b, R_b)$, we
have
 \begin{equation}\label{finRdef}
   R^3 = \frac {3\mu_2} {8\sqrt{\pi}\; B_2\rho_b} \text{erfi} (M/\mu_2)
 \end{equation}
 for each of them, and each can be solved for $B_2$, so that:
 \begin{equation}
   \label{ProfFitB2Eq}
   \frac{3 \mu_2 \text{erfi}(M_a/\mu_2)}{8 \sqrt{\pi}\; R_a^3 \rho_b}
   = B_2 =
   \frac{3 \mu_2 \text{erfi}(M_b/\mu_2)}{8 \sqrt{\pi}\; R_b^3 \rho_b}.
 \end{equation}
 This is solved numerically for $\mu_2$, and $B_2$ follows.

   The compensation radius $R_c$ is determined by the condition that the
mean density out to $M$ equals the background density:
 \begin{equation}
   \rho_b = \frac{M_c}{4 \pi R(M_c)^3 / 3} =
   \frac{2 B_2 \rho_b M_c}{\sqrt{\pi}\; \mu_2 \text{erfi}(M_c/\mu_2)}.
 \end{equation}
 This is solved for $M_c$ numerically, and $R_c$ follows from
(\ref{finRdef}).

 \subsection{Profiles inspired by astronomy.}
 \label{ProfInspAstr}

We looked for density profiles that are considered realistic by astronomers. As
it turned out, there is no general agreement as to which profile best describes
observations, and no generally accepted definition of the radius of a galaxy
cluster exists.  However (see the Acknowledgements) the following `universal
profile' is one of the more commonly used formulas for density vs. distance
profiles \cite{NaFW1995, NaFW1996, NaFW1997}:
 \begin{equation}\label{NaFWprof}
 \rho(\zeta) = \rho_b \delta \frac 1 {(\zeta/\zeta_s)(1 + \zeta/\zeta_s)^2},
 \end{equation}
 where $\rho_b$ is the average density in the Universe, $\delta$ is a
dimensionless factor and $\zeta_s$ is a scale distance. (This is a Newtonian
formula, in which $\zeta$ is the Euclidean distance.) According to the authors of
Refs. \cite{NaFW1995, NaFW1996, NaFW1997}, this profile applies for $\zeta$
changing by two orders of magnitude.

   For our procedure, we need the density given as a function of mass, 
which cannot be found analytically for the above profile.
 \footnote{Note that the relation between mass and density in the 
 L-T model, eq. (\ref{rhoLT}), is very different from the 
 flat-space relation $\text{d}M/\text{d}\zeta = 4\pi\rho \zeta^2$. We will 
carry over to relativity the equation $\rho = \rho(M)$ resulting from the 
Newtonian relation with the hope that it is a good approximation at low 
densities and for small radii from the centre. However, a completely 
 self-consistent approach would require 
 re-interpretation of all the relevant astronomical observations against 
the 
 L-T model.}
 The calculation $\rho(\zeta) \rightarrow M(\zeta) \rightarrow \zeta(M) 
\rightarrow \rho(M)$ can always be done numerically, but it is more 
instructive to have exact explicit formulae. 

 We therefore approximate $M(\zeta)$ separately in the ranges $\zeta \ll 
\zeta_s$ and $\zeta \gg \zeta_s$. 

   For $\zeta \ll \zeta_s$ a Taylor series around $\zeta = 0$ up to $\zeta^2$ gives
 \begin{equation}\label{massatcent}
   M(\zeta) \approx 2\pi\rho_b\delta \zeta_s \zeta^2 := \beta^2 \zeta^2
   ~~~~\rightarrow~~~~ \zeta \approx \sqrt{M}/\beta.
 \end{equation}
 which converts the profile (\ref{NaFWprof}) to:
 \begin{equation}\label{profatcent}
   \rho_1(M) = \rho_b \beta \delta \zeta_s
   \frac 1 {\sqrt{M}\left(1 + \frac {\sqrt{M}}{\beta \zeta_s} \right)^2}.
 \end{equation}
 Like the original profile, this one has the unpleasant property that the density
becomes infinite at $M \to 0$, so we modify it to
 \begin{equation}\label{corprofatcent}
 \rho_2(M) = \frac {\rho_b \beta \delta \zeta_s} {(\epsilon + \sqrt{M})\left(1 +
\frac {\sqrt{M}} {\beta \zeta_s} \right)^2}.
 \end{equation}
 where $\epsilon$ is small compared to $\beta \delta \zeta_s$, and this can be
integrated to give $R(M)$.

   When $\zeta \gg \zeta_s$, the approximation
 \begin{equation}\label{massfar}
 \begin{split}
 M(\zeta) \approx 4 \pi\rho_b\delta \zeta_s \ln\left(1 + \zeta/\zeta_s\right) := \gamma
\ln\left(1 + \zeta/\zeta_s\right), \\
 ~~~~\rightarrow~~~~ \zeta = \zeta_s\left(\text{e}^{M/\gamma} - 1\right),
 \end{split}
 \end{equation}
 substituted into (\ref{NaFWprof}) gives
 \begin{equation}\label{corprofar}
   \rho_4(M) = \rho_b
   \frac {\delta} {\text{e}^{M/\gamma} - 1 + \nu} \text{e}^{-2M/\gamma}
 \end{equation}
 where $\nu$ has again been added into the denominator to permit a
 non-divergent central density.  The corresponding $R^3(M)$ is still
elementary.
 \footnote{This profile is meant for large $\zeta$ only. However, all our 
numerical codes cover $M = 0$, so we added $\nu$ to avoid modifying the 
codes.} 

   The profiles (\ref{corprofatcent}) and (\ref{corprofar}) were the starting
point for our considerations. The profiles actually used in numerical examples
were various modifications of these (see Sec. \ref{profit}), done in order to
better fit our profiles to observational data.

 \subsection{The initial density profile.} \label{initprof}

   We assume that the condensed object (model of a galaxy cluster) was created out
of a small localized initial density or velocity perturbation, superimposed on
the homogeneous spatially flat Friedmann background. During the evolution, the
perturbation was increasing in density amplitude and in mass, and thus was
effectively drawing more mass into the condensation region out of the
surrounding homogeneous region%
 \footnote{The
 L-T model is general enough to describe a situation where the final 
profile is also just a bump on a perfectly homogeneous background, but 
this picture does not fit well with the popular image of mass being 
accreted onto the initial condensation. The ``bump on a smooth 
background'' at $t_2$ would imply that the whole infinite background 
adjusted its density to the central bump, which would imply propagating 
the density wave to an infinite distance in a finite time.}%
 \footnote{In this way, we also avoid a numerical difficulty: the volume 
that is numerically monitored must be finite. The ``condensation 
surrounded by a rarefaction'' model has the advantage that the region 
outside the rarefaction remains strictly Friedmannian for all time and it 
is enough to monitor it only at its edge. In the ``bump on a smooth 
background'' model, the velocity perturbation spreads out to infinity; 
only the 
 mass-density at $t_2$ is Friedmannian outside the bump (and even this 
homogeneity gets destroyed by evolution, both backward and forward in time 
 --- it is only momentary).}%
 .

   At $t = t_1$, the profile need not be compensated. For an uncompensated
profile, it is good if it is localized (i.e. the perturbation is zero for $M >
M_1$, where $M_1$ is the assumed mass of the initial perturbation). Then the
definitions of the radius and mass of the perturbation are straightforward.

We choose for this profile the cosine shape
 \begin{equation}\label{inirhodef}
 \rho(M) =
 \left\{ \begin{array}{ll}
   B_1 (1 + \cos(\pi M / M_1)) + \rho_b \qquad
      & \text{~~for~} M \in [0, M_1]; \\
   \rho_b = \text{const}  &  \text{~~for~} M \geq M_1
 \end{array} \right.
 \end{equation}
 with $B_1 > 0$.  Then $\rho(0) = \rho_b (1 + A)$ is the maximal value of
density, with $\Delta\rho/\rho_b = A = 2 B_1 / \rho_b$ being the density
amplitude from astronomical data, while $M_1$ is the total mass of the initial
perturbation.

   The radius $R(M)$ is given from eq. (\ref{RcubedInt}) by
 \begin{equation} \label{iniRdefin}
 \begin{split}
  &R^3(M) = \frac{3M_1} {2\pi^2 \sqrt{\rho_b\left(\rho_b + 2B_1\right)}} \times \\
  &\times \arctan\left[\sqrt{\frac {\rho_b} {\rho_b + 2{B_1}}} \tan \left(\frac
  {\pi M} {2M_1}\right) \right] \quad \text{for~} M \leq M_1,
 \end{split}
 \end{equation}
 \begin{equation}\label{iniRdefout}
 \begin{split}
 R^3(M) = \frac {3M_1} {4\pi \sqrt{\rho_b\left(\rho_b + 2B_1\right)}} + \frac 3
{4\pi \rho_b} \left(M - M_1\right) \\
 \text{for~} M \geq M_1,
 \end{split}
 \end{equation}
 where the first term on the right in (\ref{iniRdefout}) is the common value of
$R^3$ at $M = M_1$.

   We did one run with this kind of initial profile ($\rho$i4 in Table
\ref{InFnDenProfileTab}) in order to demonstrate more clearly the point we  made
in Paper I: that the mass of the ``seed'' structure existing at time $t_1$ can
be much smaller than the mass of the galaxy cluster into which it will evolve.
The other initial profiles are either flat (see Table \ref{RunList} and Sec.
\ref{Results} for reasons why) or go into the background only asymptotically.

 \section{The evolution from a given velocity distribution to a given density
distribution.}\label{veltoden}

 \setcounter{equation}{0}

   The quantity that is a measure of the velocity distribution of the dust
in an L-T model is
 \begin{equation}\label{defb}
 b = R_{,t}/M^{1/3}.
 \end{equation}
 This is constant in any Robertson-Walker model, so its nonconstancy is a
measure of the velocity inhomogeneity.

   Suppose we wish to adapt an L-T model to a given initial velocity distribution
$b = b_1(r)$  at $t = t_1$, and to a given density distribution $\rho =
\rho_2(r)$ at $t = t_2$. This is a different set of data from the one considered
in paper I, and so the existence of such an evolution has to be proven. The
functions appearing along the way are different, but the overall mathematical
scheme is essentially the same. As before, we will mostly be using the mass $M$
as the radial coordinate, and in each case we will assume that at the initial
instant $t_1$ the configuration is expanding, so
 \begin{equation}\label{expcon}
 R_{,t}(t_1, M) > 0 \Longrightarrow b_1(M) > 0.
 \end{equation}
 Analogous reasoning can be carried out for matter that is initially collapsing,
but this situation is just covered by the time reverse of the method given here
and is not relevant for the problem of structure formation.

 \subsection{Hyperbolic evolution.}

   We have, for $E > 0$:
 \begin{equation}\label{dertReq}
R_{,t} = \frac M {\sqrt{2E}R} \sqrt {(1 + 2ER/M)^2 - 1}.
 \end{equation}
 If
 \begin{equation}\label{defb1}
 \left.R_{,t}/M^{1/3}\right|_{t = t_1} = b_1(r)
 \end{equation}
is the data, then denoting $R|_{t = t_1} = R_1$ and solving the above for $R_1$
we find
 \begin{equation}\label{defRbyb}
 R_1 = \frac {2M^{1/3}} {{b_1}^2 - 2E/M^{2/3}},
 \end{equation}
 and so from the evolution equations
 \begin{equation}\label{defetabyb}
 \begin{split}
 \eta_1 &= \text{arcosh} (1 + 2ER_1/M) \\
 &= \text{arcosh} \left( \frac {{b_1}^2 + 2E/M^{2/3}}{{b_1}^2 - 2E/M^{2/3}}
 \right),
 \end{split}
 \end{equation}
 \begin{equation}\label{deft1state}
 \begin{split}
 &\sqrt {\left( \frac {{b_1}^2 + 2E/M^{2/3}} {{b_1}^2 - 2E/M^{2/3}} \right)^2 - 1}
 - \text{arcosh} \left( \frac {{b_1}^2 + 2E/M^{2/3}} {{b_1}^2 - 2E/M^{2/3}}
 \right) \\
 &= \frac{(2E)^{3/2}} M \left(t_1 - t_B \right).
 \end{split}
 \end{equation}
 This is the equation that will be used as initial data at $t = t_1$.

   The equation at $t = t_2$ will be the same as eq. (3.3) in paper I
for $i = 2$ viz:
 \begin{equation}\label{deft2state}
 \begin{split}
 &\sqrt {\left(1 + 2ER_2/M\right)^2 - 1} - \text{arcosh} (1 + 2ER_2/M) \\
 &= \frac {(2E)^{3/2}} M \left(t_2 - t_B \right).
 \end{split}
 \end{equation}
 Here, the quantity $R_2 = R(t_2, r)$ is calculated from the given
$\rho_2(r) = \rho (t_2, r)$ using eq. (\ref{RcubedInt}).

   Just as in paper I, we introduce the symbols
 \begin{equation}\label{defxa2E>0}
 x = 2E/M^{2/3}, \qquad a_2 = R(t_2, M)/M^{1/3}.
 \end{equation}
 The set of equations to determine $E(M)$ and $t_B(M)$ is then (from
(\ref{deft1state}) and (\ref{deft2state})):
 \begin{equation}\label{eqatt1}
 \sqrt{\left(\frac {{b_1}^2 + x} {{b_1}^2 - x} \right)^2 - 1}
- \text{arcosh} \left( \frac {{b_1}^2 + x} {{b_1}^2 - x} \right)
= x^{3/2}\left(t_1 - t_B\right),
 \end{equation}
 \begin{equation}\label{eqatt2}
 \sqrt{\left(1 + a_2 x \right)^2 - 1} - \text{arcosh} \left(1 + a_2 x\right) =
x^{3/2}\left(t_2 - t_B\right),
 \end{equation}

   From (\ref{eqatt2}) we now find $t_B$:
 \begin{equation}\label{tBdef}
 t_B = t_2 - \frac 1 {x^{3/2}}\left[ \sqrt {\left(1 + a_2 x \right)^2 - 1} -
\text{arcosh} (1 + a_2x) \right],
 \end{equation}
 and substituting this in (\ref{eqatt1}) we obtain the following equation to
determine $x$:
 \begin{equation}\label{eqPhiH}
 \Phi_H(x) = 0,
 \end{equation}
 where
 \begin{equation}\label{defPhiH}
 \begin{split}
 \Phi_H(x) :=& \sqrt {\left(1 + a_2 x \right)^2 - 1}
 - \sqrt{\left(\frac {{b_1}^2 + x} {{b_1}^2 - x} \right)^2 - 1} \\
 &- \text{arcosh} (1 + a_2x)
 + \text{arcosh} \left( \frac {{b_1}^2 + x} {{b_1}^2 - x} \right) \\
 &- x^{3/2}\left(t_2 - t_1\right).
 \end{split}
 \end{equation}
 We will use the functions $b_1(M)$ and $a_2(M)$ implied by the assumed
$R_{,t1}(M)$ and $\rho_2(M)$ to find $E(M)$ and $t_B(M)$, and then to find
$\rho(t_1,M)$. This will tell us about the relative importance of the velocity
and density distributions for structure formation. In particular, with $b_1$ =
const, the $\rho(t_1, M)$ will tell how big the initial density inhomogeneity
has to be when the initial velocity distribution is exactly homogeneous, while
the final structure is given.

 \subsubsection{Conditions for solutions to exist}

   Now we have to verify whether the equation $\Phi_H(x) = 0$ determines a
value of $x$. We see that $\Phi_H(x)$ is defined in the range
 \begin{equation}\label{rangexEpos}
 0 \leq x < {b_1}^2.
 \end{equation}
 It is easily seen that $\Phi_H(0) = 0$. For $x \to {b_1}^2$, the second and
fourth term in (\ref{defPhiH}) tend to $(- \infty)$ and to $(+ \infty)$,
respectively, while all the other terms are finite.  Since
 \begin{equation}\label{limarchovsq}
 \lim_{x \to {b_1}^2} \frac {\text{arcosh} \left[\left({b_1}^2 +
x\right)/\left({b_1}^2 - x\right)\right]} {\sqrt{\left[\left({b_1}^2 +
x\right)/\left({b_1}^2 - x\right)\right]^2 - 1}} = 0,
 \end{equation}
 the second term is dominant. Hence:
 \begin{equation}\label{limPhiHb12}
 \lim_{x \to {b_1}^2} \Phi_H(x)  = - \infty.
 \end{equation}
 We also have
 \begin{equation}\label{derPhiH}
 \begin{split}
 \Phi_{H,x}(x) &= \sqrt {x} \left[ \frac {{a_2}^{3/2}} {\sqrt{2 + a_2x}} - \frac
 {2b_1} {\left({b_1}^2 - x\right)^2} - \frac 3 2 \left(t_2 - t_1\right)\right]
 \\
 &:= \sqrt{x} \lambda_H(x).
 \end{split}
 \end{equation}
 Since $\lambda_{H,x} = - \frac 1 2 {a_2}^{5/2}/(2 + a_2x)^{3/2} - 4b_1/({b_1}^2
- x)^3 < 0$, the function $\lambda_H(x)$ is strictly decreasing and may have at
most one zero. Hence, $\Phi_{H,x}$ can be positive anywhere only if
$\lambda_H(0) > 0$, i.e. if
 \begin{equation}\label{condtEpos}
 t_2 - t_1 < \frac {\sqrt{2}}3 {a_2}^{3/2} - \frac 4 {3{b_1}^3}.
 \end{equation}
 Since we consider the evolution from $t_1$ to $t_2$, it is natural to assume
$t_2 > t_1$. Then, (\ref{condtEpos}) can be fulfilled only if
 \begin{equation}\label{addcond}
 2/a_2 < {b_1}^2.
 \end{equation}
This inequality is easy to understand if we use (\ref{defRbyb}) for ${b_1}^2$
(with $R = R_1$), and recall the definition of $a_2$, eq. (\ref{defxa2E>0}).
Then (\ref{addcond}) is equivalent to
 \begin{equation}\label{addcond1}
 \frac 2{R_2} < \frac 2{R_1} + \frac {2E}M.
 \end{equation}
 This in turn implies that $R_2 > R_1$; otherwise there will be no $E > 0$
evolution from $R_1$ to $R_2$. Thus, in passing, we have proven the intuitively
obvious:

 \medskip

   {\em Conclusion:} If matter is expanding at $t = t_1$, then an $E > 0$ L-T
evolution from $R(t_1, M)$ to $R(t_2, M)$ will exist only if $R(t_2, M) >
R(t_1,M)$.\square

 \medskip

   Note that $R(t_2, M) > R(t_1,M)$ does not imply $\rho(t_2, M) < \rho(t_1, M)$,
contrary to the intuitive expectation, and in contrast to the Friedmann models.
This point is discussed in Appendix \ref{greden}.

   Eq. (\ref{condtEpos}) may be rewritten as
 $$
 a_2 > \left[\frac 92 \left(t_2 - t_1 + \frac 4{3{b_1}^3}\right)^2
\right]^{1/3},
 $$
 which means that between $t_1$ and $t_2$ the model must have
expanded by more than the $E = 0$ model would have done. (Applying the
definitions (\ref{defb1}) and (\ref{defxa2E>0}) to (\ref{ParEv}) and
(\ref{defRbyb}) for the $E = 0$ case shows $t_1 - 4/(3{b_1}^3) = t_B$ and $a_2 =
[(9/2 (t_2 - t_B)^2 ]^{1/3}$.)

   If (\ref{condtEpos}) is fulfilled, then $\Phi_H(x)$ starts as an increasing
function in some neighbourhood of $x = 0$, beyond which there is exactly one
maximum, and finally it decreases to $- \infty$ at $x = {b_1}^2$, thus passing
(only once) through a zero value somewhere between the maximum and $x =
{b_1}^2$. Hence, eqs. (\ref{condtEpos}) and (\ref{addcond}) are a necessary and
sufficient condition for the existence of an $E
> 0$ L-T evolution connecting the initial state (defined by the velocity
distribution (\ref{defb1})) to the final state defined by the density
distribution $\rho_2(M)$. As before, the resulting model will have to be
checked for the possible existence of shell-crossings and regular maxima or
minima.

 \subsection{Elliptic evolution; the final state in the expansion phase.}

   Assuming that the initial state at $t = t_1$ is in the expansion phase of the
Universe, we find, similarly as in (\ref{dertReq}), that with $E < 0$
 \begin{equation}\label{dertRnegE}
 R_{,t} = \frac M{\sqrt{-2E} R} \sqrt{1 - (1 + 2ER/M)^2},
 \end{equation}
 and then, using (\ref{defb1}) we obtain (\ref{defRbyb}) once again. Since this
time $E < 0$, the inequality $2M^{1/3}/R + 2E/M^{2/3} \geq 0$ has to be
fulfilled, but this is identical to $2M/R + 2E \geq 0$ that has to hold for an
$E < 0$ evolution anyway. In place of (\ref{defetabyb}) -- (\ref{deft1state})
we obtain at $t = t_1$
 \begin{equation}\label{defetabybnegE}
 \eta = \arccos
\left( \frac {{b_1}^2 + 2E/M^{2/3}} {{b_1}^2 - 2E/M^{2/3}} \right),
 \end{equation}
 \begin{equation}\label{deft1statenegE}
 \begin{split}
 &\arccos \left( \frac {{b_1}^2 + 2E/M^{2/3}} {{b_1}^2 - 2E/M^{2/3}} \right)
 - \sqrt{1 - \left(\frac {{b_1}^2 + 2E/M^{2/3}} {{b_1}^2 - 2E/M^{2/3}}\right)^2}
 \\
 &= \frac {(-2E)^{3/2}} M \left(t_1 - t_B\right).
 \end{split}
 \end{equation}

   Since we assumed the final state at $t = t_2$ to be in the expansion phase,
too, the equation at $t = t_2$ is the same as eq. (3.15) in paper I
for $i = 2$:
 \begin{equation}\label{deft2statenegE}
 \begin{split}
 &\arccos \left(1 + 2ER_2/M\right) - \sqrt{1 - \left(1 + 2ER_2/M\right)^2} \\
 &= \frac {(- 2E)^{3/2}} M \left(t_2 - t_B\right).
 \end{split}
 \end{equation}
 We introduce $a_2$ by (\ref{defxa2E>0}) and
 \begin{equation}\label{defxE<0}
 x = - 2E/M^{2/3},
 \end{equation}
then (\ref{deft1statenegE}) and (\ref{deft2statenegE}) become
 \begin{equation}\label{eqatt1negE}
 \arccos \left( \frac {{b_1}^2 - x} {{b_1}^2 + x} \right)
- \sqrt{1 - \left(\frac {{b_1}^2 - x}
{{b_1}^2 + x}\right)^2} = x^{3/2}\left(t_1 - t_B\right),
 \end{equation}
 \begin{equation}\label{eqatt2negE}
 \arccos\left(1 - a_2x\right) - \sqrt{1 - \left(1 - a_2x\right)^2} =
x^{3/2}\left(t_2 - t_B\right).
 \end{equation}
 From (\ref{eqatt2negE}) we have
 \begin{equation}\label{deftBnegE}
 t_B = t_2 - \frac 1 {x^{3/2}} \left[ \arccos \left(1 - a_2 x \right) - \sqrt {1
- \left(1 - a_2 x \right)^2}\right],
 \end{equation}
 and substituting this in (\ref{eqatt1negE}) we obtain
 \begin{equation}\label{eqPhiX}
 \Phi_X(x) = 0,
 \end{equation}
 where
 \begin{equation}\label{defPhiX}
 \begin{split}
 \Phi_X(x) :=& \sqrt{1 - \left(\frac{{b_1}^2 - x} {{b_1}^2 + x}\right)^2} -
 \sqrt{1 - \left(1 - a_2x\right)^2} \\
 &+ \arccos\left(1 - a_2x\right)
 - \arccos \left( \frac{{b_1}^2 - x} {{b_1}^2 + x} \right) \\
 &- x^{3/2}\left(t_2 - t_1\right).
 \end{split}
 \end{equation}
 The terms containing $b_1$ do not impose any extra restriction on $x$ apart from
$x > 0$ that follows from the definition (\ref{defxE<0}). Since $a_2 > 0$ and $x
> 0$, the restriction on $x$ imposed by the other terms is the same as in paper
I:
 \begin{equation}\label{xuplimnegE}
 x \leq 2/a_2,
 \end{equation}
 which follows from $2M/R + 2E \geq 0$ ($x = 2/a_2$ corresponds to the given
$M$-shell being exactly at the maximum of expansion at $t = t_2$).

 \subsubsection{Conditions for solutions to exist}

The calculations are similar to those in the previous subsection, they are
described in Appendix \ref{4bcalc}. Eq. (\ref{eqPhiX}) has a solution if and
only if the following two inequalities are obeyed:
 \begin{equation}\label{condtEneg}
 t_2 - t_1 > \frac {\sqrt{2}}3 {a_2}^{3/2} - \frac 4 {3{b_1}^3},
 \end{equation}
 which is the opposite of (\ref{condtEpos}), and
 \begin{equation}\label{condtexp}
 \begin{split}
   &t_2 - t_1 \leq \\
   &\left(a_2/2\right)^{3/2} \left[ \pi
   + \frac {b_1\sqrt{2a_2}}{a_2{b_1}^2/2 + 1}
   - \arccos \left( \frac {a_2{b_1}^2/2 - 1} {a_2{b_1}^2/2 +1}
   \right) \right].
 \end{split}
 \end{equation}
The solution is then unique. The second inequality is equivalent to eq. (3.22)
in paper I, as can be verified by writing (\ref{defRbyb}) as $b_1 = \sqrt{2/a_1
- x}$ and using $x = 2/a_2$.

   It follows that the right-hand side of (\ref{condtexp}) is always greater
than the right-hand side of (\ref{condtEneg}), as proven in Appendix A to paper
I, so the two inequalities (\ref{condtexp}) and (\ref{condtEneg}) are
consistent. It follows further, as shown in paper I, that the equality in
(\ref{condtexp}) means that the final state at $t_2$ is exactly at the maximum
expansion. Note that, in consequence of $b_1 > 0, a_2 > 0$ and $\arccos
(\text{anything}) \leq \pi$, the right-hand side of (\ref{condtexp}) is always
positive, and so, given $t_1$, values of $t_2$ obeying (\ref{condtexp}) always
exist.

   The set of inequalities (\ref{condtEneg}) -- (\ref{condtexp}) constitutes the
necessary and sufficient condition for the existence of an $E < 0$ L-T
evolution between the initial state at $t_1$ specified by the velocity
distribution (\ref{defb1}) and the final state at $t_2$ specified by the density
distribution $\rho(t_2, M) = \rho_2(M)$, such that the final state is still in
the expansion phase.

 \subsection{Elliptic evolution; the final state in the recollapse phase.}

   We assume now that the initial state at $t = t_1$ is again in the expansion
phase, so eqs. (\ref{dertRnegE}) -- (\ref{deft1statenegE}) still apply.
However, the final state at $t = t_2$ is now assumed to be in the recollapse
phase. Instead of eq. (\ref{deft2statenegE}) we then get
 \begin{equation}\label{{deft2statenegEcoll}}
 \begin{split}
 &\pi + \arccos\left(-1 - 2ER_2/M\right) + \sqrt{1 - \left(1 + 2ER_2/M\right)^2}
 \\
 &= \frac {(-2E)^{3/2}}M \left(t_2 - t_B\right).
 \end{split}
 \end{equation}

   Introducing again $a_2$ by (\ref{defxa2E>0}) and $x$ by (\ref{defxE<0}), we
obtain the set of equations consisting of (\ref{eqatt1negE}) and
 \begin{equation}\label{{eqatt2negEcoll}}
 \begin{split}
 &\pi + \arccos\left(-1 - a_2 x\right) + \sqrt{1 - \left(1 - a_2 x\right)^2} \\
 &= x^{3/2} \left(t_2 - t_B\right).
 \end{split}
 \end{equation}
 The solution for $t_B$ is now
 \begin{equation}\label{soltBcoll}
 \begin{split}
 &t_B = t_2 \\
 &- \frac 1{x^{3/2}} \left[\pi + \arccos\left(-1 + a_2x\right) +
\sqrt{1 - \left(1 - a_2 x\right)^2}\right],
 \end{split}
 \end{equation}
 and substituting this in (\ref{eqatt1negE}) we get
 \begin{equation}\label{PhiCeq}
 \Phi_C(x) = 0,
 \end{equation}
 where
 \begin{equation}
   \label{PhiCdef}
 \begin{split}
   \Phi_C(x) :=& \sqrt{1 - \left( \frac {{b_1}^2 - x} {{b_1}^2 + x}
      \right)^2} + \sqrt{1 - \left(1 - a_2x\right)^2} \\
   &+ \pi - \arccos \left( \frac {{b_1}^2 - x}{{b_1}^2 + x} \right)
      + \arccos \left(-1 + a_2x\right) \\
   &- x^{3/2}\left(t_2 - t_1\right).
 \end{split}
 \end{equation}
 As before, we have $x > 0$ by definition and $x \leq 2/a_2$.

 \subsubsection{Condition for solutions to exist}

The details of the calculation are shown in Appendix \ref{4ccalc}. Eq.
(\ref{PhiCeq}) has a solution if and only if
 \begin{equation}\label{condtcoll}
 \begin{split}
 t_2 - t_1 \geq \left(a_2/2\right)^{3/2}\left[\pi + \frac
 {b_1\sqrt{2a_2}}{a_2{b_1}^2/2 + 1} \right. \\
\left. - \arccos \left( \frac {a_2{b_1}^2/2 - 1} {a_2{b_1}^2/2 + 1} \right)
\right],
 \end{split}
 \end{equation}
and then the solution is unique. This is the necessary and sufficient condition
for the existence of an $E < 0$ L-T evolution between the initial state at
$t_1$ specified by the velocity distribution (\ref{defb1}) and the final state
at $t_2$ specified by the density distribution $\rho(t_2, M) = \rho_2(M)$, such
that the final state is in the recollapse phase.

 \subsection{A note on the meaning of parameters.}\label{mop}

   It should be noted that the values of the time coordinate $t_1$ and $t_2$,
at which the initial and final states are specified, do not play
individual roles in the calculations. The meaningful parameter is $(t_2 -
t_1)$; it is this value that determines $E(M)$, and then the corresponding
$t_B$ is calculated. The ``age of the Universe" at the two instants then
follows, being $(t_1 - t_B)$ and $(t_2 - t_B)$, respectively. This
conclusion applies as well to paper I; and the same is still true in the
Friedmann limit, where $x$, $a_1$, $a_2$ and $b_1$ are no longer functions
of $M$, but just constants. Hence, the physical input data for the
procedure, both here and in paper I, are the initial and final
distributions of density or velocity and the time-interval between them,
$(t_2 - t_1)$. The individual values of $t_1$ and $t_2$ do not have a
physical meaning.

 \section{The evolution from a given velocity distribution to another velocity
distribution.}

 \label{veltovel}

 \setcounter{equation}{0}

   For completeness, we shall also consider the case when both the initial and the
final state are defined by velocity distributions, even though this does not
seem to be directly useful for astrophysics.

 \subsection{Hyperbolic evolution.}

Eq. (\ref{deft1state}) still applies. Defining $x = 2E/M^{2/3}$ as in the
previous cases, we again obtain eq. (\ref{eqatt1}), while instead of
(\ref{eqatt2}) we now have
 \begin{equation}\label{veleqatt2}
 \sqrt{\left(\frac {{b_2}^2 + x} {{b_2}^2 - x}\right)^2 - 1} - \text{arcosh}
\left(\frac {{b_2}^2 + x} {{b_2}^2 - x}\right) = x^{3/2}\left(t_2 - t_B\right),
 \end{equation}
 where
 \begin{equation}\label{defb2}
 \left.b_2 = R_{,t}/M^{2/3}\right|_{t = t_2}.
 \end{equation}
 Since the hyperbolic expansion cannot be reversed to become collapse, and since
we assumed $b_1 > 0$ (expansion rather than collapse at $t_1$), it follows that
$b_2 > 0$ is a necessary condition here.

   From (\ref{eqatt1}) and (\ref{veleqatt2}) we obtain two equations
 \begin{equation}\label{veltB}
 \begin{split}
 t_B =& t_i - \frac 1 {x^{3/2}} \left[
 \sqrt{\left(\frac {{b_i}^2 + x} {{b_i}^2 - x}\right)^2 - 1} - \text{arcosh}
 \left(\frac {{b_i}^2 + x} {{b_i}^2 -x}\right)\right], \\
 & i = 1,2,
 \end{split}
 \end{equation}
 so $x$ is the solution of
 \begin{equation}\label{chiHeq}
 \chi_H = 0,
 \end{equation}
 where
 \begin{equation}
   \label{defchiH}
 \begin{split}
   \chi_H(x) :=& \sqrt{\left(\frac {{b_2}^2 + x} {{b_2}^2 - x}\right)^2
   - 1} - \sqrt{\left(\frac {{b_1}^2 + x} {{b_1}^2 - x}\right)^2 - 1} \\
   &- \text{arcosh} \left(\frac {{b_2}^2 + x} {{b_2}^2 - x}\right) +
   \text{arcosh} \left(\frac {{b_1}^2 + x} {{b_1}^2 - x}\right) \\
   &- x^{3/2}\left(t_2 - t_1\right).
 \end{split}
 \end{equation}
 Note that (\ref{defRbyb}) and its analogue at $t_2$ imply ${b_i}^2 - x \equiv
2/a_i > 0$, $i = 1, 2$, i.e. $x < {b_i}^2$ at both $i = 1$ and $i = 2$. This,
together with $x > 0$, guarantees that the argument of arcosh will be always
$\geq 1$, as it should be. Simultaneously, it guarantees that both square roots
will be real.

 \subsubsection{Conditions for solutions to exist}\label{condveltovel}

Details of the calculations are shown in Appendix \ref{4ccalc}.

   For further analysis, we will need to investigate the derivative of
$\chi_H(x)$, it is
 \begin{equation}\label{dchi}
 \begin{split}
 \frac {\text{d}} {\text{d} x} \chi_H(x) &= \sqrt{x}\left[\frac {2b_2} {\left({b_2}^2
 - x\right)^2} - \frac {2b_1} {\left({b_1}^2 - x\right)^2} - \frac 3 2 \left(t_2
 - t_1\right)\right] \\
 &:= \sqrt{x} \mu_H(x).
 \end{split}
 \end{equation}
 We also have to know  which of ${b_1}^2$, ${b_2}^2$ is greater. As expected, it
turns out that in consequence of the assumptions made ($t_2 > t_1$ and
$\left.R_{,t}\right|_{t = t_1} > 0$, i.e. $b_1 > 0$), ${b_1}^2$ {\it must} be
greater, or else the $E > 0$ evolution between the two states will not exist
  --- see Appendix \ref{grevelo}.

   Therefore we take it for granted that
 \begin{equation}\label{greb1}
 b_1 > b_2,
 \end{equation}
i.e. that at every point the velocity of expansion at $t_2$ must be smaller 
than at $t_1$ at the same comoving coordinate $M$. With $b_1 > b_2$, we have
 \begin{equation}\label{limofmu}
 x < {b_2}^2, \qquad \lim_{x \to {b_2}^2} \mu_H(x) = + \infty,
 \end{equation}
 and, in a similar way as in Appendix \ref{grevelo}, it can be proven that now
$\mu'_H(x) > 0$ for all $x > 0$. Consequently, $\mu_H(x)$ may have at most one
zero, and it will have one only if $\mu_H(0) < 0$, i.e. if
 \begin{equation}\label{limoftHyp}
 t_2 - t_1 > \frac 4 3 \left(\frac 1 {{b_2}^3} - \frac 1 {{b_1}^3}\right).
 \end{equation}
 Unlike in those cases where the profile was specified by density, the
time-difference between the initial and the final state must be sufficiently
{\it large}. The inequality becomes more intelligible when it is rewritten as
 \begin{equation}\label{limoftdup}
 {b_2}^3 > \frac {{b_1}^3} {1 + \frac 3 4 {b_1}^3\left(t_2 - t_1\right)},
 \end{equation}
 which means that an $E > 0$ evolution between the two states will exist provided
the velocity of expansion at $t_2$ is greater than the velocity of expansion of
the $E = 0$ model, for which $b^3 = 4/[3 (t - t_B)]$, giving an equality
in (\ref{limoftdup}).

   Since the range of $x$ is limited to $(0, {b_2}^2)$, finding $x$ numerically
will pose no problems.

 \subsection{Elliptic evolution, the final state in the expansion phase.}

   Expansion at $t_2$ is equivalent to $b_2 > 0$.  The set of equations to
define $x = -2E/M^{2/3}$ now consists of (\ref{eqatt1negE}) and its
analogue at $t_2$, obtained by the substitution $(b_1, t_1) \to (b_2,
t_2)$. Hence:
 \begin{equation}\label{vetovetB}
 \begin{split}
 t_B =& t_i - \frac 1 {x^{3/2}}\left[
 \arccos\left(\frac {{b_i}^2 - x} {{b_i}^2 + x}\right) - \sqrt{1 - \left(\frac
 {{b_i}^2 - x} {{b_i}^2 + x}\right)^2}\right], \\
 & i = 1,2,
 \end{split}
 \end{equation}
 and $x$ satisfies
 \begin{equation}\label{eqchix}
 \chi_X(x) = 0,
 \end{equation}
 where
 \begin{equation}
   \label{defchix}
 \begin{split}
   \chi_X(x) :=&
      - \sqrt{1 - \left(\frac {{b_2}^2 - x} {{b_2}^2 + x}\right)^2}
      + \sqrt{1 - \left(\frac {{b_1}^2 - x} {{b_1}^2 + x}\right)^2} \\
   &+ \arccos\left(\frac {{b_2}^2 - x} {{b_2}^2 + x}\right)
      - \arccos\left(\frac {{b_1}^2 - x} {{b_1}^2 + x}\right) \\
      &- x^{3/2}\left(t_2 - t_1\right).
 \end{split}
 \end{equation}
 This time there is no additional limitation imposed on $x$ by eqs.
(\ref{vetovetB}) -- (\ref{defchix}) --- the square roots and the arccos will be
well-defined for every value of $x > 0$. The relation between $x$ and $b_i$ is
$x = 2M^{1/3}/\left[R_i\left({b_i}^2 + 1\right)\right]$, and so indeed any value
of $x > 0$ is allowed, since $b_i$ can have values between $0$ at maximum
expansion and $\infty$ at the Big Bang, while $R_i$ can be arbitrarily large or
small, depending on the initial data. However, as always in the elliptic case,
$E \geq -1/2$ has to hold, $E = -1/2$ indicating a maximum in the spatial
section, and this will have to be checked while solving (\ref{defchix}) for $x$.

 Finding $x$ numerically will be no problem also here, since the natural
variables to work with are $y_i = \left({b_i}^2 - x\right)/\left({b_i}^2 +
x\right)$, both of which have the limited range $(-1 , 1)$ for $x > 0$.

 \subsubsection{Conditions for solutions to exist}

   We have:
 \begin{equation}\label{propchix}
 \chi_X(0) = 0, \qquad \lim_{x \to \infty}\chi_X(x) = - \infty,
 \end{equation}
 \begin{equation}\label{derchix}
 \begin{split}
 \chi'_X(x) &= \sqrt{x} \left[\frac {2b_2} {\left({b_2}^2 + x\right)^2} - \frac
{2b_1} {\left({b_1}^2 + x\right)^2} - \frac 3 2 \left(t_2 - t_1\right)\right] \\
&:= \sqrt{x}\mu_X(x).
 \end{split}
 \end{equation}
 \begin{equation}\label{propderchix}
 \Longrightarrow \qquad \chi'_X(0) = 0, \qquad \lim_{x \to \infty}\chi'_X(x) = -
\infty.
 \end{equation}

   It is known from the evolution equations (\ref{EllEv}) that if $b_1 > 0$ and
$t_2 > t_1$, then the velocity of expansion at $t_2$ must be smaller than $b_1$.
However, just as in the previous cases, this can be proven also from the
properties of the function $\chi_X(x)$, and the proof is given in Appendix
\ref{smallveloatt2}. Therefore we will take it for granted that
 \begin{equation}
   b_2 < b_1
 \end{equation}
 in the following. We find
 \begin{equation}\label{dermux}
 \mu'_X(x) = - \frac {4b_2} {\left({b_2}^2 + x\right)^3} + \frac {4b_1}
{\left({b_1}^2 + x\right)^3},
 \end{equation}
 and so $\mu'_X(0) < 0$, i.e. $\mu_X(x)$ is decreasing in a neighbourhood of $x
= 0$. It keeps decreasing as long as $\mu'_X$ is negative, i.e. as long as $x <
x_m$, where $x_m$ is the (unique) solution of $\mu'_X(x) = 0$:
 \begin{equation}\label{minmux}
 \begin{split}
   &x_m = \left(b_1b_2\right)^{1/3}\frac {{b_1}^{5/3} - {b_2}^{5/3}}
   {{b_1}^{1/3} -{b_2}^{1/3}} \equiv \left(b_1b_2\right)^{1/3} \times \\
   &\left({b_1}^{4/3} + b_1{b_2}^{1/3} + {b_1}^{2/3}{b_2}^{2/3}
   + {b_1}^{1/3}b_2 + {b_2}^{4/3}\right).
 \end{split}
 \end{equation}
 From here we find
 \begin{equation}\label{bisqx}
 {b_i}^2 + x_m =  {b_i}^{1/3}\frac {{b_1}^2 - {b_2}^2} {{b_1}^{1/3} -
{b_2}^{1/3}},
 \end{equation}
 and so the value of $\mu_X(x_m)$ is
 \begin{equation}\label{muxatxm}
 \begin{split}
 \mu_X(x_m) =& 2\frac {\left({b_1}^{1/3} - {b_2}^{1/3}\right)^2} {\left({b_1}^2 -
{b_2}^2\right)^2}\left({b_2}^{4/3} - {b_1}^{4/3}\right) \\
 &- \frac 3 2 \left(t_2
- t_1\right) < - \frac 3 2 \left(t_2 - t_1\right) < 0
 \end{split}
 \end{equation}
 (in consequence of $b_2 < b_1$). For $x > x_m$, $\mu'_X(x)$ is positive, and so
$\mu_X(x)$ is increasing, but $\lim_{x \to \infty}\mu_X(x) = - \frac 3 2 (t_2 -
t_1)$ is still negative. Hence, in summary, $\mu_X(x)$ is decreasing from the
value $2/{b_2}^3 - 2/{b_1}^3 - \frac 3 2 (t_2 - t_1)$ at $x = 0$ (which may be
positive or negative) to a minimum at $x = x_m$ which is negative, and then
keeps increasing, but remains negative for all $x > x_m$. Thus, from
(\ref{derchix}) it follows that the function $\chi_X$ is necessarily decreasing
for $x > x_p \geq 0$, where $x_p$ is the zero of $\chi'_X(x)$. Consequently,
the equation $\chi_X(x) = 0$ will have a positive solution only if $x_p > 0$,
i.e. if there is a neighbourhood of $x = 0$ in which $\chi_X(x)$ is increasing
(recall: $\chi_X(0) = 0$). Thus, the necessary and sufficient condition for the
existence of a positive solution of $\chi_X(x) = 0$ is $\mu_X(0) > 0$, i.e. the
opposite of (\ref{limoftHyp}):
 \begin{equation}\label{{limoftell}}
 t_2 - t_1 < \frac 4 3 \left(\frac 1 {{b_2}^3} - \frac 1 {{b_1}^3}\right),
 \end{equation}
 which now implies that the expansion between $t_1$ and $t_2$ must have been
slower than it would be in an $E = 0$ model.

 \subsection{Elliptic evolution, the final state in the collapse phase.}

   Finally, we shall consider the $E < 0$ evolution between two velocity profiles
for the case when the final state is in the collapse phase, $b_2 < 0$.

For the initial state, we can still use eqs. (\ref{dertReq})
 -- (\ref{deft1state}). The only difference is that this time $b_2$ comes out
negative:
 \begin{equation}\label{negb2}
 b_2 = - \sqrt{\frac {2M^{1/3}} {R_2} + \frac {2E} {M^{2/3}}},
 \end{equation}
 but the analogue of (\ref{defRbyb}) has the same form as before:
 \begin{equation}\label{defRbybatt2coll}
 R_2 = \frac {2M^{1/3}} {{b_2}^2 - 2E/M^{2/3}}.
 \end{equation}

   Hence,
 \begin{equation}\label{eqeta2'}
 \begin{split}
 t_B =& t_2 - \frac{1}{x^{3/2}} \left[ \pi + \arccos\left(\frac {x - {b_2}^2} {x +
{b_2}^2}\right) \right.\\
 &+ \left.\sqrt{1 - \left(\frac {x - {b_2}^2} {x + {b_2}^2}\right)^2}
\right],
 \end{split}
 \end{equation}
with (\ref{vetovetB}) still applicable at $t_1$, and from (\ref{eqeta2'}), the
equation to determine $x = -2E/M^{2/3}$ is:
 \begin{equation}\label{evoltovoldef}
 \chi_C(x) = 0,
 \end{equation}
 where
 \begin{equation}
   \label{evoltovoleq}
 \begin{split}
   \chi_C(x) :=&
      \sqrt{1 - \left(\frac {{b_1}^2 - x} {{b_1}^2 + x}\right)^2}
      - \sqrt{1 - \left(\frac {x - {b_2}^2} {x + {b_2}^2}\right)^2} \\
   &+ \pi + \arccos \left(\frac {x - {b_2}^2} {x + {b_2}^2}\right)
      - \arccos \left(\frac{{b_1}^2 - x} {{b_1}^2 + x}\right) \\
   &- x^{3/2}\left(t_2 - t_1\right).
 \end{split}
 \end{equation}

 \subsubsection{Conditions for solutions to exist}

   We have
 \begin{equation}\label{chicat0}
 \chi_C(0) = 2\pi, \qquad \lim_{x \to \infty} \chi_C(x) = - \infty,
 \end{equation}
 and so $\chi_C(x) = 0$ is guaranteed to have a solution for $x > 0$. The
surprising result is thus:

 \medskip

   {\it If the final state is in the collapse phase, so that $b_2 < 0$ then an $E <
0$ evolution exists between {\bf any} pair of states $(t_1, b_1)$ and $(t_2,
b_2)$.}

 \medskip

   Consistently with (\ref{chicat0}), we obtain:
 \begin{equation}\label{derchic}
\chi'_C(x) = - \frac {2b_1\sqrt{x}} {\left({b_1}^2 + x\right)^2} + \frac 
{2{b_2}^3} {\left({b_2}^2 + x\right)^2\sqrt{x}} - \frac 3 2 \sqrt{x}\left(t_2 - 
t_1\right),
 \end{equation}
 which, in consequence of $b_2 < 0$, is negative for all $x > 0$, and so the
solution of $\chi_C(x) = 0$ is unique.

   The initial point for solving $\chi_C(x) = 0$ numerically can be conveniently
found as follows: each square root and each arccos-term in (\ref{evoltovoleq})
is not larger than 1 at all values of $x > 0$, and where two of them equal 1,
the two others are strictly smaller than 1. Hence, their sum is strictly
smaller than 4, and
 \begin{equation}\label{limonchic}
 \chi_C(x) < 4 + \pi - x^{3/2}\left(t_2 - t_1\right) := \tilde{\chi_C}(x).
 \end{equation}
 Consequently, $x_Z < x_A$, where $x = x_Z$ is the solution of $\chi_C(x) = 0$,
and $x = x_A$ is the solution of $\tilde{\chi_C}(x) = 0$:
 \begin{equation}\label{approxx}
 x_A = \left[\frac {4 + \pi} {\left(t_2 - t_1\right)}\right]^{2/3}.
 \end{equation}

 \section{An initial velocity amplitude consistent with the CMB observations.}
 \label{ivelpro}

 \setcounter{equation}{0}

   Recent observations of the power spectrum of the CMB radiation
\cite{Padman96, WhiCoh02, HuDo02} show that the maximal value of the
temperature anisotropy, $\Delta T$, is approximately $80\mu K$, and is
attained for perturbations with the wavenumber $l \approx 200$, that
corresponds to the angular size of nearly 1 degree \cite{HuDo02, Wripa,
Hupa1, Hupa2}. (The numbers in the graphs in Refs. \cite{Wripa}
 -- \cite{Hupa2} imply that the ``angular size'' meant here is the radius
rather than diameter of the perturbation.) Consequently, one might take
$80\mu K$ as the upper limit of temperature variation and $1^{\circ}$ as
the upper limit on the angle subtended by the radius of the perturbation.
However, both these parameters far exceed the scale needed to model the
formation of a galaxy cluster, and even more so for a single galaxy.

   For simplicity we take the $k = 0$ Friedmann model as the background.
The relation between the physical size at recombination, and the angle on the 
CMB sky was given in paper I.

   As shown in the table in paper I, the $1^{\circ}$ angle on the CMB sky is
more than 10 times the angle occupied by the mass that will later become an
Abell cluster of galaxies, and about 250 times the angle occupied by the mass of
a single galaxy. In order to get the right estimate for a galaxy cluster, one
has to go down to the angular scales below 0.1$^{\circ}$. In this region of the
power spectrum observational data are sparse and carry big error bars (see Refs.
\cite{HuDo02, Hupa1}), but the amplitude is approx. 25$\mu K$.  At the angular
scales corresponding to single galaxies, 0.004$^{\circ}$, there is no good data.

   One way to get an upper bound on the fluctuations of velocity, is by
assuming the observed temperature variation is entirely due to local
matter motion at recombination causing a Doppler shift.  Now, the
cosmological redshift
 \begin{equation}
   \frac{T_r}{T_0} = \frac{S_0}{S_r} = 1 + z
 \end{equation}
 is due to the expansion of the space intervening between the emission
point at recombination and the observation point at us today.  A velocity
fluctuation at recombination
 --- i.e. a fluctuation of the velocity of the emitting fluid relative to
the average background flow
 --- superimposes an additional Doppler redshift
 \begin{equation}
   \left. \frac{\Delta T}{T} \right|_r
   = \frac{\Delta \nu}{\nu} = \frac{\Delta v}{c}
 \end{equation}
 which is unchanged by the subsequent cosmological redshift of both $T$
and $T + \Delta T$.  Therefore $\Delta T / T \approx 10^{-5}$ implies
$\Delta v / c \approx 10^{-5}$.

   To properly estimate the magnitude of velocity fluctuations right after
recombination, we use a paper by Dunsby \cite{Duns97}.  His eq (84) relates
the observed temperature fluctuations to a mode expansion of the velocity
perturbations%
 \footnote{
 The velocity used here is the emitting fluid velocity relative to the
normals to the constant temperature surfaces.
 }%
 .  From this and the various definitions elsewhere in the
paper, we find, for a single dominant mode of wavelength $\lambda$ and
relative velocity $\Delta v$,
 \begin{equation}
   \Delta v / c = \frac{\Delta T}{T} \frac{16 \pi^3 S_r^2}
   {3 (S_{,t})_r^3 \lambda^3}
 \end{equation}
 and, assuming a wavelength scale containing an Abell cluster mass at
recombination, ${\sim}5.87$~kpc,
 \begin{equation}
   \frac{\Delta T}{T} = \frac{25 \mu K}{2.72 K}
   ~~~~\rightarrow~~~~ \Delta v / c = 0.001895.
 \end{equation}
 This is distinctly more than was estimated by assuming $\Delta T$ was
entirely due to a Doppler shift, and the reason is that the effects (on
the observed $\Delta T/T$) of density, velocity and intrinsic temperature
perturbations partially cancel each other.

   This must now be recalculated as a fluctuation in the expansion
velocity, since our measure of velocity is $b = R_{,t}/M^{1/3}$, which is
constant in space in the Friedmann limit.

   We use the $k = 0$ FLRW dust model to estimate $R_{,t}$ at recombination.
For this model, $E = 0$ and $R(t, r) = r S(t)$,
 \begin{eqnarray}
   S & = & \left[ \frac{9 M_0 \left(t - t_B\right)^2}{2} \right]^{1/3},
 \end{eqnarray}
 and the past null cone is
 \begin{equation}
   r_{pnc} = \left( \frac{6}{M_0} \right)^{1/3} \left[
   \left(t_0 - t_B\right)^{1/3} - \left(t - t_B\right)^{1/3} \right].
 \end{equation}
 Thus
 \begin{equation}
 \begin{split}
   R_{,t}|_{pnc} &= 2 \left[ \left( \frac{t_0 - t_B}{t - t_B} \right)^{1/3}
                            - 1 \right]
                 = 2 \left[ \sqrt{\frac{S_0}{S}}\; - 1 \right] \\
                 &= 2 \left[ \sqrt{1 + z}\; - 1 \right].
 \end{split}
 \end{equation}
 We can now write, for recombination,
 \begin{equation} \label{velampnum}
 \begin{split}
   \Delta b / b &= \Delta R_{,t} / R_{,t} = (\Delta v / c) / R_{,t} \\
   &= \frac{\Delta T}{T} \frac{16 \pi^3 S_r^2}{3 (S_{,t})_r^3 \lambda^3}
     \frac{1}{2 [ \sqrt{1 + z_r}\; - 1 ]} \\
   &= 0.001895 / (2 [ \sqrt{1 + 1000}\; - 1 ]) \\
   &= 9.896 \times 10^{-5} \approx 10^{-4}.
 \end{split}
 \end{equation}

   Returning to inhomogeneous models, what should the radius of the
velocity perturbation be?  The initial density profile considered in paper
I had an initial fluctuation of small mass which later accreted more mass
to form the condensation.  But with a velocity perturbation, accretion
would be caused by a small initial inwards perturbation over the whole
final mass.

   Nothing is known from observations about the possible profiles of the
initial velocity perturbation. Lacking this, we are free to choose the
profile that will be easy to calculate with.

 \section{Profiles \& Fitting} \label{profit}

 \setcounter{equation}{0}

 \subsection{Geometrical Units}

   For convenience of computation, geometrical units were employed in
order to avoid extremely large numerical values.  If a mass $M_G$ is
chosen as the geometrical unit of mass, then the geometrical length and
time units are $L_G = G M_G / c^2$ and $T_G = G M_G / c^3$.  It was
convenient to define different units for different sized structures.  The
units used for Abell clusters, and for voids are summarised in table
\ref{GeomU}.

 \begin{center}
 ------------------------------ \\
 Table \ref{GeomU} goes here \\
 ------------------------------ \\
 \end{center}

 \subsection{Background Model.}

 \setcounter{equation}{0}

We use the $k = 0$ Friedmann model as our background or reference model, and 
all our compensation radii are calculated with respect to this model. It is 
chosen primarily for simplicity, and any other Friedmann model would do.

For this model, with $M$ as radial coordinate, the areal radius is
 \begin{equation}
   R_b = \left( \frac{9 M t^2}{2} \right)^{1/3},
 \end{equation}
 where subscript b indicates the background rather than L-T functions or
values, and the density is
 \begin{equation}
 \label{rho_b(t)}
   8 \pi \rho_b = \frac{4}{3 t^2}.
 \end{equation}
 The velocity (rate of change of $R$) is
 \begin{equation}
   (R_{,t})_b = \left( \frac{4 M}{3 t} \right)^{1/3},
 \end{equation}
 so that scaled velocity becomes
 \begin{equation}
   b_b = \frac{(R_{,t})_b}{M^{1/3}}
          = \left( \frac{4}{3 t} \right)^{1/3}.
 \end{equation}
 We could replace $t$ with $t - t_B$ in all of the above, but we took
$t_B = 0$ for our background model.

 \subsection{Choice of Density Profile at $t_2 = 1.4\cdot10^{10}$~yr}

 \subsubsection{An Abell Cluster}
 \label{AbellClusterProfileChoices}

   The widely used `universal profile' (UP) for the variation of density
with radius in condensed structures (e.g. \cite{NaFW1997}), discussed in
section \ref{ProfInspAstr} was found not suitable because of its divergent
central density, and also it does not convert to $\rho(M)$ in any nice
analytic form%
 \footnote{
 We checked whether this profile would encounter a black hole horizon $R =
2M$. The mass within radius $R$ is
 $$
   M = \frac{4 \pi}{3} \rho_b \delta R_s^3 \left(
       \ln \left( 1 + \frac{R}{R_s} \right)
       - \frac{1}{\left( 1 + \frac{R}{R_s} \right)} \right),
 $$
 so while the function $R - 2M$ could in principle have a root, it does
not have one for reasonable astronomical parameter values.}%
 .  We therefore sought a profile that is similar in the region for which
data are known, but had a finite --- if large --- central density,
is expressed as $\rho(M)$ and has an analytic $R(M)$.

   To determine the parameters of a profile, we fitted it to the
 mass-radius data from \cite{MVFS1999} for A2199 \& A496, mostly A2199
They give the mass at 0.2 Mpc and at 1 Mpc for each cluster, estimated
from fitting a model to the ROSAT X-ray luminosity data (see table
\ref{AbellData}), so this allows the determination of two parameters.  We
also attempted to control the maximum density and the compensation radius
by trying profiles with a total of 4 parameters, but these adjustments did
not have a large effect overall.

 \begin{center}
 ------------------------------ \\
 Table \ref{AbellData} goes here \\
 ------------------------------ \\
 \end{center}

\begin{figure*}[h]
 \begin{center}
 \parbox{14cm}{
 \includegraphics[scale = 0.6]{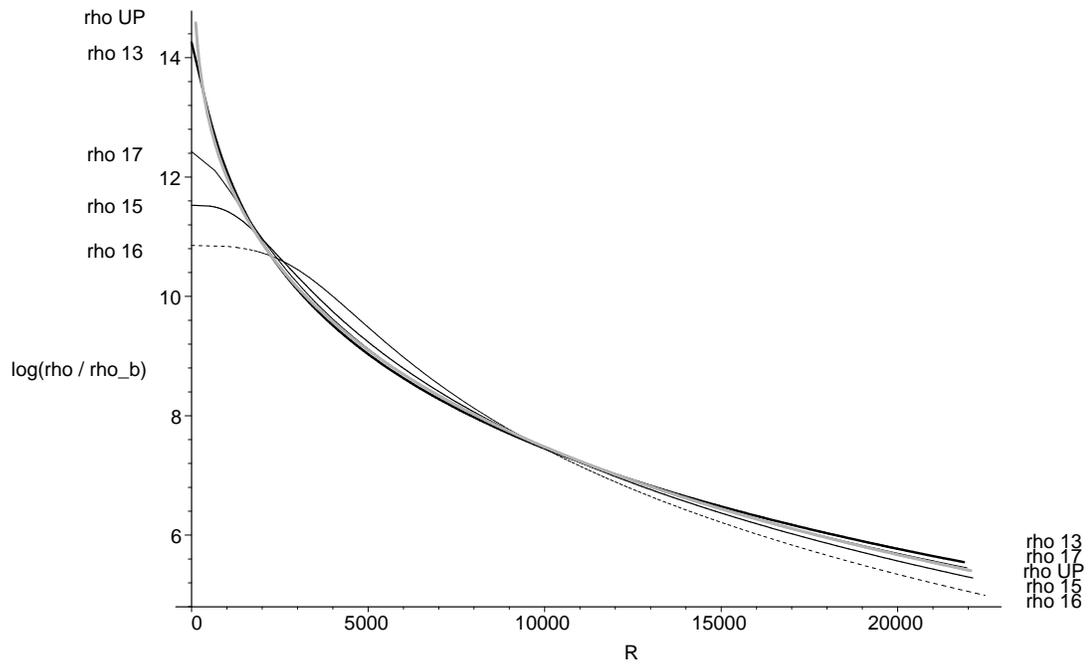}
 \caption{
 \label{ACProfCompFig}
 \footnotesize
 Comparison of density profiles used for Abell clusters.  The curves are
identified by the numbers given in table \ref{ACProfileTab}.  Although
each curve is matched to the same pair of $(R, M)$ points (see table
\ref{AbellData}), this only ensures the {\it average} density within each
of those two radii is the same for all profiles, not the actual density.
 }
 }
 \end{center}
 \end{figure*}

   We considered quite a variety of functional forms for our profiles.
For a number of them we fixed the parameters by fitting to the mass-radius
data.  A comparison of four of these fits with the UP (also fitted to the
same data) is shown in figure \ref{ACProfCompFig}, and in table
\ref{ACProfileTab}.  Several runs were done with each of these.

 \begin{center}
 ------------------------------ \\
 Table \ref{ACProfileTab} goes here \\
 ------------------------------ \\
 \end{center}

 \begin{center}
 ------------------------------ \\
 Figure \ref{ACProfCompFig} goes here \\
 ------------------------------ \\
 \end{center}

 \subsubsection{A Void}

\begin{figure*}[h]
 \begin{center}
 \parbox{14cm}{
 \includegraphics[scale = 0.6]{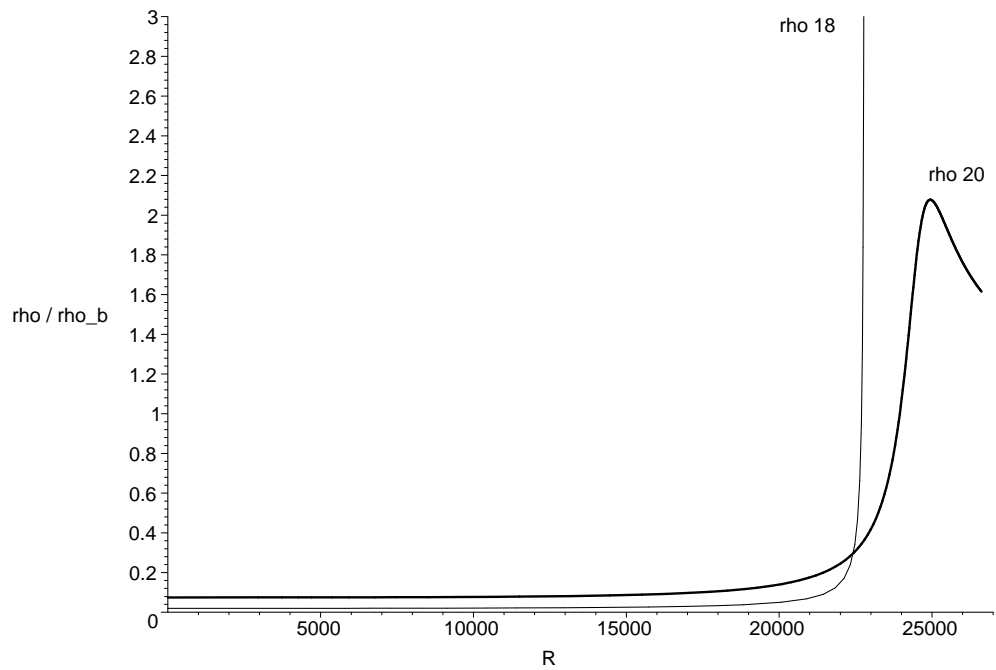}
 \caption{
 \label{VoidProfCompFig}
 \footnotesize
 Comparison of density profiles used for voids.  The curves are
identified by the numbers given in table \ref{VoidProfileTab}.
 }
 }
 \end{center}
 \end{figure*}

The main feature of a void is a low density region surrounded by walls of 
higher than background density.  Our initial attempt to model a void was to 
choose a profile with $\rho(M)$ increasing, but this was problematic because we 
did not have good control over how high the density became before we reached 
the compensation radius.  Thus we designed a profile with a density maximum at 
the wall, decreasing to background at large $M$. In this way no unreasonable 
density values are encountered even if the compensation radius is well beyond 
the density maximum.  The profiles tried are summarised in table 
\ref{VoidProfileTab} and figure \ref{VoidProfCompFig}.

 \begin{center}
 ------------------------------ \\
 Table \ref{VoidProfileTab} goes here \\
 ------------------------------ \\
 \end{center}

 \begin{center}
 ------------------------------ \\
 Figure \ref{VoidProfCompFig} goes here \\
 ------------------------------ \\
 \end{center}

 \subsection{Choice of Velocity \& Density Profiles at $t_1 = 10^5$~yr}

\begin{figure*}[h]
 \begin{center}
 \parbox{14cm}{
 \includegraphics[scale = 0.6]{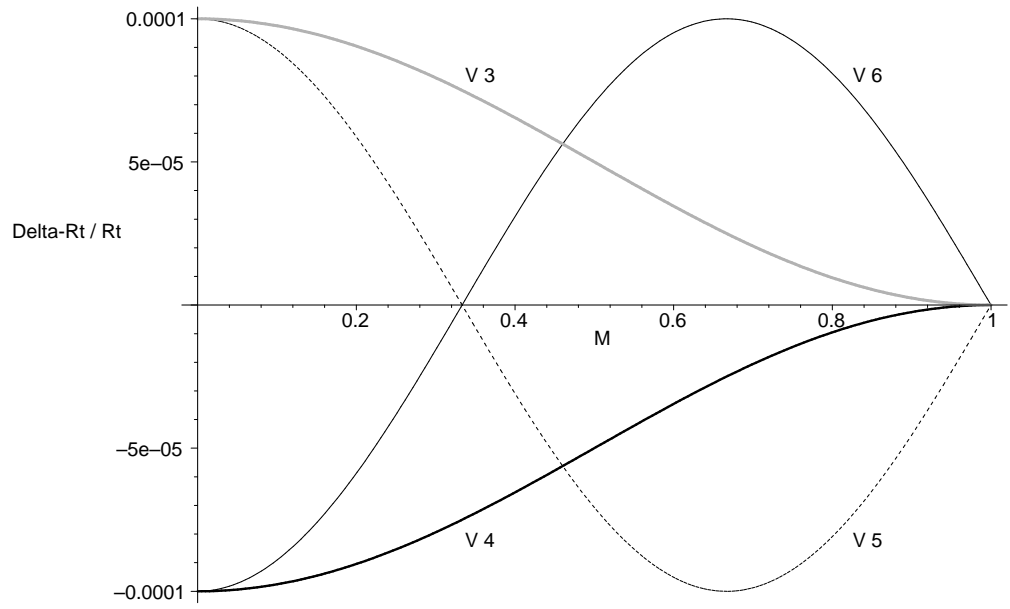}
 \caption{
 \label{VelProfCompFig}
 \footnotesize
 Comparison of initial velocity profiles at recombination.  The curves
are identified by the numbers given in table \ref{VelProfileTab}.
 }
 }
 \end{center}
 \end{figure*}

   For velocity fluctuations at recombination, we merely choose a harmonic
wave of very low amplitude, with either 1 or 1.5 wavelengths across the
diameter of interest.  The boundary of the region of interest is the
radius that will become the compensation radius at time 2.  See table
\ref{VelProfileTab} and figure \ref{VelProfCompFig} for the details.

 \begin{center}
 ------------------------------ \\
 Table \ref{VelProfileTab} goes here \\
 ------------------------------ \\
 \end{center}

 \begin{center}
 ------------------------------ \\
 Figure \ref{VelProfCompFig} goes here \\
 ------------------------------ \\
 \end{center}

   A few runs were done from initial density profiles, which are given in
table \ref{InFnDenProfileTab}.

 \begin{center}
 ------------------------------ \\
 Table \ref{InFnDenProfileTab} goes here \\
 ------------------------------ \\
 \end{center}

 \section{Programs}

 \setcounter{equation}{0}

   The foregoing procedure to solve for the L-T functions $E(M)$ and
$t_B(M)$ from given profiles $R_{,t1}(M)$ and $\rho_2(M)$, was coded in a
set of Maple \cite{MapleURL} programs that also re-derived and verified
the formulas, and fitted the profiles to the data.  The programs to evolve
the resulting L-T model and plot the results were written in MATLAB
\cite{MATLABURL}.  Unlike the Maple programs, the latter did not need
large changes from the Paper I versions, but were in any case refined and
made more flexible, and in particular were adjusted to handle times where
the big crunch has already happened in parts of the model.%
 \footnote{These will be useful in modelling the formation of black holes.}
 Also the density to density solving program (Maple) was modified to 
handle both $\rho_1 > \rho_2$ and $\rho_2 \geq \rho_1$. 

   As before, slightly different variables from those used in the foregoing
algebra were more convenient for the numerics.  In particular, for the
velocity to density case, in place of $x$, $a_2$, $b_1$ and $t_i$, we use
$y = x a_2$, $\beta = {b_1}^2 a_2$ and $z_i = t_i/{a_2}^{3/2}$.  For the
velocity to velocity case we use $y = x/{b_2}^2$, $\gamma =
{b_1}^2/{b_2}^2$ and $z_i = t_i {b_2}^3$. Expressions for the following
limits at the origin are also needed,
 \begin{equation}
 \begin{split}
   a_i(0) &= \lim_{M \rightarrow 0} \frac{R(t_i, M)}{M^{1/3}}
            = \left( \frac{6}{\kappa \rho(t_i, 0)} \right)^{1/3} \\
           &= \frac{2}{{b_i}^2(0) \mp x(0)},
 \end{split}
 \end{equation}
 \begin{equation}
   b_i(0) = \lim_{M \rightarrow 0} \frac{R_{,t}(t_i, M)}{M^{1/3}}
            = \epsilon \sqrt{\frac{2}{a_i(0)} \pm x(0)},
 \end{equation}
 where the upper sign is for the hyperbolic case and the lower one for the
elliptic case; $\epsilon = +1$ is for expansion, and $\epsilon = -1$ for
collapse.  Since $x(0)$ comes from the solution procedure, we have an
expression for $a_i(0)$, whether velocity or density profiles are given,
but for $b_i(0)$ when the velocity profile is given, the limit must be
calculated using the explicit velocity profile function.  Alternatively,
the velocity may be given in the form $b_i(M)$.

   For reconstructing the model evolution from the solution functions $x$
and $t_B$, the formulas in Paper I still suffice.

 \section{Runs}

 \setcounter{equation}{0}

   The runs that we did are listed in table \ref{RunList}, which also lists
the relevant figure numbers where appropriate, and the significance of
each run.

 \begin{center}
 ------------------------------ \\
 Table \ref{RunList} goes here \\
 ------------------------------ \\
 \end{center}

 \begin{center}
 ------------------------------ \\
 Run Figures go here \\
 ------------------------------ \\
 \end{center}

 \section{Results}\label{Results}

   A run is defined by specifying one of the initial profiles, $\rho_1$ \&
$R_{,t1}$, and one of the final profiles, $\rho_2$ \& $R_{,t2}$.  The
results of a particular run are the two L-T functions $E(M)$ and $t_B(M)$
 --- the local energy/geometry function and the local bang time function
 --- that determine the L-T model.  The model so defined may then be evolved
over any time range, or its whole lifetime, and in particular, it can be
verified that the given profiles are reproduced. In addition, of the 4
profiles $\rho_1$, $R_{,t1}$, $\rho_2$ and $R_{,t2}$, the two that were
not used to define the run are also determined by $\{E(M), t_B(M)\}$, and
are of interest, especially the initial one.

\begin{figure*}[h]
 \begin{center}
 \parbox{14cm}{
 \includegraphics[scale = 0.5]{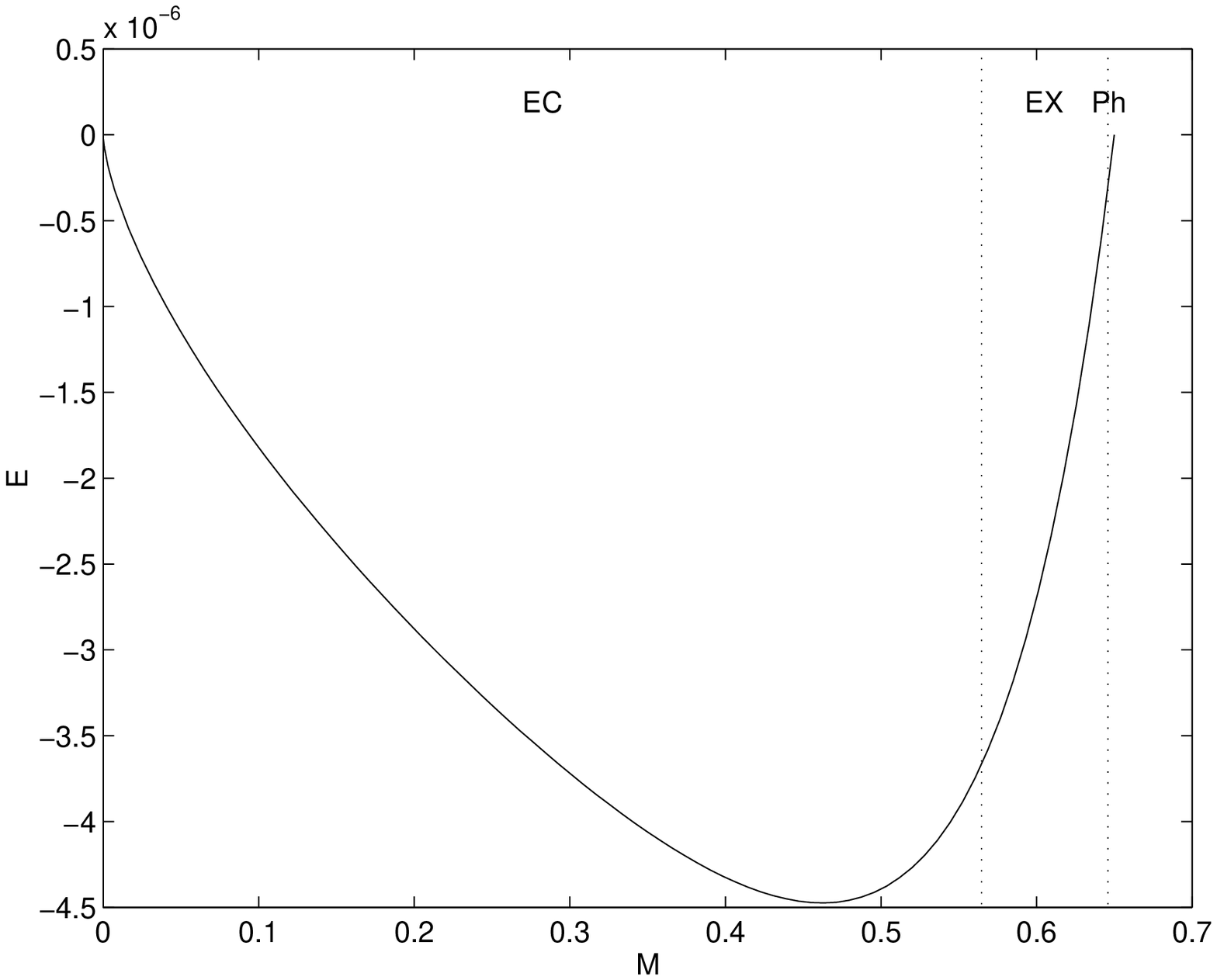}
 ${}$ \\[-1mm]
 \includegraphics[scale = 0.5]{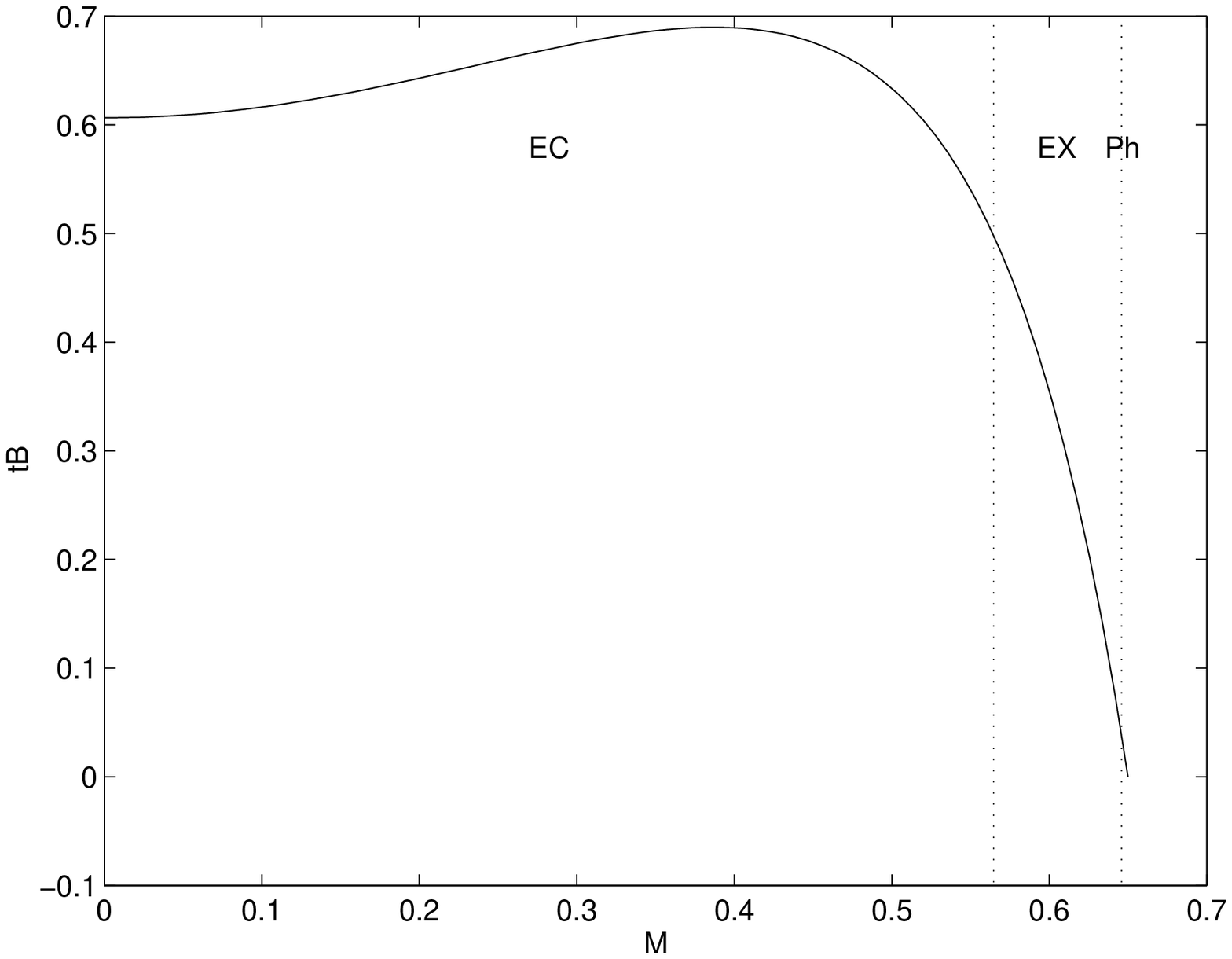}
 ${}$ \\[-1mm]
 \hspace*{-4mm}
 \includegraphics[scale = 0.5]{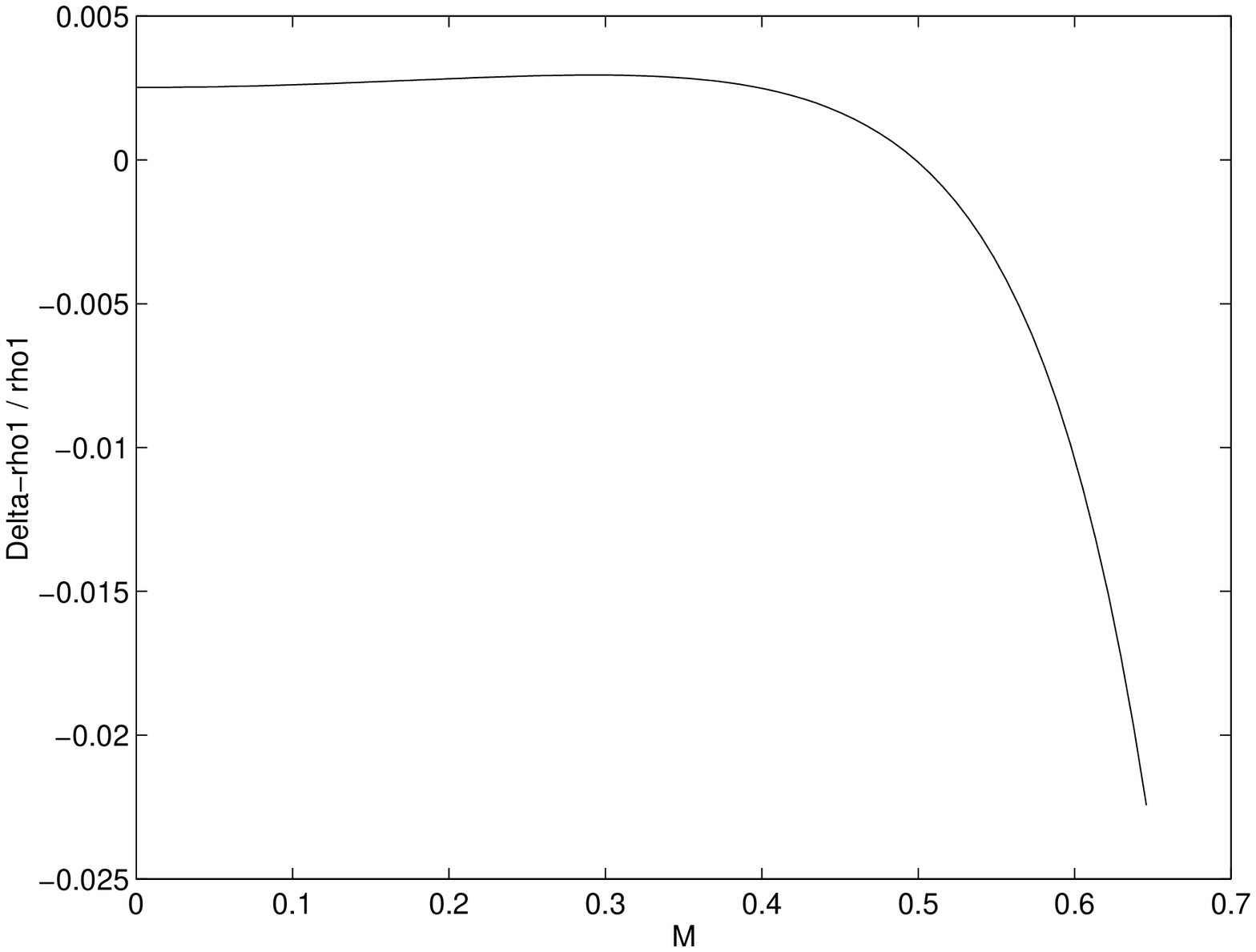}
 \caption{
 \label{V4D16fig}
 \footnotesize
 $E(M)$, $t_B(M)$ and the $\rho_1(M)$ fluctuation for run
$V$i4$\rho$f16, in which a region of initially low expansion rate evolves
into an Abell cluster.  In this and other figures, ``H", ``EX", ``EC",
``Pe" and ``Ph" indicate regions that are hyperbolic, elliptic and still
expanding at time $t_2$, elliptic and recollapsing at time $t_2$, elliptic
but within the range for a series expansion about the parabolic model, and
hyperbolic but within the range for a series expansion about the parabolic
model.
  }
 }
 \end{center}
 \end{figure*}

\begin{figure*}[h]
 \hspace*{-10mm}
 \parbox{17.5cm}{
 \hspace*{-124mm}
 \includegraphics[scale = 0.3]{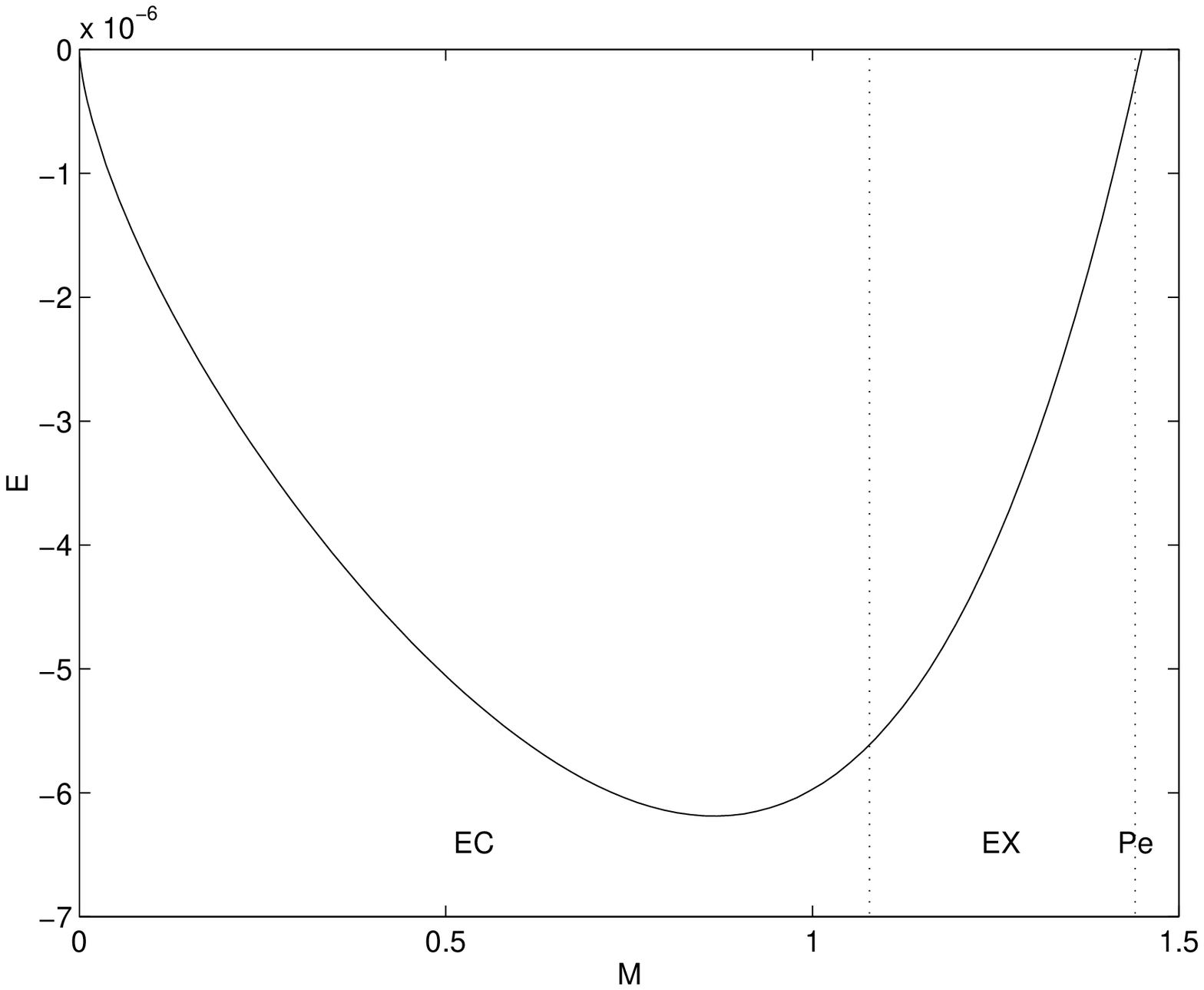}
 ${}$ \\[-43mm]
 \hspace*{0mm}
 \includegraphics[scale = 0.3]{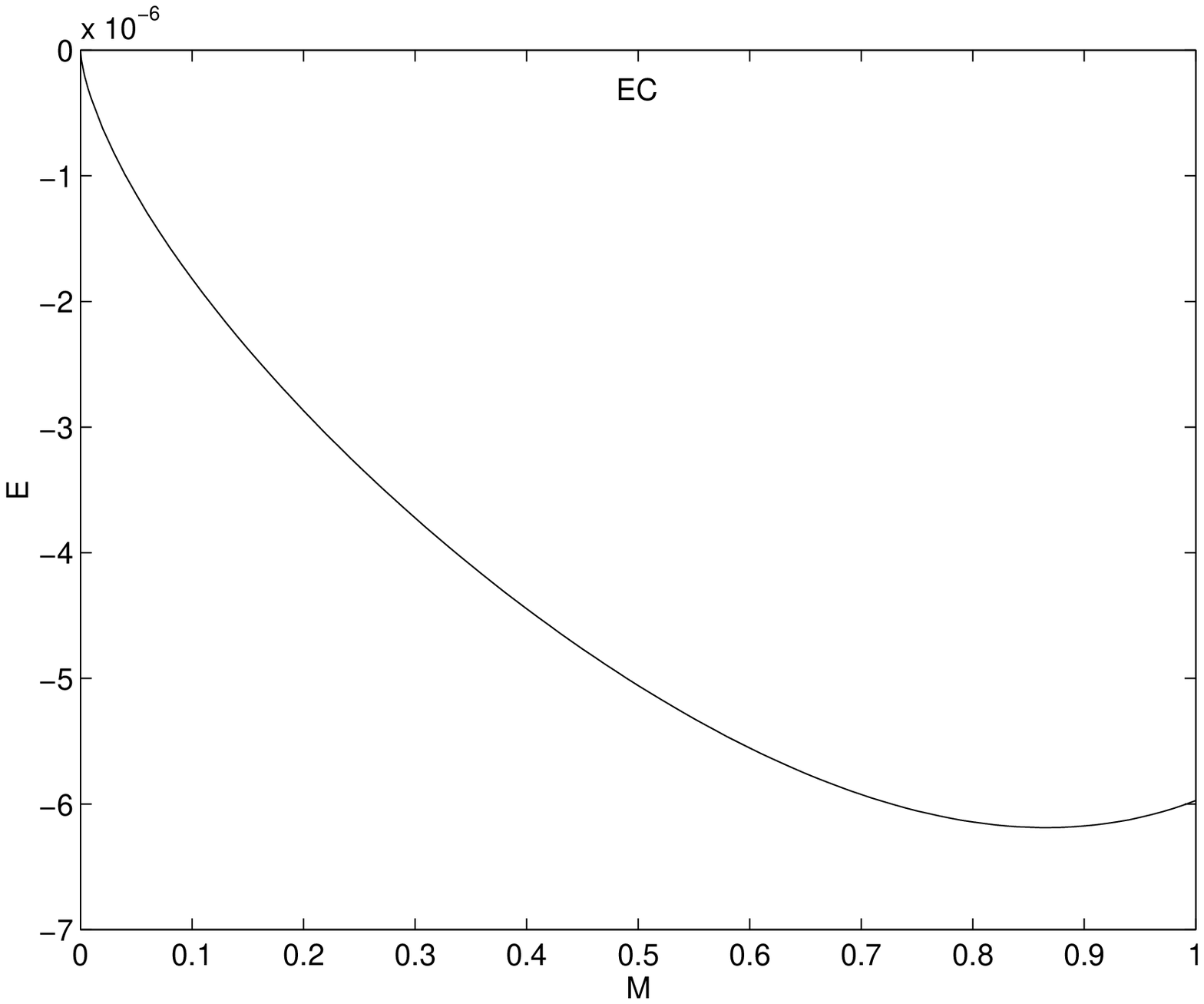}
 ${}$ \\[-43mm]
 \hspace*{124mm}
 \includegraphics[scale = 0.3]{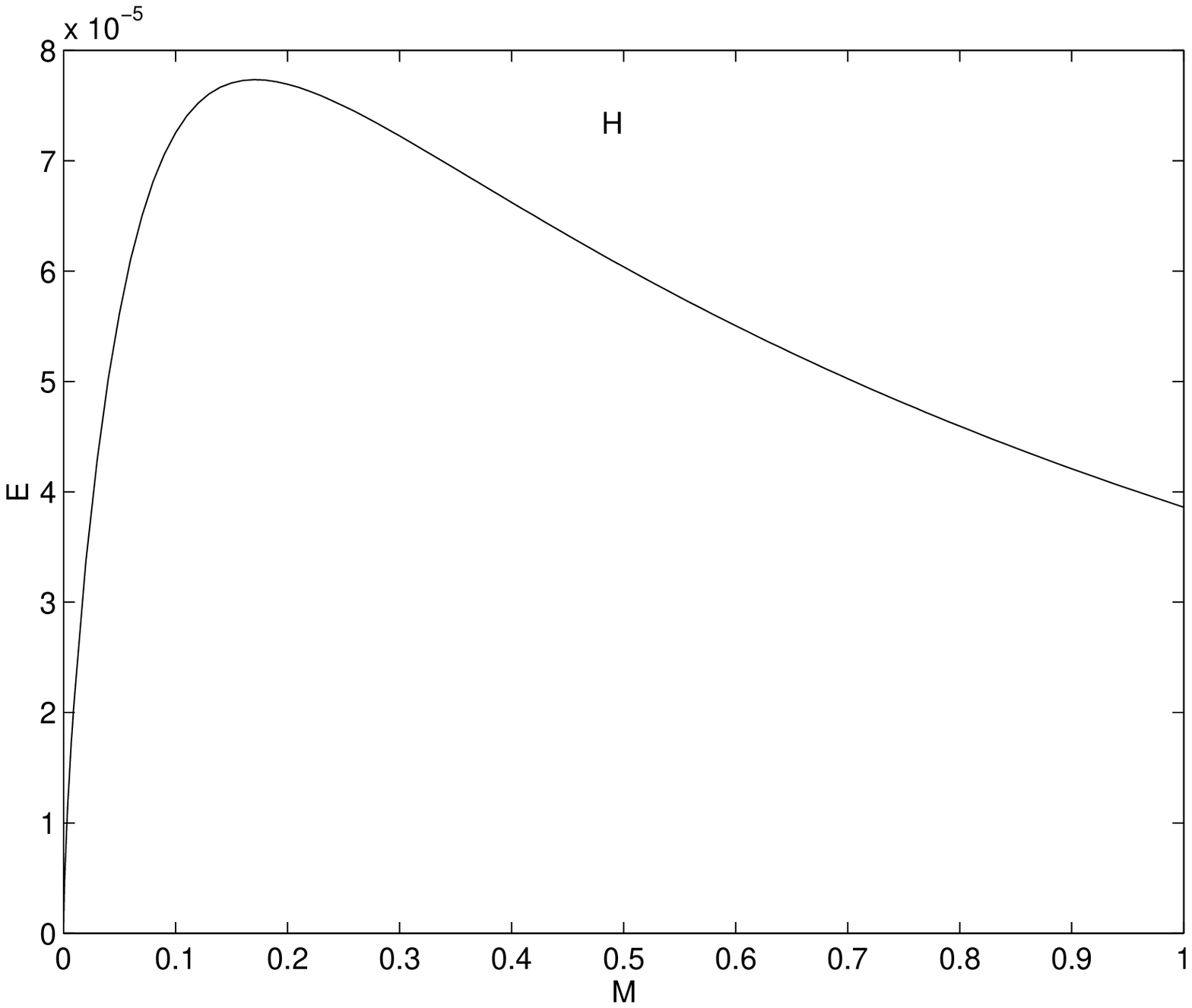}
 ${}$ \\[-1mm]
 \hspace*{-124mm}
 \includegraphics[scale = 0.3]{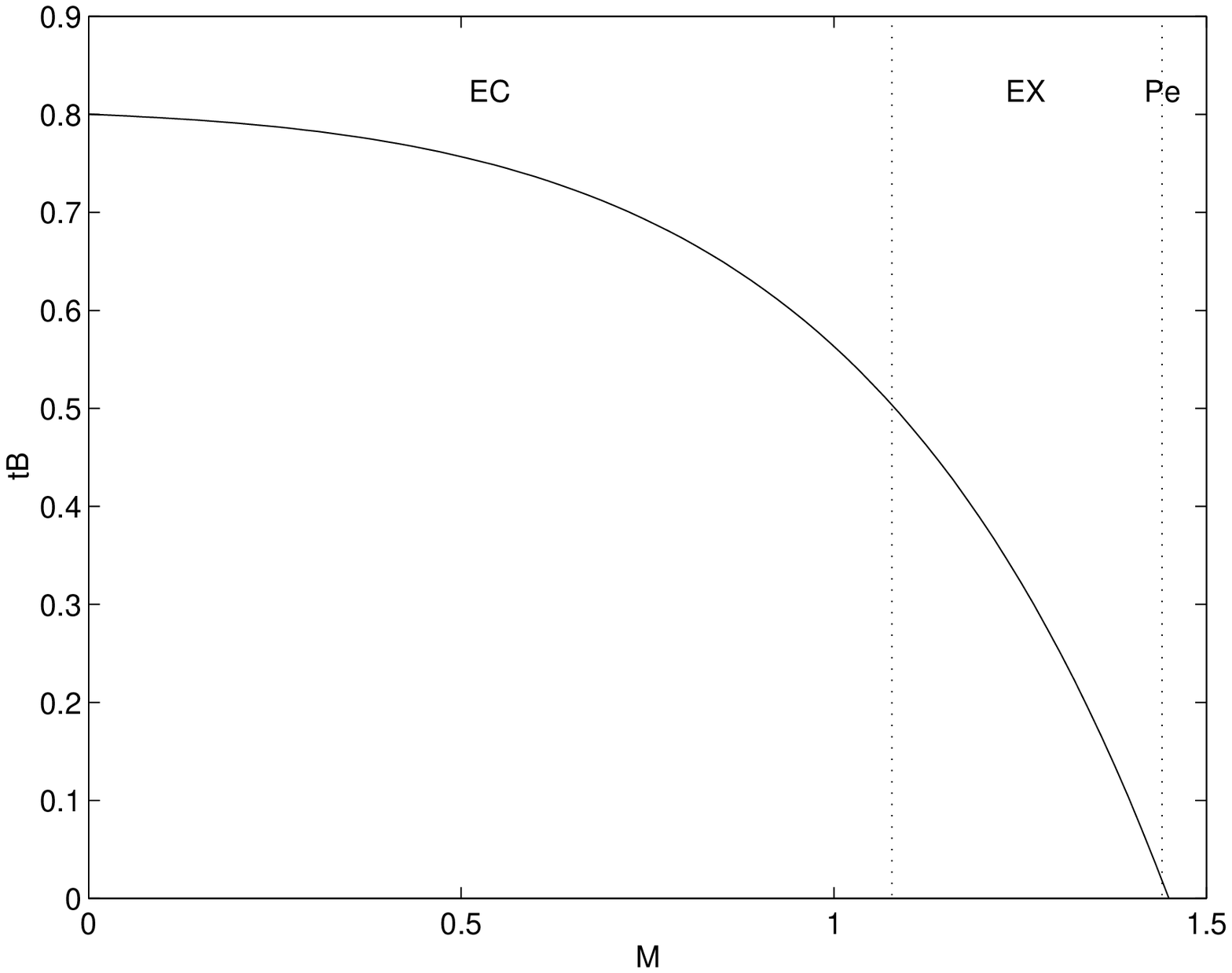}
 ${}$ \\[-42mm]
 \hspace*{0mm}
 \includegraphics[scale = 0.3]{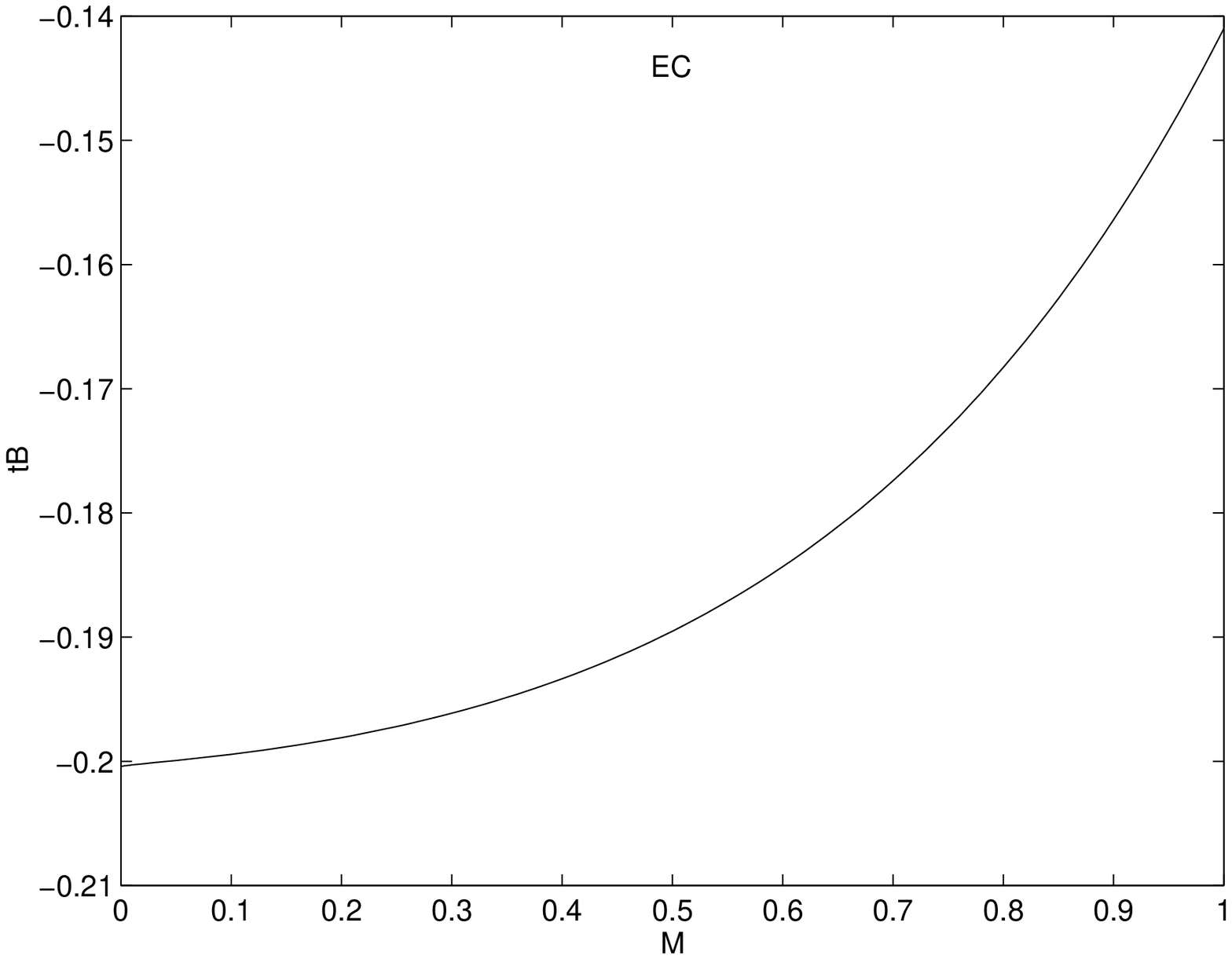}
 ${}$ \\[-42mm]
 \hspace*{123mm}
 \includegraphics[scale = 0.3]{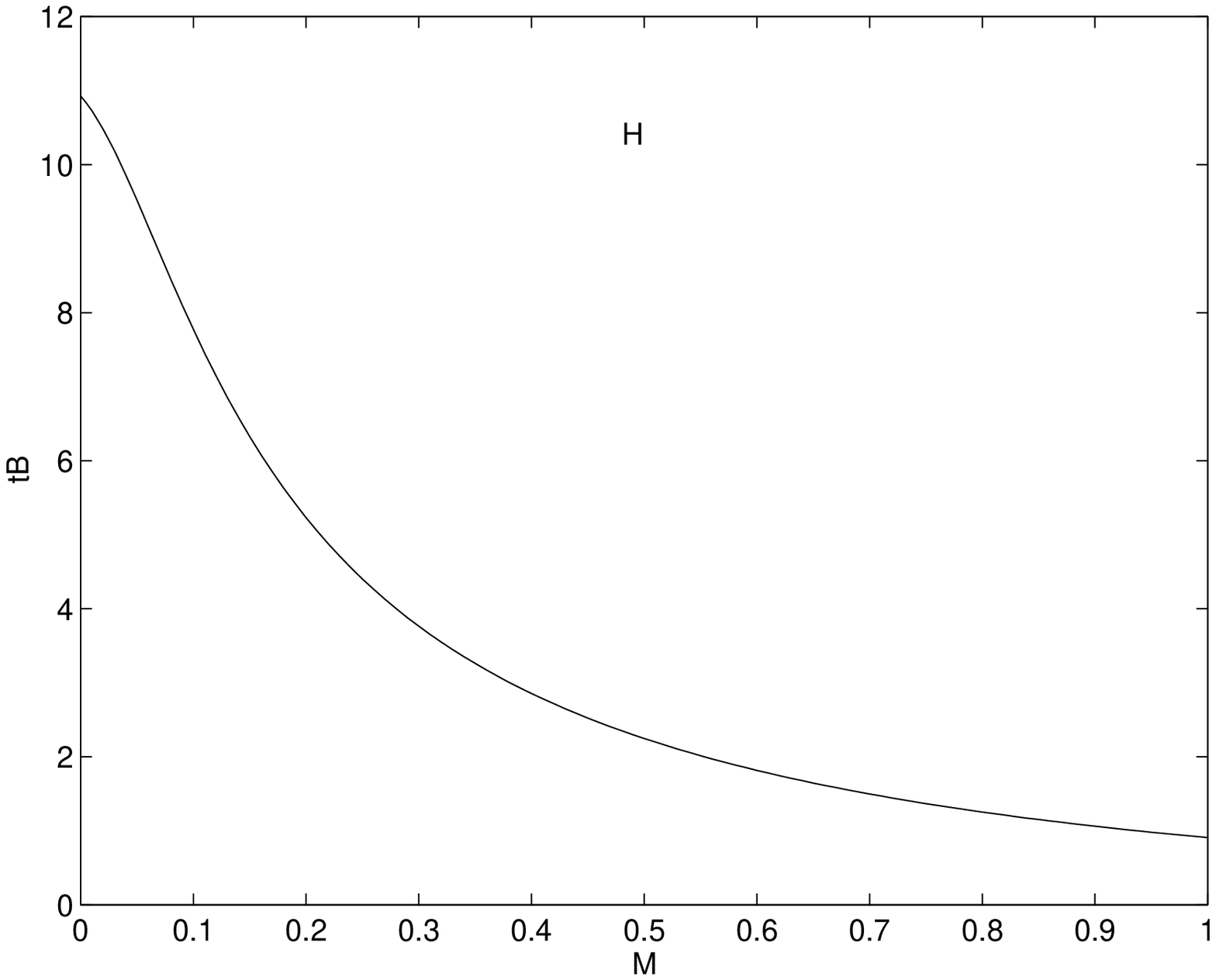}
 ${}$ \\[-1mm]
 \hspace*{-124mm}
 \includegraphics[scale = 0.3]{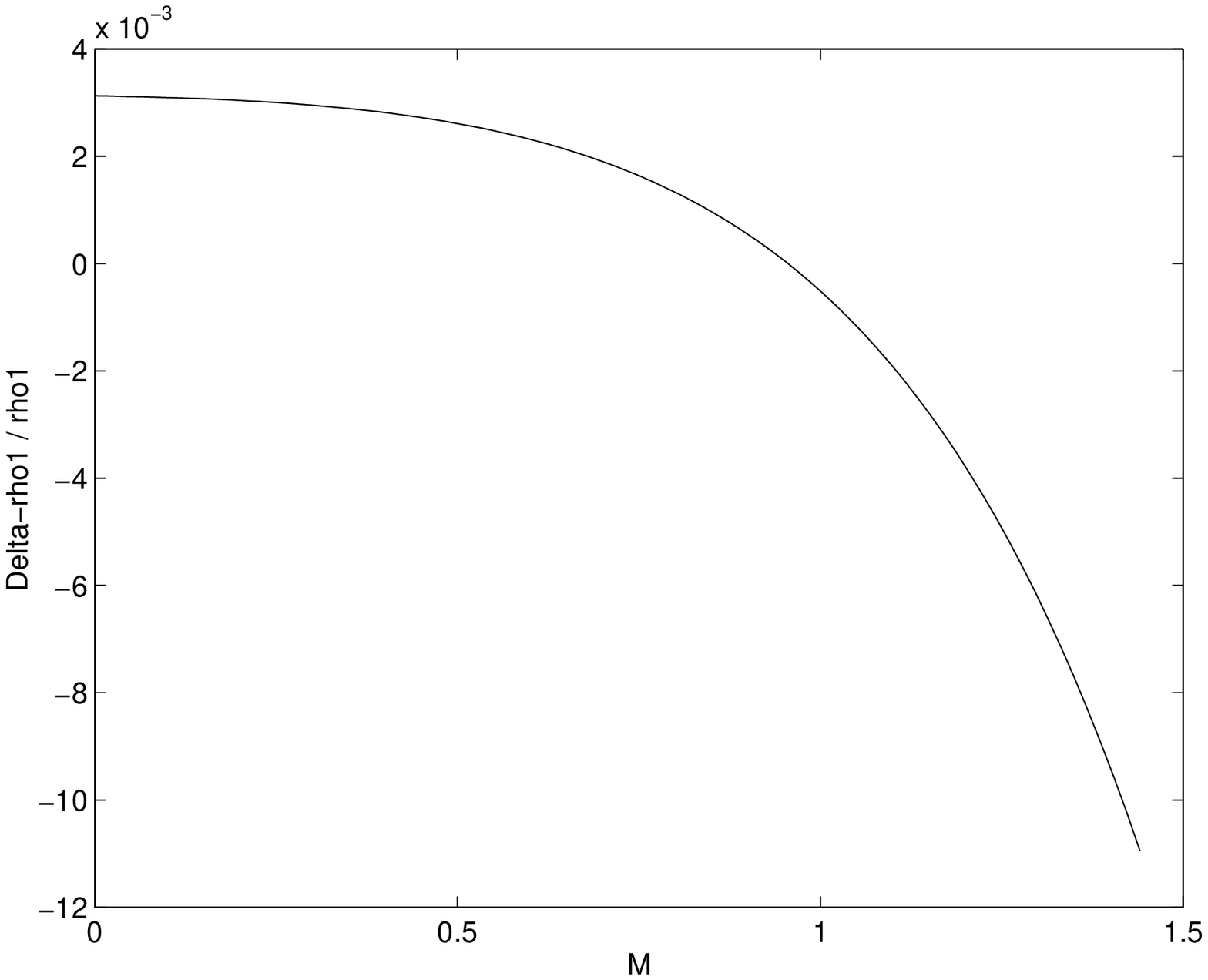}
 ${}$ \\[-44.5mm]
 \hspace*{1mm}
 \includegraphics[scale = 0.3]{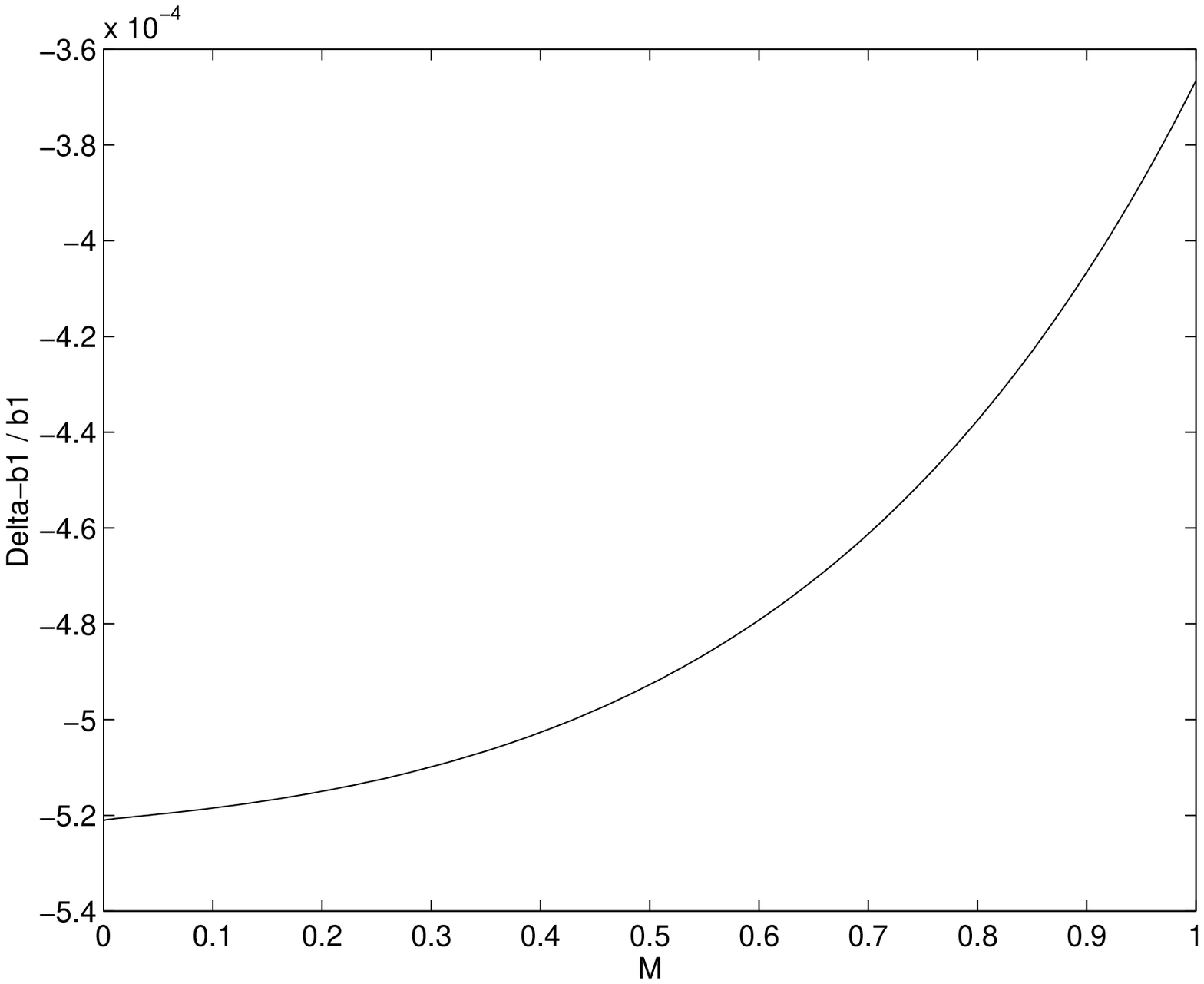}
 ${}$ \\[-43mm]
 \hspace*{121mm}
 \includegraphics[scale = 0.3]{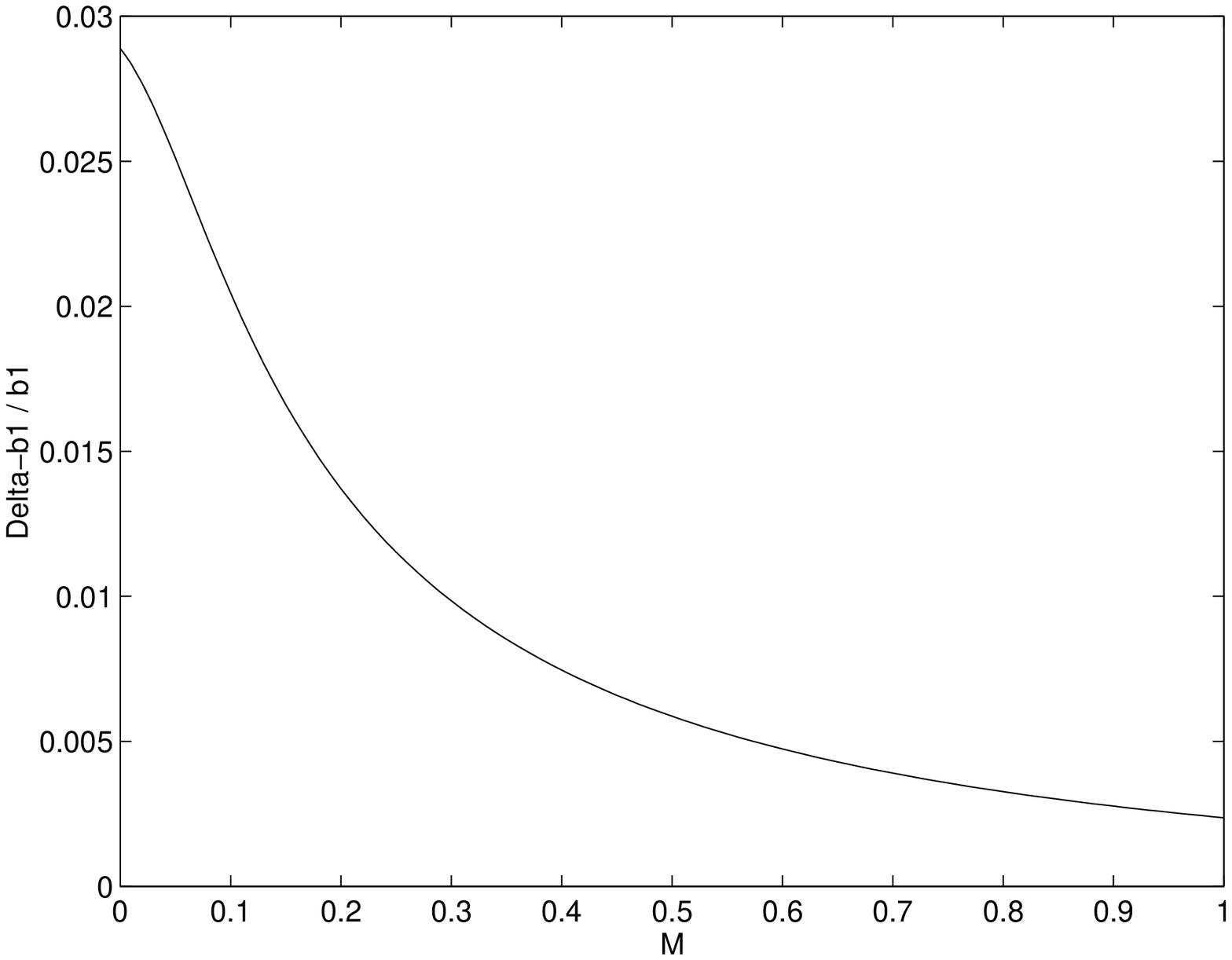}
 \\[1mm]
 \hspace*{-124mm}
 \includegraphics[scale = 0.3]{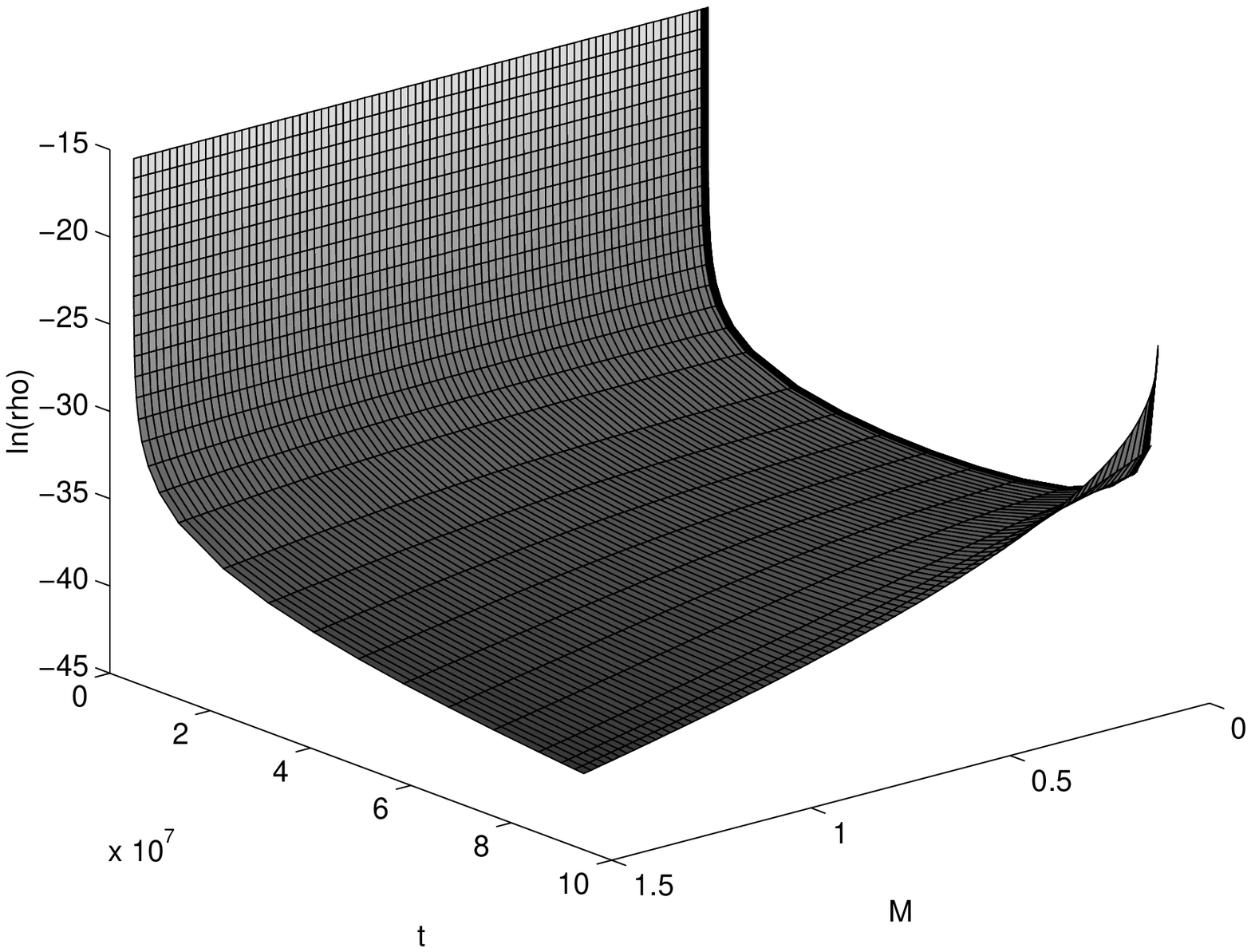}
 ${}$ \\[-42mm]
 \hspace*{2mm}
 \includegraphics[scale = 0.3]{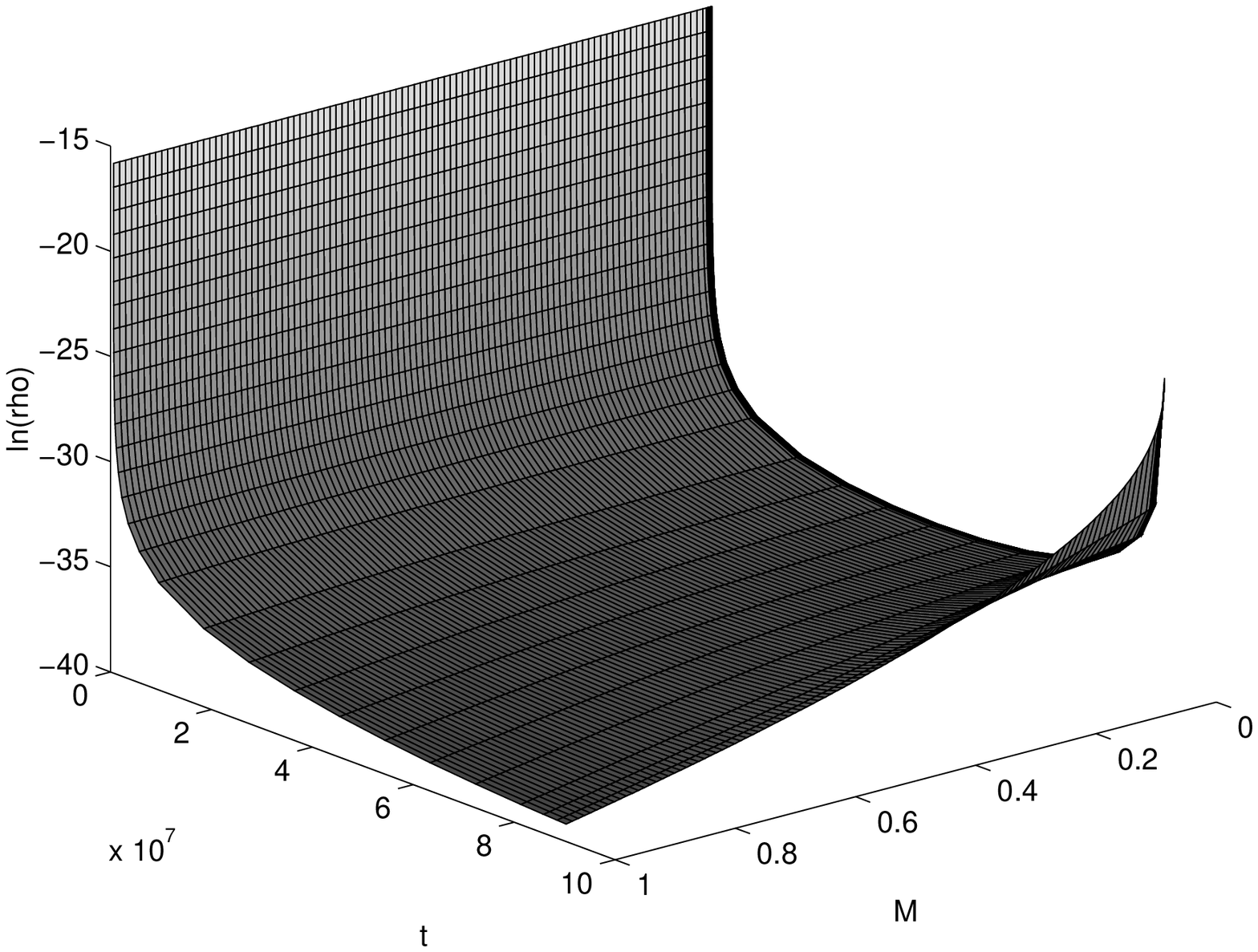}
 ${}$ \\[-42mm]
 \hspace*{122mm}
 \includegraphics[scale = 0.3]{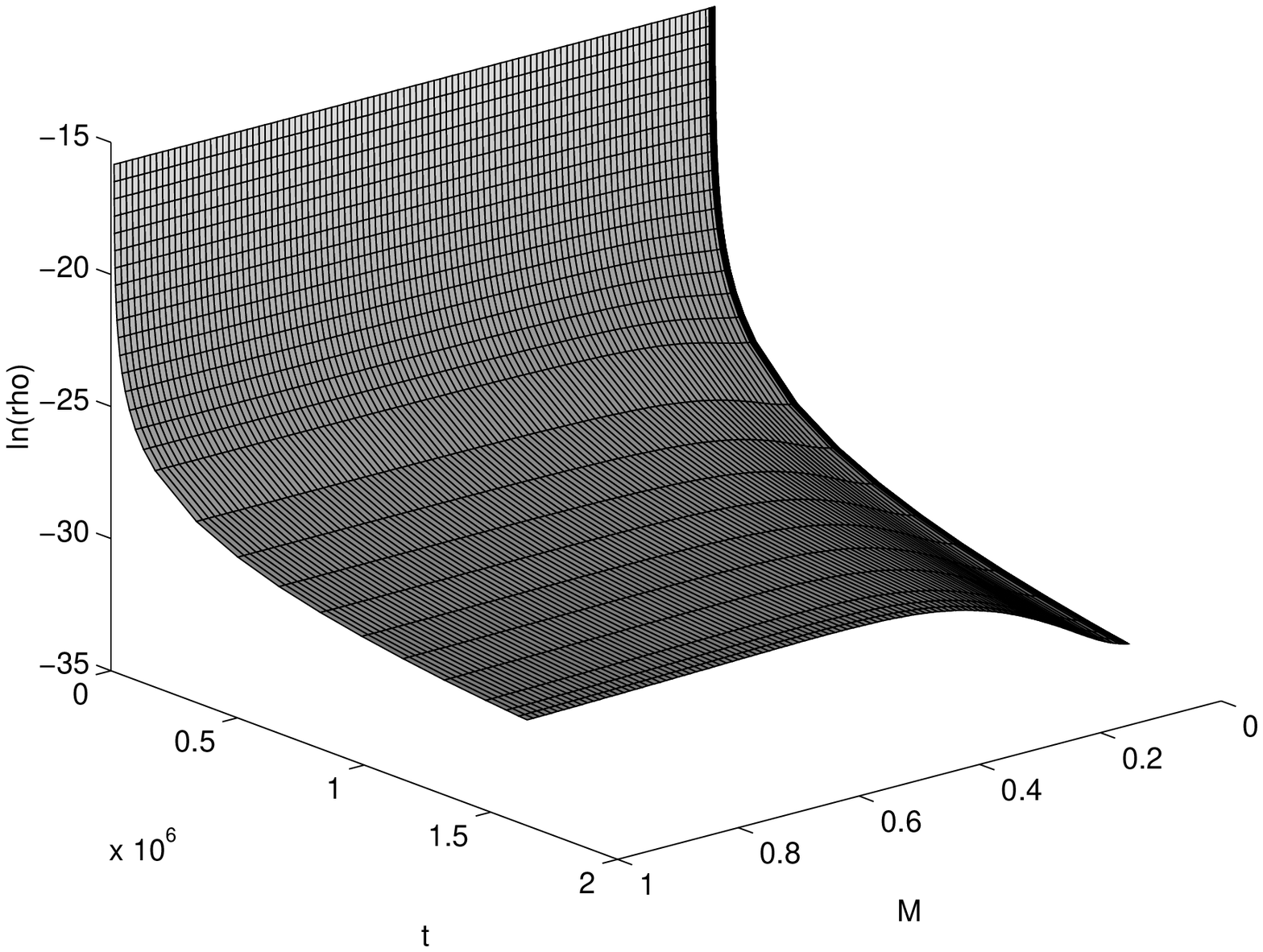}
 \caption{
 \label{V0D13D0D13D0D20fig}
 \footnotesize
 $E(M)$, $t_B(M)$ the $\rho_1(M)$ or $R_{,t1}(M)$ fluctuation, and the
$\rho(t, M)$ evolution (top to bottom) for runs $V$i0$\rho$f13,
$\rho$i0$\rho$f13 and $\rho$i0$\rho$f20 (left to right).  The left and
middle columns show the pure density and pure velocity fluctuations needed
at recombination to create an Abell cluster today.  The middle and right
columns show the difference in the pure velocity fluctuations needed at
recombination to create an Abell cluster and a void today.
 }
 }
 \end{figure*}

\begin{figure*}[h]
 \begin{center}
 \parbox{14cm}{
 \includegraphics[scale = 0.7]{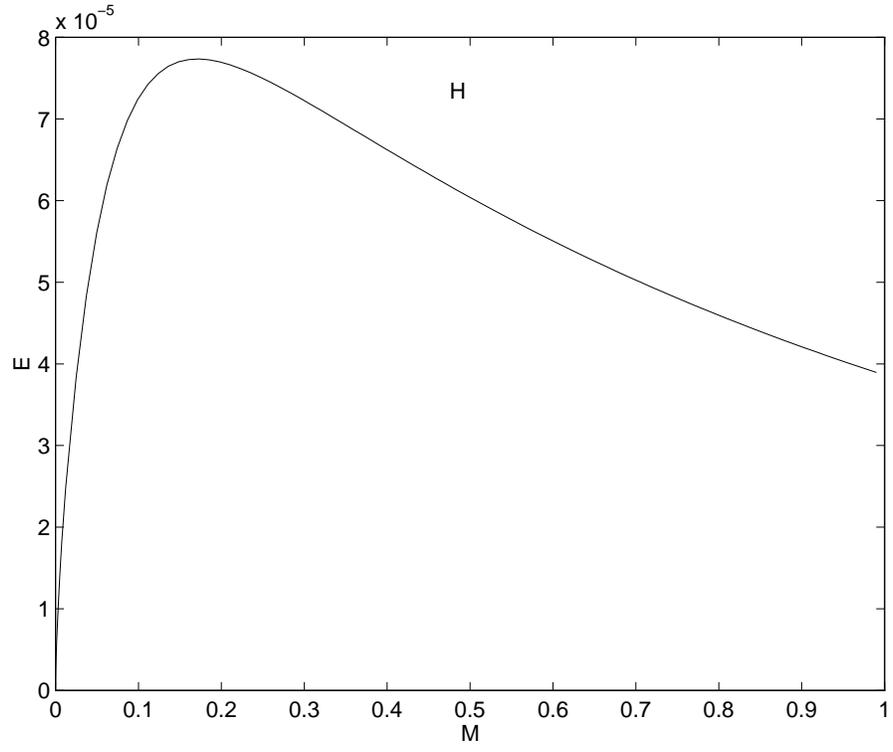}
 \caption{
 \label{V3D20EMfig}
 \footnotesize
 $E(M)$ for run $V$i3$\rho$f20, which successfully creates a present day
void.  Though shell crossings are inevitable, they don't occur for a long
time after the present.
 }
 }
 \end{center}
 \end{figure*}

\begin{figure*}[h]
 \begin{center}
 \parbox{14cm}{
 \includegraphics[scale = 0.7]{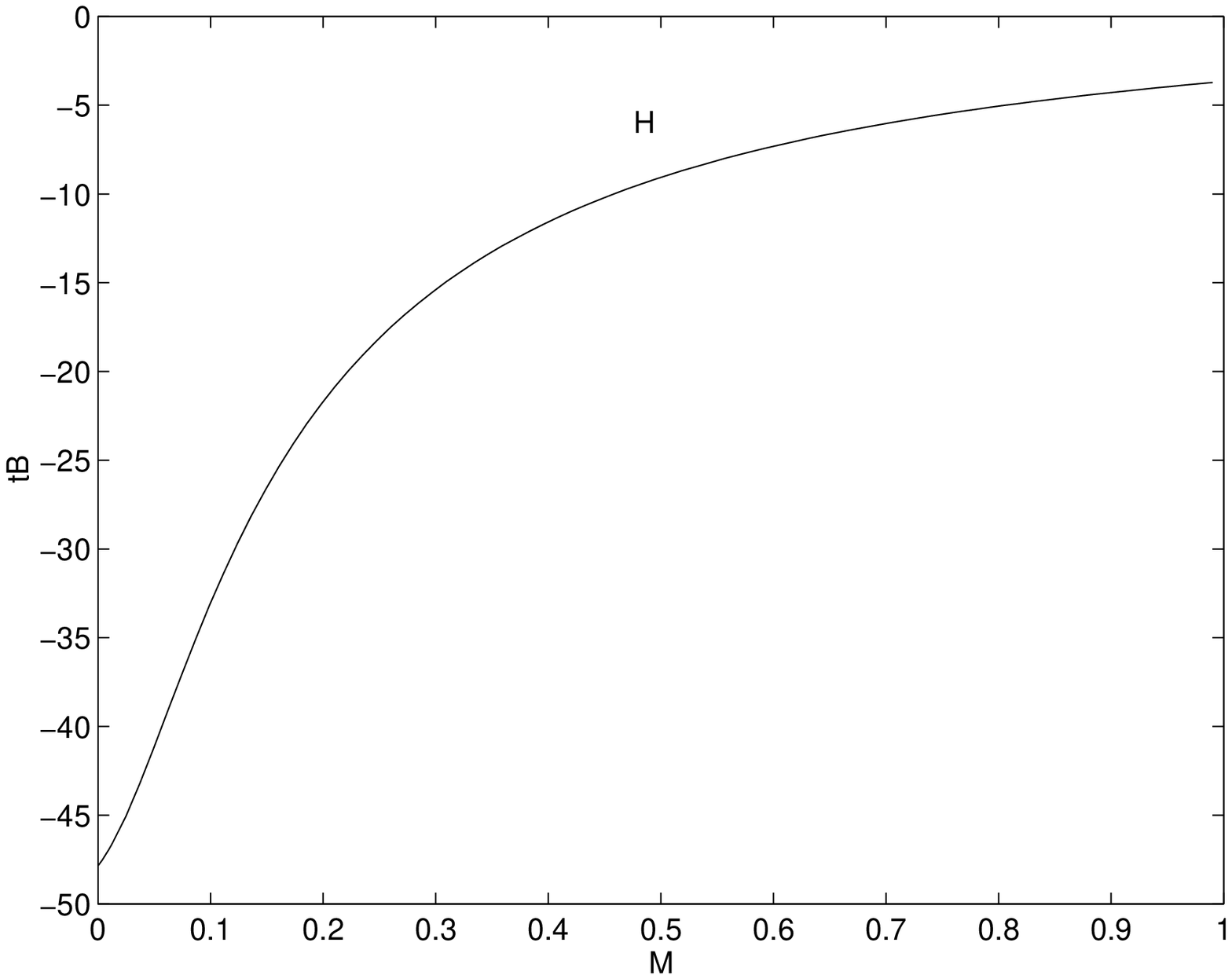}
 \caption{
 \label{V3D20tBMfig}
 \footnotesize
 $t_B(M)$ for the void run $V$i3$\rho$f20.
 }
 }
 \end{center}
 \end{figure*}

\begin{figure*}[h]
 \begin{center}
 \parbox{14cm}{
 \includegraphics[scale = 0.7]{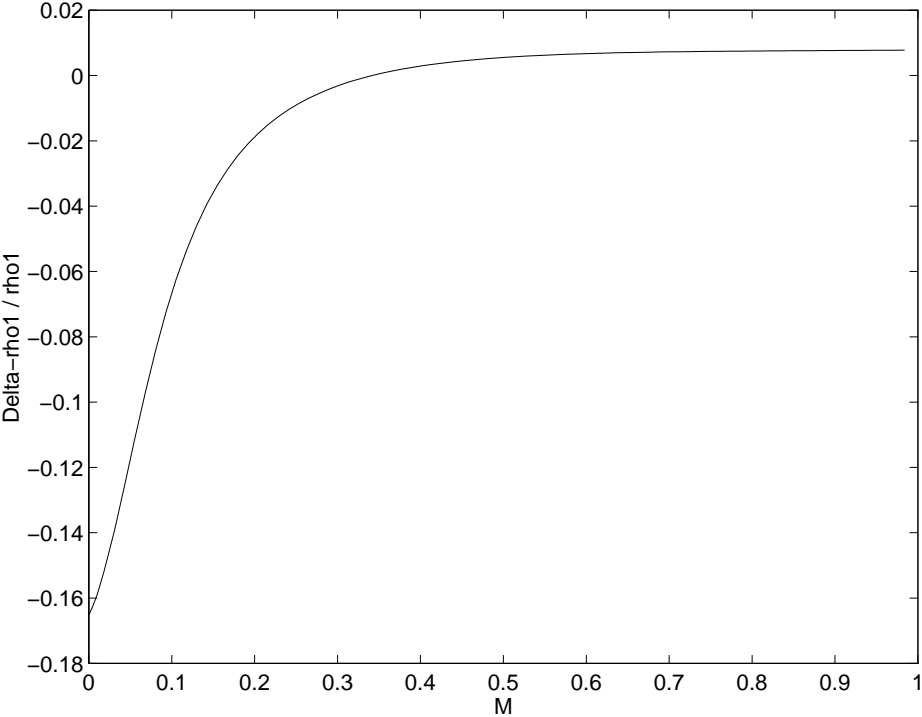}
 \caption{
 \label{V3D20D1Mfig}
 \footnotesize
 $\rho_1(M)$ fluctuation for the void run $V$i3$\rho$f20.  The amplitude
of this fluctuation is too large, so the initial velocity profiles need
 fine-tuninng.
 }
 }
 \end{center}
 \end{figure*}

\begin{figure*}[h]
 \begin{center}
 \parbox{14cm}{
 \includegraphics[scale = 0.7]{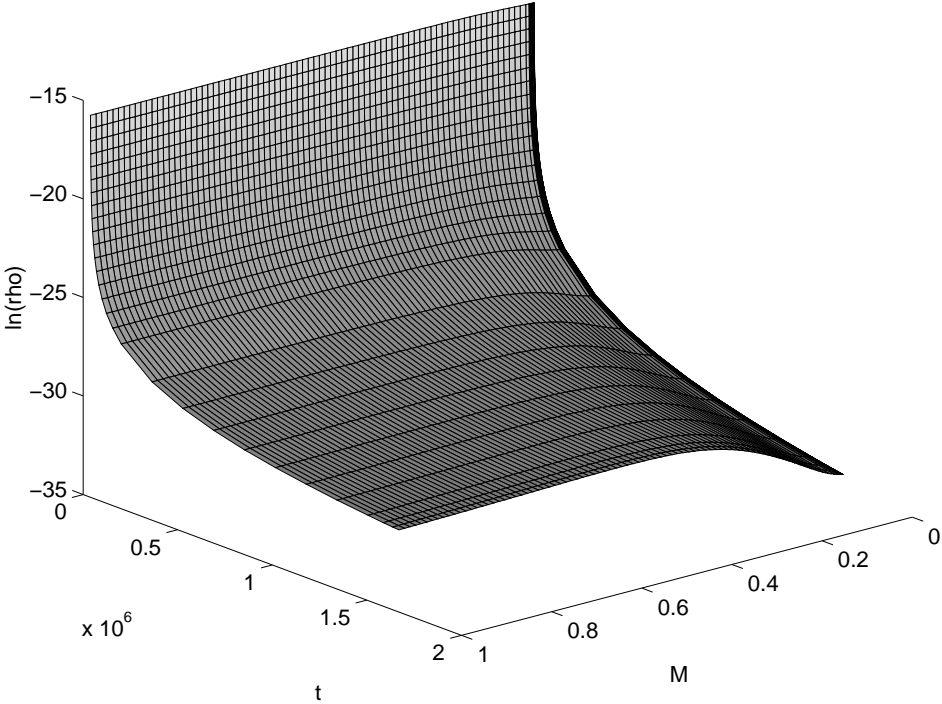}
 \caption{
 \label{V3D20Dsffig}
 \footnotesize
 $\rho(t, M)$ evolution for the void run $V$i3$\rho$f20.  Note that
$\rho_2(M)$ is the same profile as $\rho_2(R)$ shown in figure
\ref{VoidProfCompFig}, though it looks quite different.  The reason is that
there is very little mass in the void interior,
so a large increase in distance corresponds to a small increase in mass.
 }
 }
 \end{center}
 \end{figure*}

\begin{figure*}[h]
 \hspace*{-10mm}
 \parbox{17.5cm}{
 \hspace*{-90mm}
 \includegraphics[scale = 0.45]{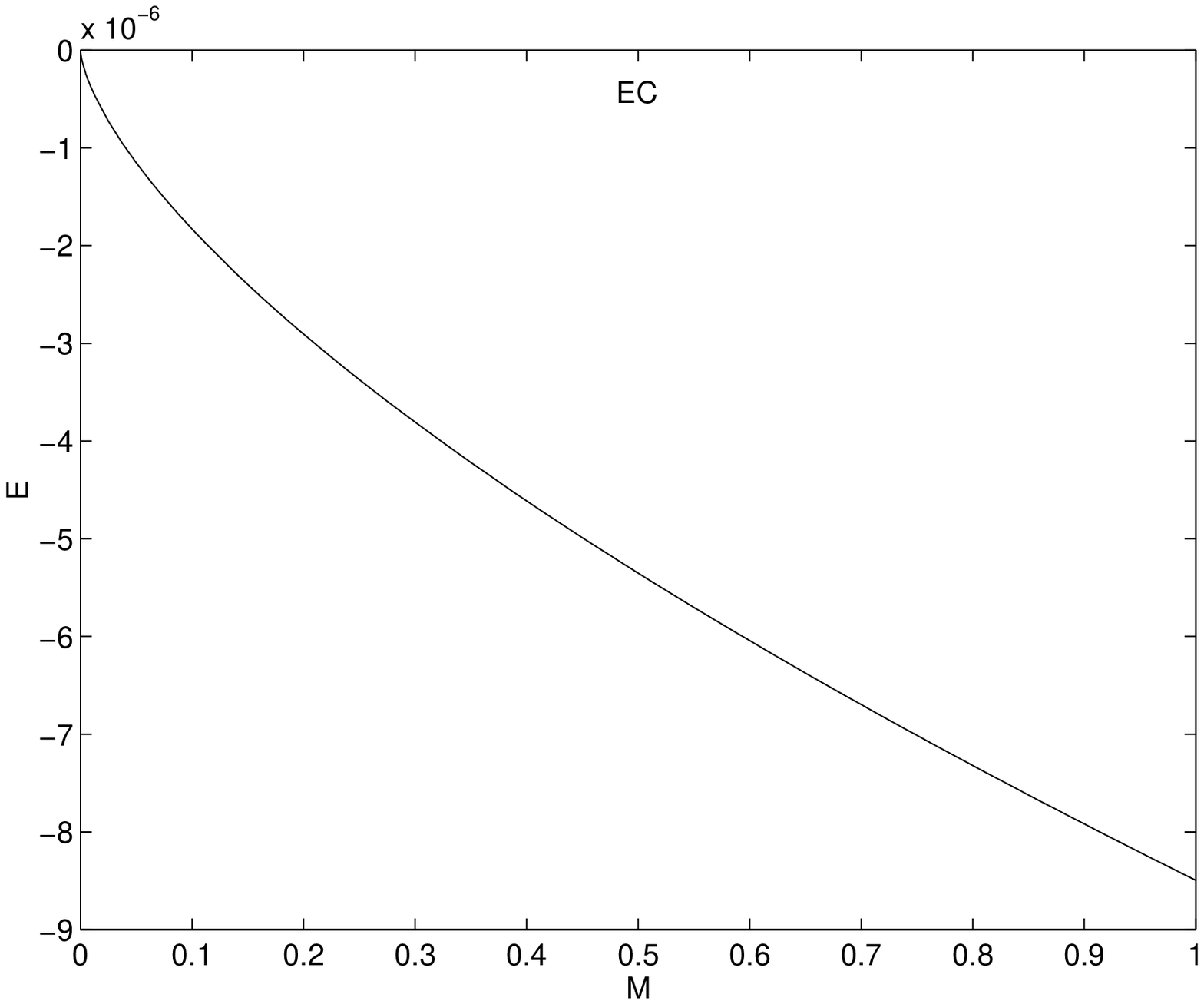}
 ${}$ \\[-62.5mm]
 \hspace*{90mm}
 \includegraphics[scale = 0.45]{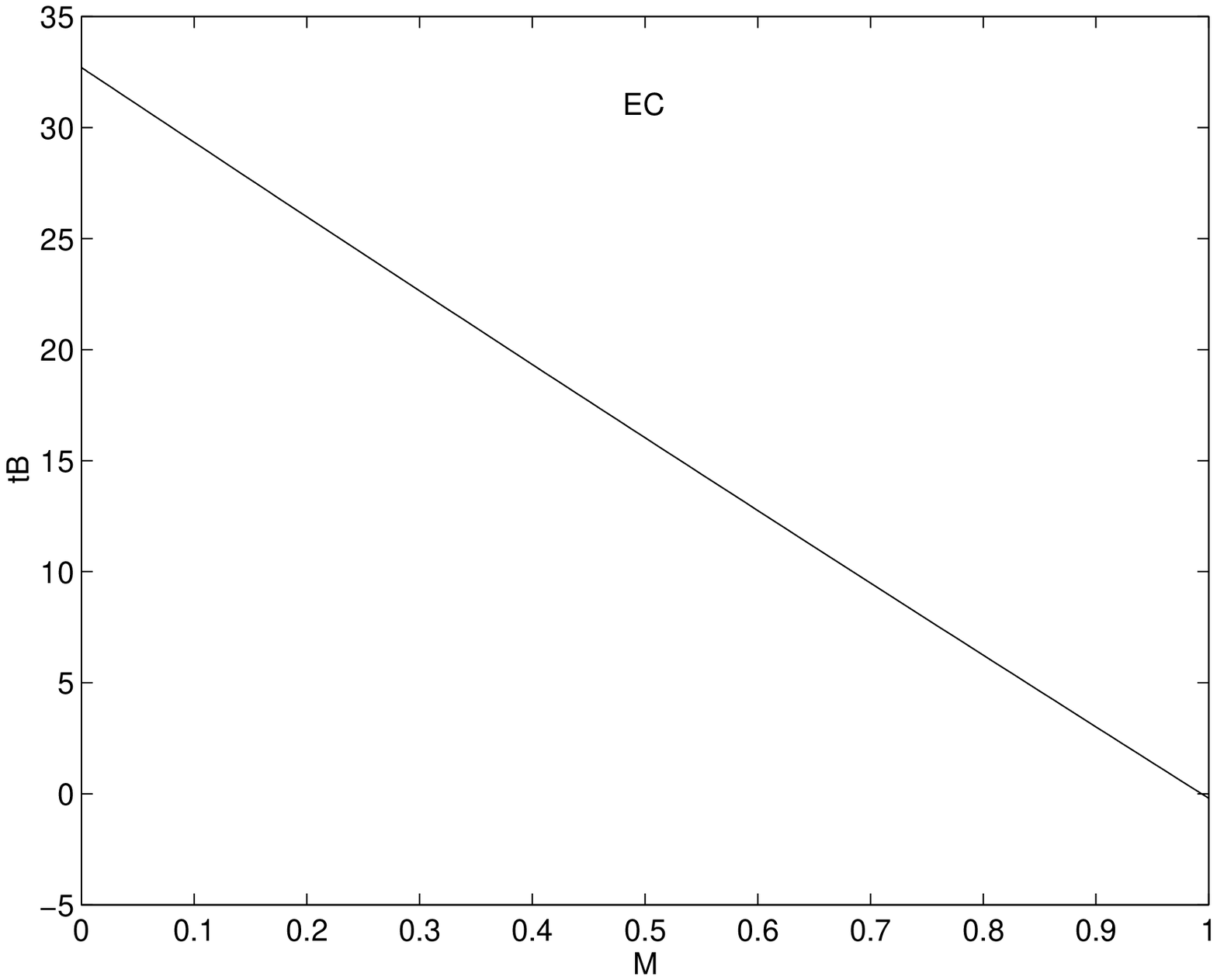}
 ${}$ \\[6mm]
 \hspace*{-92mm}
 \includegraphics[scale = 0.45]{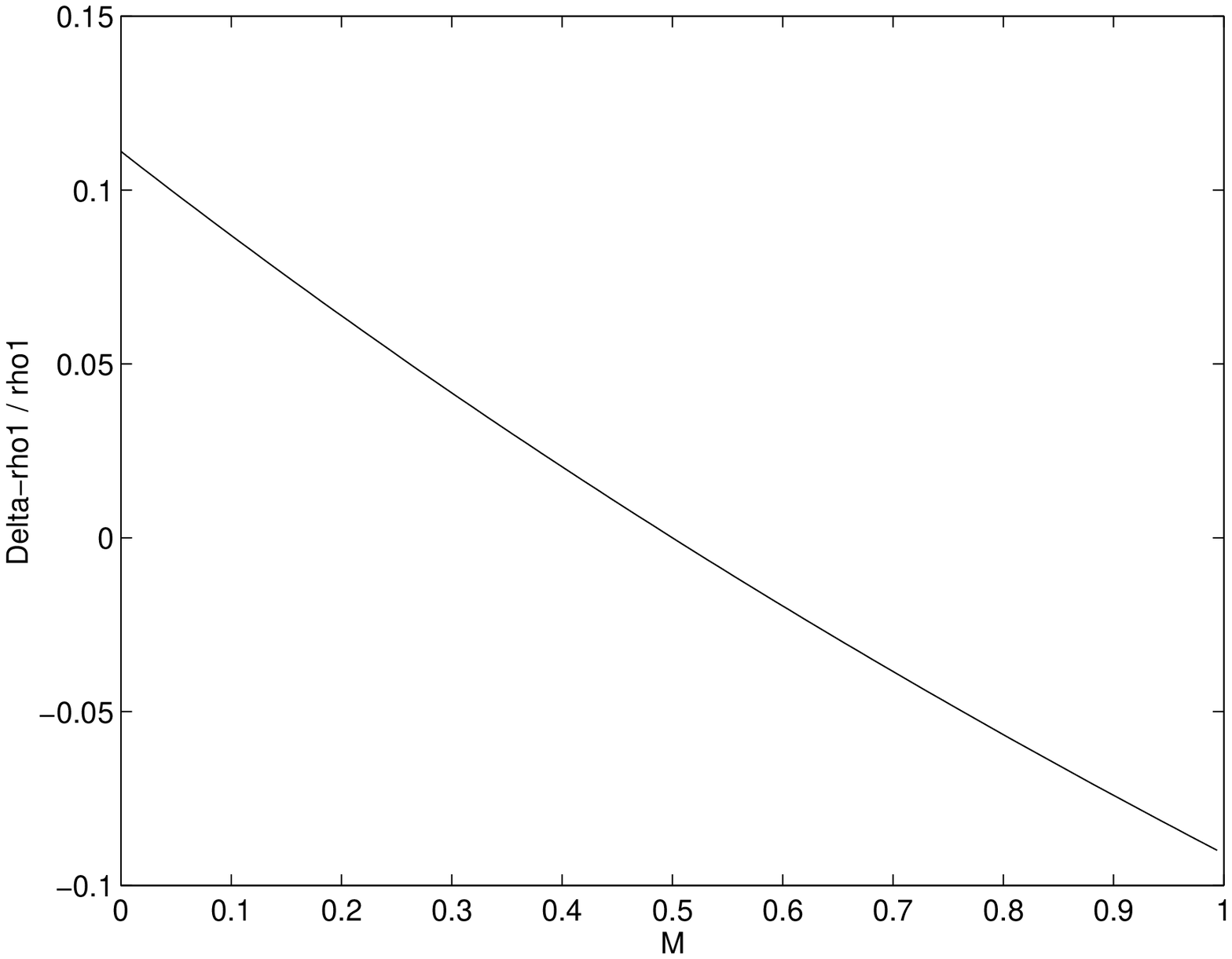}
 ${}$ \\[-65mm]
 \hspace*{88mm}
 \includegraphics[scale = 0.45]{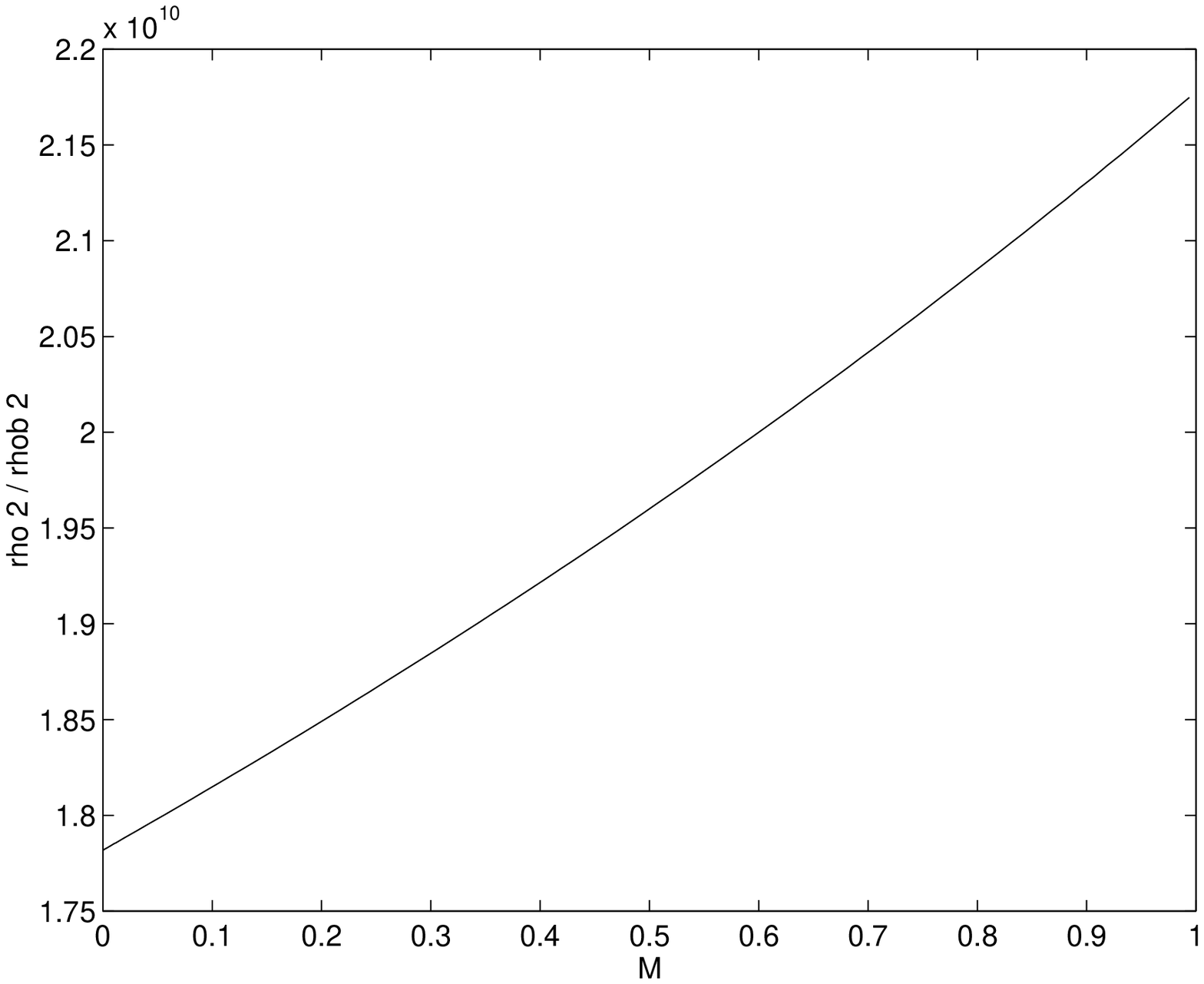}
 \caption{
 \label{D3D21fig}
 \footnotesize
 $E(M)$, $t_B(M)$ and the $\rho_1(M)$ and $\rho_2(M)$ fluctuations
for run $\rho$i3$\rho$f21.  The densities $\rho_1$ \& $\rho_2$ at times $t_1$ \&
$t_2$ are of comparable magnitude (which makes $\rho_2$ around $2 \times
10^{10}$ times the parabolic background at $t_1$).  In particular, $\rho_1 >
\rho_2$ at the centre, $M = 0$, and $\rho_2 > \rho_1$ at the edge, $M = 1$.
 }
 }
 \end{figure*}

 \begin{itemize}

 \item
The primary Abell cluster forming run, $V$i4$\rho$f16 (fig. \ref{V4D16fig}), 
used an improved present day density profile based on observational results, 
and also started from an initial velocity profile that was expanding less fast 
than the background at the centre, which is what one would expect to evolve 
into a condensation, see figs. \ref{VelProfCompFig} and \ref{V4D16fig}.  Though 
an Abell cluster model was generated in Paper I, the final profile was not so 
carefully chosen, and it was evolved from an initial density profile.  However, 
the evolution of $\rho$, $R$ and $R_{,t}$ is visually very similar to that 
given in paper I.  The main differences are at $t_1$. Several runs with 
different initial velocity profiles gave quite similar results --- with only 
$\rho_1$ and $t_B$ changing significantly.

 \item
The primary void forming run, $V$i3$\rho$f20 (figs. \ref{V3D20EMfig},
\ref{V3D20tBMfig}, \ref{V3D20D1Mfig}, \ref{V3D20Dsffig}), used an initial state 
that was expanding faster than the background at the centre, which is what one 
would expect to evolve into a low density region.  Using $\rho$f20 for the 
final density profile was particularly successful, as the required density 
profile develops smoothly, and the density is still decreasing at every point 
even at the present time, so there is no sign of imminent shell crossings.

   It is evident from the plots of $E$ \& $t_B$ and the conditions for no
shell crossings \cite{HeLa1985}, that there must be shell crossing at some
time, but these are long before recombination, and well after the present
time.  A significant feature is the relatively low central density at the
initial time.  The evolution of voids in an L-T model has been extensively
discussed by Sato and coworkers, see a summary in Ref. \cite{Sato1984}.
They, too, found that it necessarily leads to a shell crossing at some
time.

   Our original attempt, using $\rho$18, created a shell crossing (where the
density diverges and then goes negative) so soon after $t_2$, that the
model evolution program could only generate smooth density profiles up to
$0.98 t_2$.

 \item
 The case of a pure density perturbation, $V$i0$\rho$f13, was investigated
by choosing a flat initial $b_1$ evolving to a cluster.  This shows what
initial density perturbation is needed at recombination to form the
cluster by today
 --- a central overdensity surrounded by an underdensity, of order $10^{-
3}$.

   Similarly, the case of a pure velocity perturbation, $\rho$i0$\rho$f13,
created by choosing a flat initial $\rho_1$ evolving to a cluster, shows
what initial velocity perturbation is needed
 --- a general underexpansion of order $10^{-4}$, decreasing outwards.

   The latter is contrasted with the pure velocity perturbation needed to
create a void in $\rho$i0$\rho$f20
 --- a general overexpansion of order $10^{-2}$, decreasing outwards. (See
fig. \ref{V0D13D0D13D0D20fig} for all 3 runs.)

   The rather large size of these perturbations tends to confirm that
perturbations in both density and velocity are needed.

 \item
 Our entire approach
 --- specifying profiles for an L-T model at 2 different times
 --- demonstrates that two L-T models with identical $\rho(M)$ [and
consequently $R(M)$] at a given time, or identical $R_{,t}(M)$ at a given
time, can have quite different evolutions.  This is shown by comparing
runs $\rho$i0$\rho$f13 \& $\rho$i0$\rho$f20 (both in fig
\ref{V0D13D0D13D0D20fig}); also by runs $V$i4$\rho$f16 \& $V$i4$\rho$f0;
and in a small way by the set of 4 runs that all end with $\rho$f0.  Thus,
knowledge of only the density (or velocity) profile at one time does not
determine a model's evolution at all.

 \item
 To make clearer the effect of varying the initial velocity profiles, a
set of runs was done with all 4 initial $V$ profiles and the homogeneous
final $\rho$ profile.  We found $t_B$ has almost the same shape as the
velocity perturbation $b_1$; $E$ is strongly influenced by $b_1$, except
that it starts from zero at $M = 0$, and returns close to zero at $M =
M_c$; and the $\rho_1$ perturbation is also strongly dictated by $b_1$,
but turns away from zero towards $M = M_c$.

 \item
 Run $\rho$i3$\rho$f21 (fig. \ref{D3D21fig}) demonstrates the density
profile inversion that \cite{MusHel01} proved was not only possible but
likely, and our method makes it easy to obtain such evolutions.

   This same run also shows a model can evolve from lower to higher density
in one region and higher to lower in another; and that our method can
handle such occurrences.

 \item
 For the two times we have been considering --- $t_1$ at recombination and
$t_2$ at the present day --- the shape of the resultant $E(M)$ in any
given run is dominated by the final density profile, while both initial
and final profiles have a noticeable effect on the resultant $t_B(M)$.

 \item
 We have found that, when choosing the initial velocity fluctuation with an
appropriate magnitude, it is quite difficult to keep the initial density
fluctuation (which is not chosen) small enough, and
 vice-versa.  The main run of Paper I, which imposed a final density
profile of similar scale to $\rho$f13, but less centrally concentrated,
and a $3 \times 10^{-5}$ density fluctuation at $t_1$, was quite
successful in this regard, having a resulting velocity variation that was
within $3 \times 10^{-5}$ over much of its range, only reaching $8 \times
10^{-4}$ in the outer regions that evolved into vacuum.  This sensitivity
to profile shape, and better way of choosing initial profiles could be
investigated further.

 \item
The compensation radius was close to a parabolic point in every run. This must 
in fact be the case to high approximation under the conditions we imposed. 
Since we are using a $k = 0$ Friedmann model as our reference or background 
model, then any L-T model which has the same $M$ within the same $R$ at the 
same $t - t_B$ {\it must} be locally the same (same $R(t,M_c)$) as the 
background, and therefore parabolic.  By definition, the compensation radius is 
where $M$ \& $R$ are the same as in the background, but in general our method 
does {\it not} ensure $t - t_B$ matches there. At the beginning of the 
calculation, we do not know $(t - t_B)$ yet, it is one of the results of the 
procedure. As explained is sec. \ref{mop}, the individual values of the time 
coordinate have no meaning.
 \footnote{Indeed, all physical quantities calculated are invariant under 
the transformation $t \to t +$ const.}
 Only the differences, $(t_2 - t_1)$, $(t_2 - t_B)$, $(t - t_B)$, etc, 
appear in the calculations and are physical parameters. Nevertheless, we 
assume that our $t_2$ equals $(t_2 - t_B)$ in the background. In 
consequence, $t_B = 0$ in the background, while in general $t_B(M_c) \neq 
0$ in the perturbation.  However, in practice, our use of tiny deviations 
from the background value at recombination $t_1$, ensures that the jump in 
$t_B$ can only be very slight.  This is what ensures a close-to-parabolic 
model at $M_c$. 

 \item   As a result of the above, in each of the numerical examples 
$E(M)$ has the same sign in the whole range. However, the method discussed 
here and in Paper I can be freely applied to models in which $E(M)$ 
changes sign at some $M$ 
 --- see Paper I for an example.

 \end{itemize}

 \section{Conclusions}

 \begin{itemize}

 \item
 It is traditional to think of spacetimes as being specified by functions
such as components of the metric and its derivative, or density and
velocity, given on some initial surface.  We have taken the new approach
of specifying one function on an initial surface and another on a final
surface, and demonstrated that, in the case of the L-T model, an evolution
from one to the other can be found.  This approach is better suited to
feeding the observational data into the model functions.

 \item
 In particular, the spherically symmetric evolution from a given initial
velocity (or density) profile to another given density (or velocity)
profile, in any of the four possible combinations, can always be found.
The solution is a determination of the arbitrary functions that
characterise the model.  While there is no guarantee that the resulting
model will be free of shell crossings, it is a simple matter to check the
arbitrary functions once calculated, and our experience has shown that
reasonable choices of the profiles keep any shell crossing well before the
initial time or well after the final time.

 \item
The solution method has been programmed, and a variety of runs have demonstrated
the practicality of the method. Models of a void and of an Abell cluster have 
been generated, and shown to have well behaved evolutions between the initial 
and final times.  It is noteworthy that the void model was able to concentrate 
matter into a wall without the formation of any shell crossings up to the 
present time and for well after it.

 \item
The effect on the solution of varying the initial and final profiles was
effectively illustrated. Some other previously known features of the L-T model, 
such as the possibility that clumps could evolve into voids, were highlighted 
in the results section by producing examples.

 \end{itemize}

 \appendix

 \renewcommand{\theequation}{\thesection.\arabic{equation}}

 \section{Extensions of Paper I
 --- the evolution between two density profiles.} \label{extend}

 \setcounter{equation}{0}

 \subsection{Models with Both Larger and Smaller Densities at Later Times}

It is clear from the method of Paper I that, for any $r$ value, whatever 
solution is found for $\rho_2 < \rho_1$, its time reverse ($t \rightarrow t_1 + 
t_2 - t$, $t_B \rightarrow t_1 + t_2 - t_B$) will solve the case when the 
density values are interchanged. But could both $\rho(t_1, r_a) > \rho(t_2, 
r_a)$ and $\rho(t_1, r_b) < \rho(t_2, r_b)$ occur at different $r$ values $r_a$ 
\& $r_b$ in the same model?  It is obvious that a model with adjacent expanding 
and collapsing hyperbolic regions must have severe shell crossings, unless they 
were separated by a neck \cite{Hell1987}. But the latter puts the two 
hyperbolic regions on either side of a wormhole, so they don't really 
communicate.  On the other hand it seems entirely possible that two worldlines 
in an elliptic region could display such behaviour, or even a hyperbolic region 
outside an elliptic region.

   In Paper I, the condition $\rho_2(M) \leq \rho_1(M)$ was imposed merely
to ensure $R_2 \geq R_1$, which allowed us to know there was always a bang in
the past, and hence a well defined $t_B$.  However, we note that this condition
did not exclude other equally unreasonable possibilities, such as an elliptic
region outside a hyperbolic region (see \cite{HeLa1985}).

   Thus, if we relax this requirement, and allow all $\rho_1 > 0$ and all
$\rho_2 >0$ profiles, we merely have to note which of $R_1$ \& $R_2$ is larger
before generating the "forwards" or "backwards" solution.

   We in any case have to check whether the conditions for no shell
crossings \cite{HeLa1985} are satisfied, or whether a regular maximum or
minimum has been reached, and this merely adds one more thing to check
 --- whether the bang and crunch functions are sufficiently continuous.

 \subsection{Including regions of zero density in the profiles}

 \subsubsection{Transient zeros in the density}

   Given the expression for the density, (\ref{rhoLT}) it is clear that if
$M_{,r} = 0$ for a particular $r$, but $E_{,r}$ and $(t_B)_{,r}$ are not zero,
the density will be zero there for all time.  Is it possible that $\rho = 0$ at
isolated events, or on non-comoving worldsheets?

   Assuming $M_{,r} \neq 0$ and $0 < R < \infty$, the only way for this to
happen would be for $R_{,r}$ to diverge without changing sign, while $M_{,r}$
remains finite.  We immediately see that divergent $R_{,r}$ makes $g_{rr}$
divergent, which suggests bad coordinates at the very least. However, for all
$E$ values we may write \cite{HelLak84, Barn1970}
 \begin{equation}
 \begin{split}
   \frac{R_{,r}}{R} =& \left( \frac{M_{,r}}{M} - \frac{E_{,r}}{E} \right) R \\
      &- \left[ (t_B)_{,r} + \left( \frac{M_{,r}}{M}
      - \frac{3 E_{,r}}{2 E} \right) (t - t_B)
      \right] R_{,t}
   \end{split}
 \end{equation}
 and it is clear that $t_{B,r} R_{,t}$ is only divergent on the bang or
crunch, while $(t - t_B) R_{,t}$ is zero there, because $R_{,t} \sim (t -
t_B)^{-1/3}$ at early times.  So one possibility is that $t_1$ intersects a
non-simultaneous bang while $t_2$ does not intersect the crunch at the same $r$
value, or vice versa.  This of course means the constant time surface cannot be
extended beyond this point, which is not really satisfactory.  The only other
possibility is for $E_{,r}$ or $t_{B,r}$ to diverge.%
 \footnote{If we allow $M_{,r}$ to diverge, then we need $R'$ to diverge
faster, and the argument is pretty much the same.}%
 But this would give a comoving zero of density, and a coordinate
transformation would make $E_{,r}$ and/or $t_{B,r}$ finite and $M_{,r}$ zero.

   Thus we conclude that, apart from the case of the initial or final
surface intersecting the bang or crunch, the density along a particle worldline
cannot be zero at one time and
 non-zero at another.  Thus zeros of density have to be permanent and
comoving.

 \subsubsection{$\rho(M) = 0$ at a single $M$ value}

   To look at the question of whether zero density at a single $M$ value can
be accommodated in the methods of Paper I, we first determine what kind of zero
we might expect in $\rho(M)$.  Working in flat 3-d space, let
 \begin{equation}
   \rho(R) \approx A |R - R_z|^n , ~~~~~~~~A, n \text{ const \&} > 0
 \end{equation}
 near a zero in $\rho$.  Then
 \begin{eqnarray}
   |M - M_z| & = & 4 \pi \int_{R_z}^R \rho(R') {R'}^2 \, dR'   \nonumber \\
             & = & 4 \pi A \left( \frac{|R - R_z|^{n+3}}{n + 3}
                  + \frac{2 R_z |R - R_z|^{n+2}}{n + 2} \right.\nonumber \\
             & & + \left.\frac{R_z^2 |R - R_z|^{n+1}}{n + 1} \right),
 \end{eqnarray}
 so, to lowest order
 \begin{eqnarray}
   |M - M_z| & \approx & 4 \pi A \, \frac{R_z^2 \, |R - R_z|^{n+1}}{n + 1},\\
   |R - R_z| & \approx & \left( \frac{(n + 1) |M - M_z|}{4 \pi A R_z^2}
                         \right)^{1/(n+1)},
 \end{eqnarray}
 which means
 \begin{equation}
   \rho(M) \approx A \left( \frac{(n + 1) |M - M_z|}{4 \pi A R_z^2}
                         \right)^{n/(n+1)},
 \end{equation}
 and we notice $n/(n + 1) < 1$ always.

   If we use this as an approximation near a point zero in the full curved
spacetime expression, we find
 \begin{equation}   \label{Rintegral}
 \begin{split}
   R^3 - R_z^3 &= \int_{M_z}^M \frac{6}{\kappa \rho(M')} \, dM' \\
               &= \frac{6}{\kappa B} (n + 1) |M - M_z|^{1/(n + 1)},
 \end{split}
 \end{equation}
 which is indeed well determined.  We thus conclude that no modification
of the Paper I method is needed in principle, though some extra coding might be
needed if the integration had to be done numerically.  (With all the profiles
tried so far, Maple did symbolic integrations to get $R_i^3$.)

 \subsubsection{$\rho = 0$ over an extended region}

   We have already seen that zeros in the density have to be permanent and
comoving, and this obviously applies to extended vacuum regions.  In any case,
spherically symmetric vacuum is Schwarzschild and must remain so.

   In order to allow the possibility that $\rho(t_i, r)$ is zero over a
finite range, the choice $\rho_i = \rho_i(M)$ must be abandoned as all points
in the zero density region will have the same mass, so $M$ will be a degenerate
coordinate there.

   Whatever alternative possibilities we choose must allow us to identify
the corresponding comoving coordinate points on each of the density profiles at
$t_1$ \& $t_2$.  Thus $R$ is not a suitable coordinate, as the two $R(t_i, r)$
are different.

   Thus we instead consider the following alternatives:
 \begin{enumerate}

 \item   Specify $\rho_i = \rho_i(r)$ and $M = M(r)$ \\
 This parametric version of choosing $\rho(M)$, allows $\rho_i(r) = 0$ and
$M(r) = const$ over some range of $r$, but actually is not sufficient, as we
still have no idea how much $R_i(r)$ increases over this range
 --- how much space there is between the
 non-vacuum regions.  In other words, in such a range, neither $\rho_i(r)$
nor $M(r)$ provides a useable definition of $r$ in terms of a physical quantity.

 \item   Specify $\rho_i = \rho_i(R_i)$ \\
 in which case
 \begin{equation}
   M_i(R) = (\kappa/2) \int_0^R R_i^2 \rho_i(R) dR.
 \end{equation}
Now where $\rho_i > 0$, $M(R)$ will be different for each $R$ value, and
therefore can be used to identify corresponding points on the two profiles.  We
can then use $R_2$ or $R_1$ (or something else) as the coordinate radius $r$.
Let us say we choose $r = R_2$.  But for regions where $\rho_i = 0$, $M(R)$ will
be degenerate.  In this case, we take advantage of the fact that in vacuum 
there are no matter particles, and so we have extra freedom in choosing the 
geodesics that constitute our constant $r$ paths.  Using a linear interpolation 
between the $R$ values at the two edges of the vacuum region at each of $t_1$ 
\& $t_2$ provides the obvious choice of corresponding points, i.e. at time 
$t_2$,
 \begin{equation}
   r = R_2,
 \end{equation}
 and at time $t_1$,
 \begin{widetext}
 \begin{equation}
   r =
   \left\{
   \begin{array}{ll}
      R_{2,\text{vac,min}}
         + \frac{(R_{1,\text{vac}} - R_{1,\text{vac,min}})
                 (R_{2,\text{vac,max}} - R_{2,\text{vac,min}})}
                {(R_{1,\text{vac,max}} - R_{1,\text{vac,min}})}
      & \text{~~in vacuum,} \\
      M_2^{-1}(M_1(R_1))
      & \text{~~in non-vacuum.}
   \end{array}
   \right.
 \end{equation}
 \end{widetext}

   The original procedure for extracting the functions $E$
and $t_B$, which actually used $R_1$ \& $R_2$, goes through with the only
change that functions of $r = M$ are now functions of $r = R_2$. In particular,
in vacuum regions, each worldline has different pairs of $R_1$ \& $R_2$ values,
but the same $M$ value.

 \end{enumerate}

 \section{The evolution of $R(t, M)$ vs the evolution of $\rho(t, M)$.}

 \label{greden}

 \setcounter{equation}{0}

   Let us take the relation between $R$ and $\rho$ in the most general case,
without any assumptions, eq. (\ref{RcubedInt}).

   This means that the value of $R$ at any $M$ depends on the values of $\rho$ in
the whole range $[M_0, M]$. Also the inverse relation, (\ref{rhoRM}), is
nonlocal --- to find $\rho(t, M)$ we need to know $R$ in an open neighbourhood
of the value of $M$.

   Assume now $\rho_2 < \rho_1$ over the whole of $[M_0, M]$.  Then
 \begin{equation}\label{intRvsrho}
 \begin{split}
 &R_2^3(M) - R_1^3(M) - \left({R_0}^3(t_2) - {R_0}^3(t_1)\right) \\
 &= \frac 6{\kappa} \int^M_{M_0}\left(\frac 1{\rho_2(u)}
- \frac 1{\rho_1(u)}\right) \text{d}u > 0.
 \end{split}
 \end{equation}
 Hence, $R_2^3(M) > R_1^3(M)$ if
 $$
{R_0}^3(t_2) - {R_0}^3(t_1) + \frac 6{\kappa} \int^M_{M_0}\left(\frac 
1{\rho_2(u)} - \frac 1{\rho_1(u)}\right) \text{d}u > 0.
 $$
 This will hold if we assume, as was done in paper I, $R_0(t_i) = 0, i = 1, 2$ in
addition to $\rho_2 < \rho_1$. The assumption $M_0 = 0$, i.e. the absence of a
mass-point at $R = 0$, also made in paper I, is not needed for this purpose.

   The converse implication is simply not true: $R(t_2, M) > R(t_1, M)$ for all $M
\in [M_0, M_1]$ does not imply anything for the relation between $\rho(t_2, M)$
and $\rho(t_1, M)$. This somewhat surprising fact is easy to understand on
physical grounds. $R(t_2, M) > R(t_1, M)$ for all $M \in [M_0, M_1]$ means that
every shell of constant $M = \tilde{M}$ has a larger radius at $t_2$ than it
had at $t_1$. However, the neighbouring shells may have moved closer to
$\tilde{M}$ at $t_2$ than they were at $t_1$. If they did, then a local
condensation around $\tilde{M}$ was created that may result in $\rho(t_2,
\tilde{M})$ being larger than $\rho(t_1, \tilde{M})$. This does not happen in
the Friedmann limit, where local condensations are excluded by the symmetry
assumptions.

   A sufficient condition for $\rho_2(M) < \rho_1(M), M \in [M_0, M_1]$ is
 $$
 \frac 1 {\left({R_1}^3\right),_M} > \frac 1 {\left({R_2}^3\right),_M} \qquad
\text{for all } M \in [M_0, M_1].
 $$
 If $R,_M > 0$ for all $M \in [M_0, M_1]$ at both $t_1$ and $t_2$ (i.e. there
are no shell-crossings in $[M_0, M_1]$), then this is equivalent to
${\left({R_2}^3\right),_M} > {\left({R_1}^3\right),_M}$ for all $M \in [M_0,
M_1]$. Incidentally, this implies $R_2 > R_1$ for all $M \in [M_0, M_1]$ if
$R_2(M_0) > R_1(M_0)$.

 \section{Calculations for section IV.}

 \setcounter{equation}{0}

 \subsection{Derivation of (\ref{condtEneg}) and (\ref{condtexp}).}
 \label{4bcalc}

In eq. (\ref{defPhiX}) we have
 \begin{align}
   \Phi_X(0) &= 0,   \nonumber \\
   \label{PhiXata2}
   \Phi_X\left(2/a_2\right) =& \pi
      + \frac{b_1\sqrt{2a_2}}{a_2{b_1}^2/2 + 1}
      - \arccos \left( \frac {a_2{b_1}^2/2 - 1} {a_2{b_1}^2/2 + 1} \right)
      \nonumber \\
      &- \left(2/a_2\right)^{3/2}\left(t_2 - t_1\right),
 \end{align}
 \begin{equation}\label{derPhiX}
 \begin{split}
   \Phi_{X,x} &= \sqrt{x} \left[ \frac {{a_2}^{3/2}} {\sqrt{2 - a_2x}}
      - \frac {2b_1} {\left({b_1}^2 + x\right)^2}
      - \frac 32 \left(t_2 - t_1\right)\right] \\
      &:= \sqrt{x} \lambda_X(x).
 \end{split}
 \end{equation}
 We see that $\Phi_{X,x}(0) = 0$ and that $\lim_{x \to 2/a_2} \Phi_{X,x} = +
\infty$. Now $\lambda_{X,x}(x) = \frac 12 {a_2}^{5/2}/(2 - a_2x)^{3/2} +
4b_1/({b_1}^2 + x)^3 > 0$ for all $x \in [0, 2/a_2]$. Hence, $\lambda_X$ can
have at most one zero and it will have one only if $\lambda_X(0) < 0$, i.e. if
eq. (\ref{condtEneg}) holds.

With (\ref{condtEneg}) fulfilled, $\Phi_{X,x} < 0$ in a neighbourhood of $x =
0$, $\Phi_{X,x} = 0$ at some $x = x_{\text{min}}$, and $\Phi_{X,x} \to +
\infty$ at $x \to 2/a_2$; the latter means that the tangent to $\Phi_X(x)$ at
$x = 2/a_2$ is vertical. Consequently, $\Phi_X(x)$ itself is a decreasing
function for $x \in [0, x_{\text{min}})$, and is negative in this range, then
it is increasing for $x \in (x_{\text{min}}, 2/a_2]$. It can thus have a zero
at any $x > 0$ if and only if $\Phi_X(2/a_2) \geq 0$, i.e. if (\ref{condtexp})
is fulfilled.

 \subsection{Derivation of (\ref{condtcoll}).} \label{4ccalc}

   The relevant properties of $\Phi_C(x)$ are
 \begin{align}
   \Phi_C(0) &= 2\pi > 0, \nonumber \\
   \label{PhiCat0}
   \Phi_C(2/a_2) =& \pi + \frac {b_1\sqrt{2a_2}}{a_2{b_1}^2/2 + 1}
      - \arccos \left( \frac {a_2{b_1}^2/2 - 1} {a_2{b_1}^2/2 + 1} \right)
      \nonumber \\
      &- \left(2/a_2\right)^{3/2}\left(t_2 - t_1\right),
 \end{align}
 \begin{equation} \label{derPhiC}
 \begin{split}
   \Phi_{C,x}(x) &= - \sqrt{x} \left[ \frac{{a_2}^{3/2}} {\sqrt{2 - a_2x}}
   + \frac {2b_1} {\left({b_1}^2 + x\right)^2}
   + \frac 32 \left(t_2 - t_1\right)\right] \\
   &< 0.
 \end{split}
 \end{equation}
 Since $\Phi_C(0) > 0$ and $\Phi_{C,x} < 0$ for all $x > 0$, the function
$\Phi_C(x)$ may have (only one) zero anywhere in its range if and only if
$\Phi_C(2/a_2) \leq 0$, which translates into the opposite of (\ref{condtexp}),
i.e. eq. (\ref{condtcoll}).

 \section{The velocity at $t_2$ must be smaller than at $t_1$ in the hyperbolic
evolution.}

 \label{grevelo}

 \setcounter{equation}{0}

   We now show that ${b_1}^2 > {b_2}^2$ must hold, as stated in subsection
\ref{condveltovel}, in consequence of $t_2 > t_1$ and
$\left.R_{,t}\right|_{t = t_1} > 0$, i.e. $b_1 > 0$.  (The assumption
$R_{,t}(t_1) > 0$ is hidden already in eqs. (\ref{EllEv})
 -- (\ref{HypEv}); for $R_{,t} < 0$, $(t - t_B)$ would have to be replaced by
$(t_B - t)$ in all three places.)

   Suppose that ${b_2}^2 > {b_1}^2$, so that $x < {b_1}^2$ applies. Then
 \begin{equation}\label{limchib1}
 \lim_{x \to {b_1}^2}\chi_H(x) = - \infty,
 \end{equation}
 and in addition $\chi_H(0) = 0$ (this second property does not depend on the
sign of ${b_2}^2 - {b_1}^2$). It follows that a second zero of $\chi_H(x)$ will
exist if there exists a subset of $R^1_+$ on which $\chi_H$ is an increasing
function. From Eq. (\ref{dchi}) we see that $\chi_{H,x}(0) = 0$, $\lim_{x \to
{b_1}^2} \chi_{H,x} = - \infty$. Hence, in order that $\chi_H(x)$ may have a
zero at $x > 0$, the $\mu_H(x)$, defined in (\ref{dchi}), must be positive
somewhere in the range $x > 0$. The derivative of $\mu_H(x)$ is
 \begin{equation}\label{dermu}
 \mu'_H(x) = \frac {4b_2} {\left({b_2}^2 - x\right)^3} - \frac {4 b_1}
{\left({b_1}^2 - x\right)^3},
 \end{equation}
 and so there must exist values of $x$ such that
 \begin{equation}\label{ineqb}
 b_2\left({b_1}^2 - x\right)^3 - b_1\left({b_2}^2 - x\right)^3 > 0.
 \end{equation}
 This is equivalent to
 \begin{equation}\label{ineqb2}
 \left[\left(b_1/b_2\right)^{1/3} - 1\right]x > {b_1}^{1/3}\left({b_2}^{5/3} -
{b_1}^{5/3}\right).
 \end{equation}
 However, this is a contradiction since, with $b_1 < b_2$ and $x > 0$, the
left-hand side is negative, and the right-hand side is positive.

   Since ${b_2}^2 > {b_1}^2$ has thus led to a contradiction, the opposite must
hold. This result is intuitively obvious (for dust, expansion must slow down
with time), so the above is in fact just a consistency check.

 \section{The velocity at $t_2$ must be smaller than at $t_1$ in the elliptic
evolution (final state still expanding).}

 \label{smallveloatt2}

 \setcounter{equation}{0}

   We will prove here that the $b_2$ in eq. (\ref{defchix}) must be smaller than
$b_1$ if, as was assumed earlier, $b_1 > 0$ and $t_2 > t_1$. The proof comes
out quite simply if we look at the function $\chi_X(x)$ defined by
(\ref{defchix}) as a function of the argument $b_2$. At $b_2 = b_1$ we have:
 \begin{equation}\label{chi1x}
 \chi_X(x) = \chi_1(x) = - x^{3/2}\left(t_2 - t_1\right),
 \end{equation}
 which is obviously negative for all values of $x > 0$. At $b_2 \to \infty$ we
have
 \begin{equation}\label{chi2x}
 \begin{split}
 &\chi_X(x) = \chi_2(x) = \\
 &\sqrt{1 - \left(\frac {{b_1}^2 - x} {{b_1}^2 +
x}\right)^2} - \arccos\left(\frac {{b_1}^2 - x} {{b_1}^2 + x}\right) -
x^{3/2}\left(t_2 - t_1\right).
 \end{split}
 \end{equation}
 We find that $\chi_2(0) = 0$, $\lim_{x \to \infty}\chi_2(x) = - \infty$ and the
derivative is $\chi'_2(x) = -2b_1\sqrt{x}/({b_1}^2 + x^2)^2 - (3/2)\sqrt{x}(t_2
- t_1)$ --- obviously negative for all $x > 0$. Therefore $\chi_2(x)$
itself is negative for all $x > 0$.

   Now we calculate:
 \begin{equation}\label{derchib2}
 \frac {\partial \chi_X(x)} {\partial b_2} = - \frac {4x^{3/2}} {\left({b_2}^2 +
x\right)^2},
 \end{equation}
 which is negative for all values of $b_2$ at every value of $x > 0$. Thus,
$\chi_X(x)$ is negative for all $x > 0$ at $b_2 = b_1$, negative for all $x >
0$ at $b_2 \to \infty$ and is a decreasing function of $b_2$ at every $b_2 \in
(b_1, \infty)$ for every $x > 0$. This means that $\chi_X(x)$ is negative at
every value of $x > 0$ for any $b_2 > b_1$, and so the equation $\chi_X(x) = 0$
has no solutions in $(0, \infty)$ when $b_2 > b_1$. \square

 \begin{acknowledgements}

   We thank Stanis{\l}aw Bajtlik, Andrzej So{\l}tan and Micha{\l} Cho\-do\-row\-ski
for useful instructions on the power spectrum of the CMB radiation and on the
mass-distribution in galaxy clusters. We also thank Peter Dunsby for guidance on
estimating velocity amplitudes at recombination. CH is most grateful to the N.
Copernicus Astronomical Centre of the Polish Academy of Sciences, and the
Institute for Theoretical Physics at Warsaw University, for generous provision
of facilities and support while the majority of this work was done, and
particularly to his host Andrzej Krasi\'nski for making the arrangements and
ensuring a very enjoyable and fulfilling visit.

 \end{acknowledgements}

 \newpage

 \setcounter{secnumdepth}{0}

 \section{Tables}

 \begin{longtable*}{l|l|l|l}
 \caption{
 \label{GeomU}
 \footnotesize
 {\bf Geometrical Units.}  The sets of geometrical units used for models of
Abell clusters and for of voids.} \\
   Object & $M_G$ & $L_G$ & $T_G$ \\
 \hline
 \hline
   Abell clusters & $10^{15}~M_\odot$ & $47.84$~parsecs & $156.0$~years \\
 \hline
   Voids & $55 \times 10^{15}~M_\odot$ & $2631$~parsecs & $8582$~years \\
 \hline
 \end{longtable*}

 \begin{longtable*}{l|l|l}
 \caption{
 \label{AbellData}
 \footnotesize
 {\bf Abell Cluster Data.}
 Data from \cite{MVFS1999}, giving the observed masses within 0.2~Mpc and
1~Mpc deduced from
 X-ray observations, which is used to fit the profiles in table
\ref{ACProfileTab} to observations.
 } \\
   Abell Cluster & $M$(0.2 Mpc) & $M$(1 Mpc) \\
 \hline
 \hline
   A 2199 & $(0.65 \pm 0.11) \times 10^{14}~M_\odot$
               & $(2.9 \pm 0.3) \times 10^{14}~M_\odot$ \\
 \hline
   A 496  & $(0.47 \pm 0.10) \times 10^{14}~M_\odot$
               & $(3.1 \pm 0.3) \times 10^{14}~M_\odot$ \\
 \hline
 \end{longtable*}

 \begin{longtable*}[h]{l|l|l|l|l}
 \caption{
 \label{ACProfileTab}
 \footnotesize
 {\bf Abell cluster profiles.}
 The equations and properties of the profiles shown in fig
\ref{ACProfCompFig}, that were used to model the present density
distribution in Abell clusters.  The parameters given are those that fit
the A2199 data \cite{MVFS1999}.  All
 non-dimensionless values are in geometric units
 --- see table \ref{GeomU}.} \\
   Profile & $\rho / \rho_b$ & Parameters & $R_c$ & $M_c$ \\
 \hline
 \hline
   $\rho$fUP
      & $\frac{\rho}{\rho_b} =
        \frac{\delta}{\frac{R}{R_s} \left( 1 + \frac{R}{R_s} \right)^2}$
      & \parbox[t]{4cm}{
        $\delta = 77440$,\\
        $R_s = 3457~L_G$
        }
      & $324600~L_G$ & $0.9442~M_G$ \\
 \hline
   $\rho$f13
      & $\frac{\rho}{\rho_b} = \frac{B_2 e^{-\sqrt{M}\; / \sigma_2}}
         {\left( \nu_2 + \frac{\sqrt{M}\;}{\mu_2} \right)
         \left( 1 + \frac{\sqrt{M}\;}{\mu_2}\right)}$
      & \parbox[t]{4cm}{
        $B_2 = 7774000$,\\
        $\mu_2 = 0.05304~\sqrt{M_G}\;$,\\
        $\sigma_2 = 2 \mu_2$, \\
        $\nu_2 = 5$
        }
      & $374500~L_G$ & $1.449~M_G$ \\
 \hline
   $\rho$f15
      & $\frac{\rho}{\rho_b} = \frac{B_2 e^{-M / \sigma_2}}
         {\left( \nu_2 + \frac{M}{\mu_2} \right)
         \left( 1 + \frac{M}{\mu_2} \right)}$
      & \parbox[t]{4cm}{
        $B_2 = 507500$,\\
        $\mu_2 = 0.04474~M_G$,\\
        $\sigma_2 = 2 \mu_2$, \\
        $\nu_2 = 5$
        }
      & $312400~L_G$ & $0.8416~M_G$ \\
 \hline
   $\rho$f16
      & $\frac{\rho}{\rho_b} = \frac{B_2}{1 + e^{M / \mu_2}}$
      & \parbox[t]{4cm}{
        $B_2 = 103400$,\\
        $\mu_2 = 0.04577~\sqrt{M_G}\;$
        }
      & $286600~L_G$ & $0.6500~M_G$ \\
 \hline
   $\rho$f17
      & $\frac{\rho}{\rho_b} = \frac{B_2}{1 + e^{\sqrt{M}\; / \mu_2}}$
      & \parbox[t]{4cm}{
        $B_2 = 498500$,\\
        $\mu_2 = 0.07144~M_G$
        }
      & $349800~L_G$ & $1.182~M_G$ \\
 \hline
 \end{longtable*}

 \begin{longtable*}[h]{l|l|l|l|l|l}
 \caption{
 \label{VoidProfileTab}
 \footnotesize
 {\bf Void Profiles.}  The equations and properties of the profiles shown
in fig \ref{VoidProfCompFig}, that were used to model the present density
distribution in voids.  The parameters were chosen to give a low density
region well below the background value (see $[\rho / \rho_b]_0$) within
60~Mpc radius.  All non-dimensionless values are in geometric units
 --- see table \ref{GeomU}.
 } \\
   Profile & $\rho / \rho_b$ & Parameters & $[\rho / \rho_b ]_0$
      & $R_c$ & $M_c$ \\
 \hline
 \hline
   $\rho$f18
      & $\frac{\rho}{\rho_b} = A_2 + B_2 e^{M / \mu_2}$
      & \parbox[t]{3.2cm}{
        $A_2 = 0.01$,\\
        $B_2 = 0.01$,\\
        $\mu_2 = 0.01429~M_G$,\\
        }
      & 0.01
      & $22804~L_G$ & $0.9902~M_G$ \\
 \hline
   $\rho$f20
      & $\frac{\rho}{\rho_b} =
        \frac{B_2^2 \mu_2^2 + M^2}{C_2 \mu_2^2 + (M - \mu_2)^2}$
      & \parbox[t]{3.2cm}{
        $B_2 = 0.3873$,\\
        $C_2 = 1$,\\
        $\mu_2 = 0.2475~M_G$,\\
        }
      & 0.075
      & $35850~L_G$ & $3.846~M_G$ \\
 \hline
 \end{longtable*}

 \newpage

 \begin{longtable*}[h]{l|l|l}
 \caption{
 \label{VelProfileTab}
 \footnotesize
 {\bf Initial Velocity Profiles.}
 The equations and properties of the profiles shown in fig
\ref{VelProfCompFig}, that were used to model the initial velocity
fluctuations at recombination.  The values of $M_c$ were determined by the
final density profile being used in each run.  All values are
dimensionless.
 } \\
   Profile & $R_{,t} / (R_{,t})_b$ & Parameters \\
 \hline
 \hline
   $V$i0
      & $\frac{R_{,t}}{(R_{,t})_b} = 1$
      & \\
 \hline
   $V$i3
      & $\frac{R_{,t}}{(R_{,t})_b} =
        1 + A_1 \left( 1 + \cos \left(\frac{\pi M}{M_c} \right) \right)$
      & \parbox[t]{4cm}{
        $A_1 = 5 \times 10^{-5}$, \\
        $M_c$ from $t_2$ data, \\
        } \\
 \hline
   $V$i4
      & $\frac{R_{,t}}{(R_{,t})_b} =
        1 - A_1 \left( 1 + \cos \left(\frac{\pi M}{M_c} \right) \right)$
      & \parbox[t]{4cm}{
        $A_1 = 5 \times 10^{-5}$, \\
        $M_c$ from $t_2$ data, \\
        } \\
 \hline
   $V$i5
      & $\frac{R_{,t}}{(R_{,t})_b} =
        1 + A_1 \cos \left(\frac{3 \pi M}{2 M_c} \right)$
      & \parbox[t]{4cm}{
        $A_1 = 1 \times 10^{-4}$, \\
        $M_c$ from $t_2$ data, \\
        } \\
 \hline
   $V$i6
      & $\frac{R_{,t}}{(R_{,t})_b} =
        1 - A_1 \cos \left(\frac{3 \pi M}{2 M_c} \right)$
      & \parbox[t]{4cm}{
        $A_1 = 1 \times 10^{-4}$, \\
        $M_c$ from $t_2$ data, \\
        } \\
 \hline
 \end{longtable*}

 \begin{longtable*}[h]{l|l|l|l}
 \caption{
 \label{InFnDenProfileTab}
 \footnotesize
 {\bf Initial and Final Density Profiles.}
 The equations and properties of the profiles that were used to model the
initial density fluctuations at recombination and the final structure.
 } \\
   Pro- & & & \\
   file & $\rho / \rho_b$ & Parameters & Description \\
 \hline
 \hline
   $\rho$f0
      & $\rho / \rho_b = 1$
      &
      & \parbox[t]{4cm}{
        Uniform background density at time 2.
        }\\
 \hline
   $\rho$i0
      & $\rho / \rho_b = 1$
      &
      & \parbox[t]{4cm}{
        Uniform background density at time 1.
        } \\
 \hline
   $\rho$i3
      & $\rho / \rho_b = \frac{1}{1 - A_1(1 - 2M/\mu_1)}$
      & \parbox[t]{3.5cm}{
        $A_1 = 0.1$, \\
        $\mu_1 = 1$.}
      & \parbox[t]{4cm}{
        Central overdensity plus surrounding underdensity of order $10^{-
1}$ in region of mass $10^{15}M_\odot$.  Used for the initial profile in
run $\rho$i3$\rho$f21.
        } \\
 \hline
   $\rho$i4
      & \parbox[t]{5.5cm}{
        $\rho / \rho_b = $ \\
        $ \left\{
        \begin{array}{lr}
        \multicolumn{2}{l}{
        1 + A_1 \left( 1 + \cos\left(\frac{100 \pi M}{M_c} \right)
           \right), ~~~ } \\
        & M < \frac{M_c}{100} \\
        1, & M \geq \frac{M_c}{100}
        \end{array}
        \right. $
        }
      & \parbox[t]{3.5cm}{
        $A_1 = 1.5 \times 10^{-5}$, \\
        }
      & \parbox[t]{4cm}{
        Overdensity of order $3 \times 10^{-5}$ in region of mass
$10^{13}M_\odot$, with exactly background density outside it.
        } \\
 \hline
   $\rho$f21
      & $\rho / \rho_b = \frac{1}{1 + A_2(1 - 2M/\mu_2)}
        \frac{\rho_{b1}}{\rho_{b2}}$
      & \parbox[t]{3.5cm}{
        $A_2 = 0.1$, \\
        $\mu_2 = 1$.
        }
      & \parbox[t]{4cm}{
        Central underdensity plus surrounding overdensity of order $10^{-
1}$ in region of mass $10^{15}M_\odot$, with a `background' density equal
to that at time $t_1$.  Used for the final profile in run
$\rho$i3$\rho$f21.
        } \\
 \hline
 \end{longtable*}

 \newpage

 \begin{longtable*}{l|l|p{4cm}|p{3.5cm}|p{4.1cm}}
 \caption{
 \label{RunList}
 \footnotesize
 {\bf Modelling Runs.}
 } \\
   Initial & Final   &             & Function (Fig & \\
   profile & profile & Description & no) list      & What it shows \\
 \hline
 \hline
   $\rho$i4 & $\rho$f17 & small localised density perturbation
$\rightarrow$ Abell cluster
      & \raggedright
        $E(M)$~(\ref{D4D17fig}),
        $t_B(M)$~(\ref{D4D17fig}),
        $R_{,t1}(M)$~(\ref{D4D17fig}),
        $\rho(t,M)$~(\ref{D4D17fig}),
      & Classic evolution of a small initial fluctuation with 100th final
mass into a condensation with realistic density profile at $t_2$. \\
 \hline
   $V$i4 & $\rho$f16 & Low central initial expansion rate $\rightarrow$
Abell cluster
      & \raggedright
        $E(M)$~(\ref{V4D16fig}),
        $t_B(M)$~(\ref{V4D16fig}),
        $\rho_1(M)$~(\ref{V4D16fig})
      & Evolution of low expansion region, to a realistic density
profile at $t_2$.  Graphs fairly similar to Paper I model. \\
 \hline
   $V$i0 & $\rho$f13 & Uniform initial expansion rate $\rightarrow$ Abell
cluster
      & \raggedright
        $E(M)$(\ref{V0D13D0D13D0D20fig}, left),
        $t_B(M)$(\ref{V0D13D0D13D0D20fig}, left),
        $\rho_1(M)$(\ref{V0D13D0D13D0D20fig}, left),
        $\rho(t,M)$(\ref{V0D13D0D13D0D20fig}, left),
      & Case of a pure density perturbation --- how big a perturbation is
needed for a present day structure? \\
 \hline
   $\rho$i0 & $\rho$f13 & Uniform initial density $\rightarrow$ Abell
cluster
      & \raggedright
        $E(M)$(\ref{V0D13D0D13D0D20fig}, centre),
        $t_B(M)$(\ref{V0D13D0D13D0D20fig}, centre),
        $R_{,t1}(M)$(\ref{V0D13D0D13D0D20fig}, centre),
        $\rho(t,M)$(\ref{V0D13D0D13D0D20fig}, centre),
      & Case of a pure velocity perturbation --- how big a perturbation is
needed for a present day structure? \\
 \hline
   $\rho$i0 & $\rho$f20 & Uniform initial density $\rightarrow$ Void
      & \raggedright
        $E(M)$(\ref{V0D13D0D13D0D20fig}, right),
        $t_B(M)$(\ref{V0D13D0D13D0D20fig}, right),
        $R_{,t1}(M)$(\ref{V0D13D0D13D0D20fig}, right),
        $\rho(t,M)$(\ref{V0D13D0D13D0D20fig}, right),
      & Together with $\rho$i0$\rho$f13, shows two models with the same
density profile at one time can have totally different evolutions. \\
 \hline
   $V$i3 & $\rho$f18 & High central initial expansion rate $\rightarrow$ Void
      & (None)
      & Not so good example of evolution to a void --- it is so close to a
shell crossing at $t_2$ that the evolution program fails on the last time
step.
\\
 \hline
   $V$i3 & $\rho$f20 & High central initial expansion rate $\rightarrow$ Void
      & \raggedright
        $E(M)$~(\ref{V3D20EMfig}),
        $t_B(M)$~(\ref{V3D20tBMfig}),
        $\rho_1(M)$~(\ref{V3D20D1Mfig}),
        $\rho(t,M)$~(\ref{V3D20Dsffig}),
      & Example of evolution to a void with good behaviour at $t_2$ and well
past it. The density perturbation at $t_1$ is too large by 4 orders. \\
 \hline
   $V$i3 & $\rho$f0
         & High central initial expansion rate $\rightarrow$ Uniform density
      & (None)
      & These 4 show very clearly the effect on $E(M)$ \& $t_B(M)$ of
varying initial density \\
 \cline{1-3}
   $V$i4 & $\rho$f0
         & Low central initial expansion rate $\rightarrow$ Uniform density
      &
      & or initial velocity.  Since the final density is uniform, the deviation of
$E(M)$ \& \\
 \cline{1-3}
   $V$i5 & $\rho$f0
         & High central \& low outer initial expansion rate $\rightarrow$
Uniform density
      &
      & $t_B(M)$ from their FLRW forms is
entirely due to non-uniform initial profiles. \\
 \cline{1-3}
   $V$i6 & $\rho$f0
         & Low central \& high outer initial expansion rate $\rightarrow$
Uniform density
      &
      & They also show a given density profile at one moment can have many
different evolutions. \\
 \hline
   $\rho$i3 & $\rho$f21 & Expanding underdensity $\rightarrow$ collapsing
overdensity
      & \raggedright
        $E(M)$~(\ref{D3D21fig}),
        $t_B(M)$~(\ref{D3D21fig}),
        $\rho_1(M)$~(\ref{D3D21fig}),
        $\rho_2(M)$~(\ref{D3D21fig}),
      & An example of $\rho_2 < \rho_1$ and within the same model $\rho_1
< \rho_2$.  It also provides an example of density profile inversion. \\
 \hline
 \end{longtable*}

 \newpage

 \section{Figures}

 \newpage

 \newpage

 \newpage

 \begin{figure*}[h]
 \hspace*{-10mm}
 \parbox{17.5cm}{
 \hspace*{-90mm}
 \includegraphics[scale = 0.45]{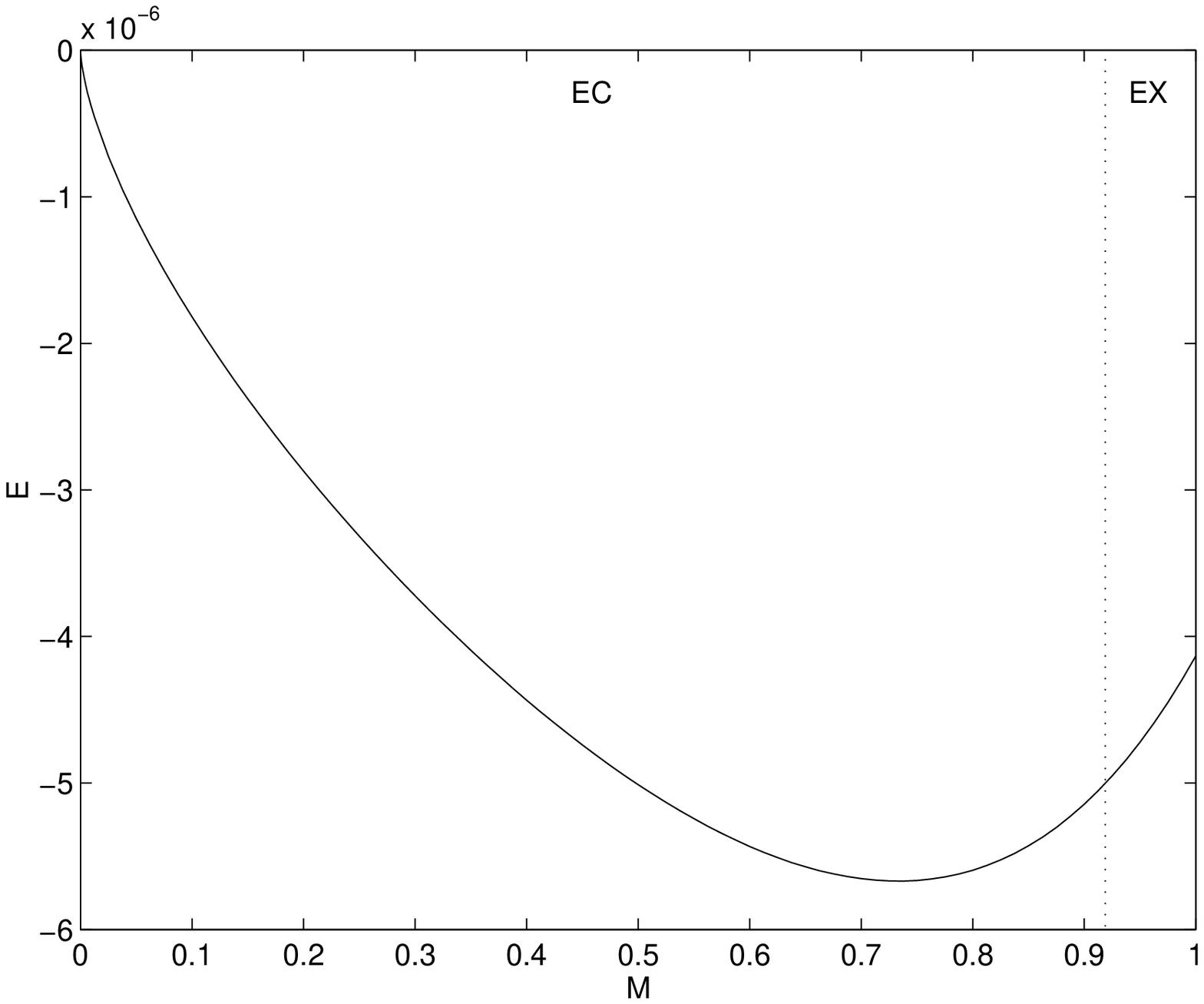}
 ${}$ \\[-62.5mm]
 \hspace*{90mm}
 \includegraphics[scale = 0.45]{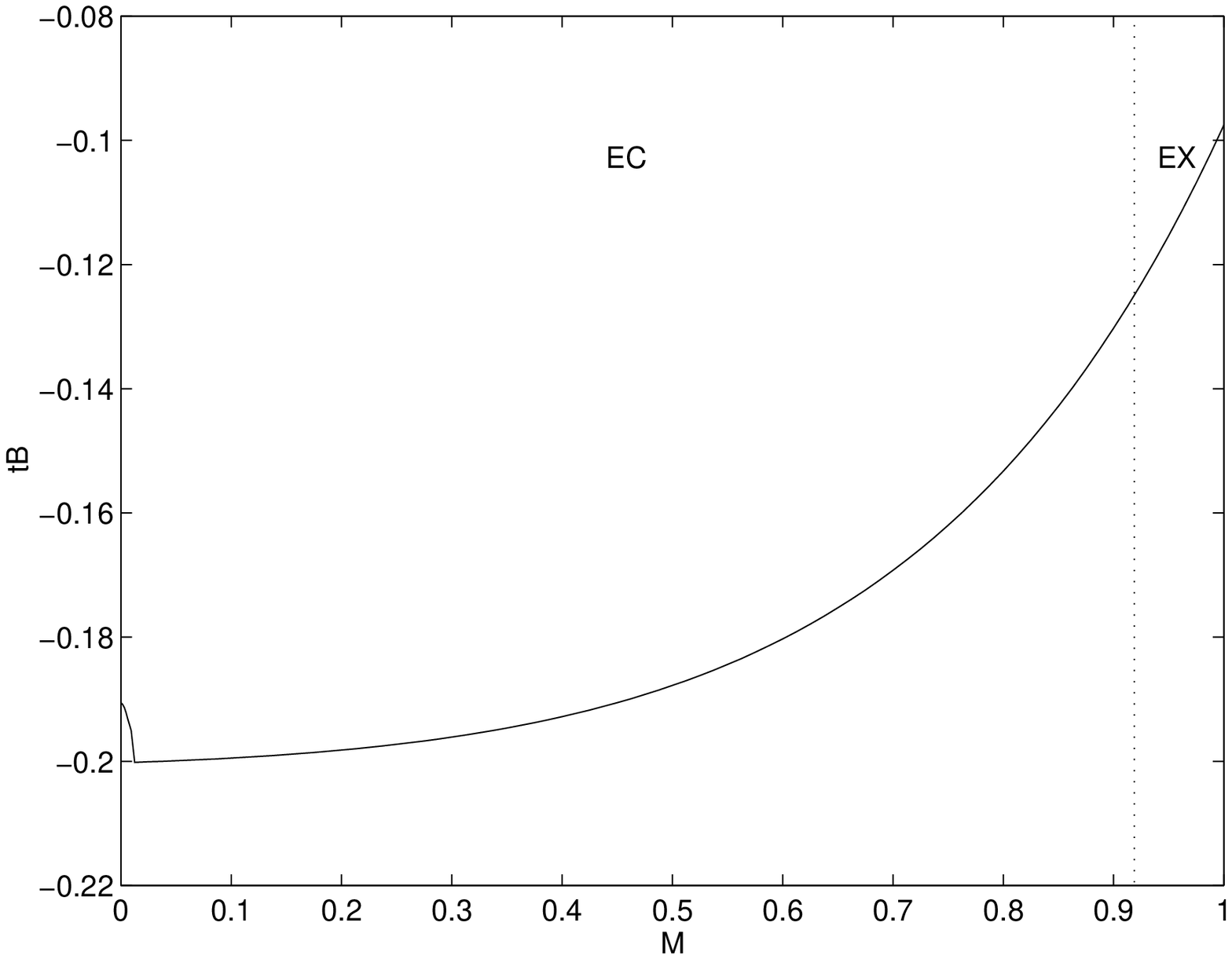}
 ${}$ \\[6mm]
 \hspace*{-90.5mm}
 \includegraphics[scale = 0.45]{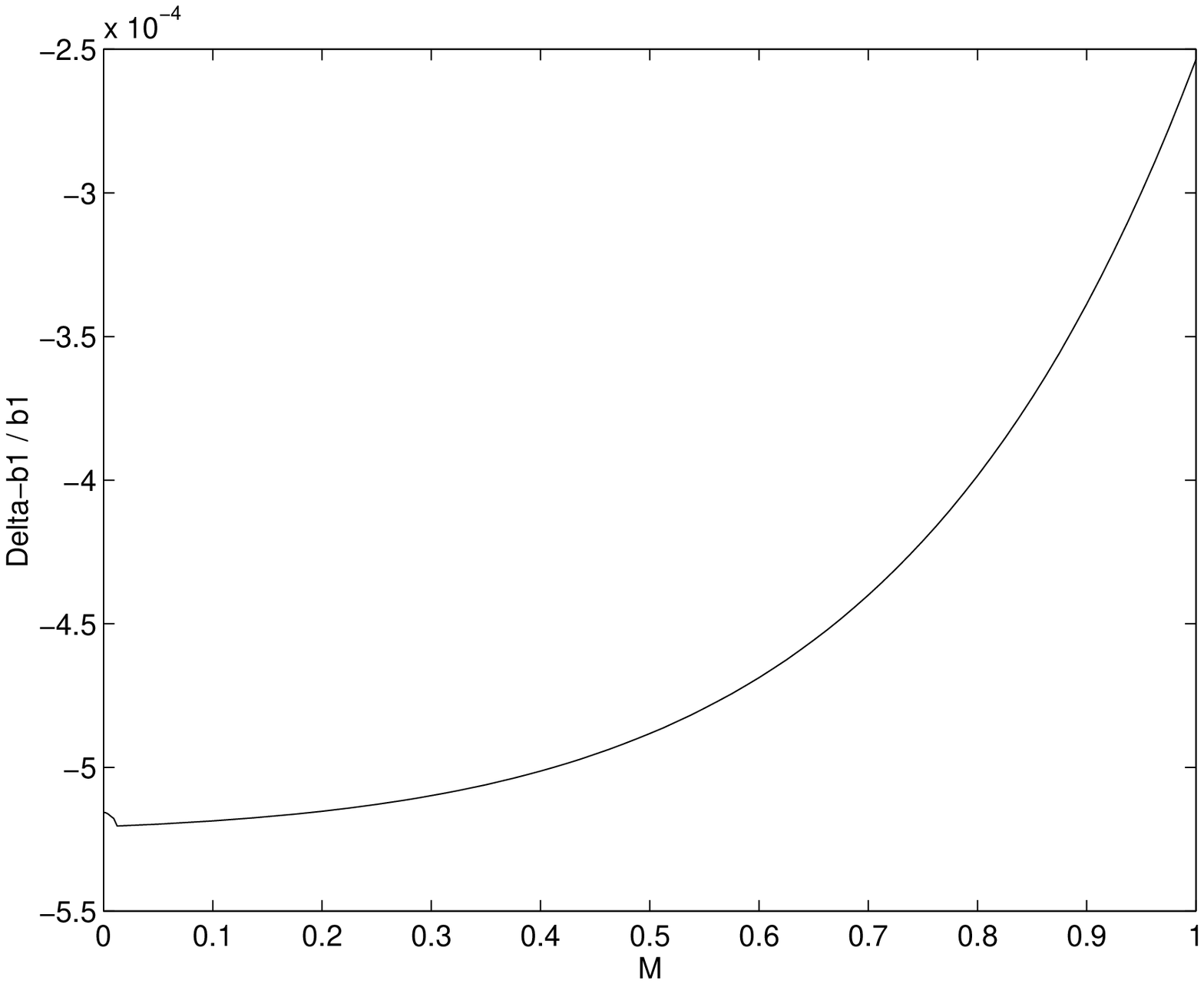}
 ${}$ \\[-63mm]
 \hspace*{90mm}
 \includegraphics[scale = 0.45]{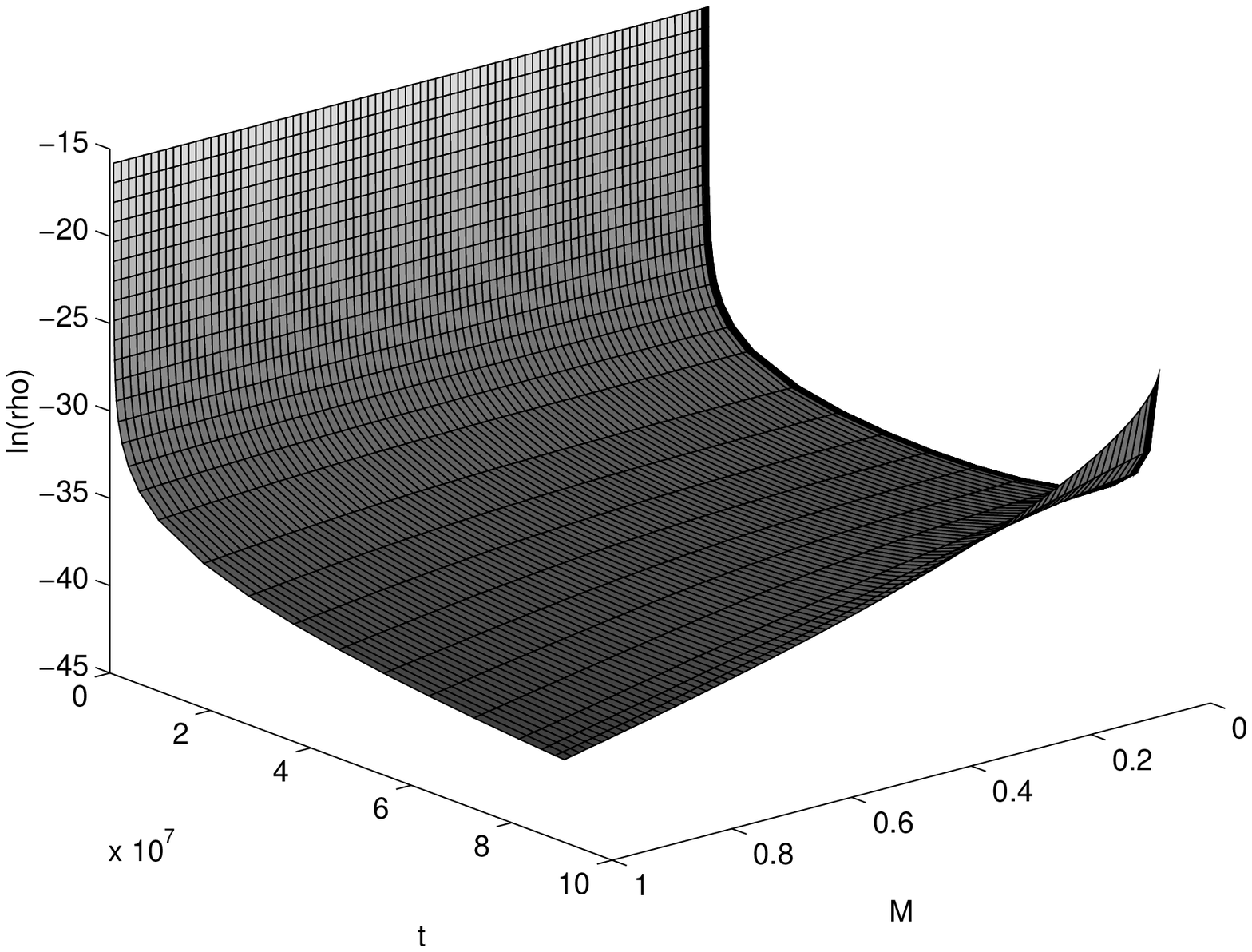}
 \caption{
 \label{D4D17fig}
 \footnotesize
 $E(M)$, $t_B(M)$, the $R_{,t1}(M)$ fluctuation, and the $\rho(t, M)$
evolution for run $\rho$i4$\rho$f17.  Notice the effect of the central
bump in $\rho_1(M)$ shows up in the $t_B$ and $R_{,t1}$ curves.  Although
$dt_B/dM > 0$ indicates there will be a shell crossing, the small
magnitude of $t_B$ ensures it will occur long before $t_1$, when the model
first becomes valid.
 }
 }
 \end{figure*}

 \newpage

 \newpage

 \newpage

 \newpage

 \newpage

 \newpage

 \newpage

\begin{thebibliography}{aa}

 \bibitem {KrHe2001} [Paper I] A. Krasi\'nski and C. Hellaby, {\it Phys. Rev.}
{\bf D65}, 023501 (2001).

 \bibitem {Lema1933} G. Lema\^{\i}tre, {\it Ann. Soc. Sci. Bruxelles} {\bf
A53}, 51 (1933); reprinted in {\it Gen. Rel. Grav.} {\bf 29}, 641 (1997).

 \bibitem {Tolm1934} R. C. Tolman, {\it Proc. Nat. Acad. Sci. USA} {\bf
20}, 169 (1934); reprinted in {\it Gen. Rel. Grav.} {\bf 29}, 935 (1997).

 \bibitem {Kras1997} A. Krasi\'nski, ``{\it Inhomogeneous Cosmological
Models}", Cambridge U P (1997), ISBN 0 521 48180 5.

 \bibitem {Hell1987} C. Hellaby, {\it Class. Q. Grav.} {\bf 4}, 635 (1987).

 \bibitem {HeLa1985} C Hellaby, K. Lake, {\it Astrophys. J.} {\bf 290}, 381
(1985) [+ erratum: {\it Astrophys. J.} {\bf 300}, 461 (1985)].

 \bibitem{HelLak84} C. Hellaby and K. Lake {\it Astrophys. J.} {\bf 282},
1-10 (1984).

 \bibitem {Barn1970} A. Barnes, {\it J. Phys.} {\bf A3}, 653 (1970).

 \bibitem{NaFW1996} J. F. Navarro, C. S. Frenk and S. D. M. White, {\it
Astrophys. J.} {\bf 462}, 563 (1996).

 \bibitem{Padman96} T. Padmanabhan, ``Cosmology and Astrophysics Through
Problems", Cambridge U P, 1996, ISBN 0-521-46783-7.

 \bibitem{WhiCoh02} M. White and J.D. Cohn {\tt astro-ph/0203120}

 \bibitem{HuDo02} W. Hu and S. Dodelson {\it Ann. Rev. Astron. Astrophys.},
(2002), {\tt astro-ph/0110414}.

 \bibitem{Wripa} E. L. Wright, at http://www.astro.ucla.edu
 /$\widetilde{\ \ }$wright/CMB-DT.html

 \bibitem{Hupa1} W. Hu, at http://background.uchicago.edu/
 $\widetilde{\ \ }$whu/physics/tour.html

 \bibitem{Hupa2} W. Hu, at http://background.uchicago.edu
 /$\widetilde{\ \ }$whu/araa/node4.html

 \bibitem{Duns97} P.K.S. Dunsby, {\it Class. Q. Grav.} {\bf 14}, 3391-3405 (1997).

 \bibitem{NaFW1995} J. F. Navarro, C. S. Frenk and S. D. M. White, {\it
Mon. Not. Roy. Astr. Soc.} {\bf 275}, 720 (1995).

 \bibitem{NaFW1997} J. F. Navarro, C. S. Frenk and S. D. M. White, {\it
Astrophys. J.} {\bf 490}, 493 (1997).

 \bibitem{MVFS1999} M. Markevitch, A. Vikhlinin, W.R. Forman and C.L.
Sarazin, {\it Astrophys. J.} {\bf 527}, 545-553 (1999).

 \bibitem{MapleURL} {\tt http://www.maplesoft.com}

 \bibitem{MATLABURL} {\tt http://www.mathworks.com}

 \bibitem{MusHel01} N. Mustapha and C. Hellaby, {\it Gen. Rel. Grav.}
{\bf 33}, 455-77 (2001).

 \bibitem{Sato1984} H. Sato, in: {\it General Relativity and Gravitation}.
Edited by B. Bertotti, F. de Felice, A. Pascolini. D. Reidel, Dordrecht 1984, p.
289.

 \end{thebibliography}
 \end{document}